\documentclass[aps,preprint,floatfix,epsfig]{revtex4}
	
\usepackage[english]{babel}
\usepackage{amsmath,amsfonts,amssymb}
\usepackage{multirow}
\usepackage{dcolumn}  

\usepackage[dvips]{graphicx}
\usepackage{epsfig}

\usepackage[T1]{fontenc}   
\usepackage[latin1]{inputenc}
\usepackage{multirow}

\usepackage[section]{placeins}

\usepackage{latexsym}
\usepackage[mathcal]{eucal}

\usepackage{units}

\usepackage{nicefrac}

\usepackage{titlesec}

\usepackage{hhline}

\usepackage{longtable}

\begin{document}

\title{Vortex-nucleus interaction in the inner crust of neutron stars}
\author{P. Avogadro$^{1,2}$, F. Barranco$^3$, R.A. Broglia$^{1,2,4}$,
and E. Vigezzi$^{2}$\\
$^1$Department of Physics, University
of Milan, Via Celoria 16, 20133, Milan, Italy\\
$^2$INFN, Sezione di Milano,
Milan, Italy\\ $^3$Escuela Tecnica Superior de Ingenieros,
Universidad de Sevilla, 41092 Camino de los Descubrimientos s/n,
Sevilla, Spain\\
$^4$The Niels Bohr Institute, University of
Copenhagen,
Blegdamsvej 17, 2100 Copenhagen, Denmark}
\bigskip
\bigskip
\begin{abstract}
The structure of a vortex in the inner crust of neutron stars is calculated 
within the framework of
quantum mean field theory taking into account
the interaction 
with the nuclei composing the Coulomb lattice. Making use of the   
results obtained with different nuclear interactions, 
the pinning energy, relevant in the study of glitches, is worked out.
Quantal size and density dependent effects are found to be important.
   
\end{abstract}
\maketitle

\section{Introduction}
According to  standard theoretical models, the inner crust of a neutron star,
where a lattice of nuclei is surrounded
by a sea of free neutrons and relativistic electrons,
lies in the  range of densities between 4 $\times 10^{11}$ and 1.6 $\times 10^{14}$ 
g/cm$^{3}$ ( 0.0014 $n_0$ and  0.57 $n_0$ , where $n_0$ is the nuclear saturation density,
$n_0= 0.16 $ fm$^{-3}$ = 2.8 10$^{14}$g/cm$^{3}$).
The first microscopic study of this system was carried out by Negele and
Vautherin. They determined, making use of a Skyrme-like functional of the
energy, the isotopic composition as well as the  distance between the
(spherical) nuclei forming the Coulomb lattice 
\cite{NegeleVaut}. 
Subsequent
studies found that for densities larger than about $n_0/3$, other shapes 
rather than spherical become favoured \cite{Pasta1}-\cite{Pasta3}. This is still a 
debated topic. In any case, in our study we shall limit ourselves to densities
$n<n_0/3$, where the assumption of spherical nuclei is considered to 
be safe. 
Another  assumption at the basis of Negele and Vautherin's  study was the 
Wigner-Seitz approximation, in which  one studies a single cell of the lattice.
Only very recently  calculations using band theory 
have been performed \cite{Chamel}. 

Studies of  neutron matter testify to the fact that  the system
is superfluid in a wide range of densities \cite{refgaps}.
This result implies that the rotation
of the star leads to the formation of vortices in the crust.
They are expected to form an array
whose average density  per unit area is given by the Onsager-Feynman formula:
\begin{equation}
\rho = 4 m \frac{\Omega}{\hbar},
\end{equation}
where $\Omega$ is the angular velocity of the star and $m$ is the 
nucleon mass. The above relation leads to an average (macroscopic)
distance between vortices
of the order of $10^{-3}$ cm for $\Omega \sim  1 $ ms$^{-1}$. 

The presence of impurities - represented, in the present case,
by the nuclei constituting the lattice - has a profound influence on 
pairing correlations and on vortex dynamics. In fact, momentum is much higher
inside the   nuclear volume   than in the surrounding neutron sea.
This fact  leads, as a rule, to a decrease of the pairing gap. 
Vortices, whose core is made out of normal matter, may thus find 
it energetically favourable to pin to the impurities, reducing 
the loss of pairing  energy.

One of the reasons for the interest in the interaction between vortex and nuclei
in the inner crust is due to its  relation with a possible  explanation for
the phenomenon of glitches, the sudden spinup of the rotation frequency observed
in many neutron stars \cite{Anderson75,vortexdyn}. 

Considerable information concerning glitches 
has been accumulated  over the years, and their properties are compatible with
the inner crust as their origin \cite{Link99}. However, a detailed comparison between
observations and theoretical models has never been attempted. This is partly due to
the fact that even the basic static interaction between
a single nucleus and a vortex has never been studied in detail with state-of-the-art
micoscopic theory, let alone the much more complex issue of vortex dynamics 
in the inner crust. Only semiclassical estimates of the 
pinning energy - the energy cost to build up  a vortex on a nucleus,
rather than  far from it -  have been available \cite{Epstein88,DonatiPizzo,Pizzonew}, while the first quantal
study of a vortex in uniform neutron matter was worked out  
only a few years ago \cite{Bulgac2003}.

In the following we present a detailed account of a microscopic 
study, based on Hartree-Fock-Bogoliubov (HFB) quantum mean field theory, 
of the structure of a vortex in the inner crust of a neutron star ( 
partial  accounts
of this work were published in refs. \cite{Avogadro_PRC,comex2,Catania}). 
In the present paper we aim at clarifying some basic qualitative features of the nucleus-vortex
interaction, which, to our knowledge, have not been considered until now (in particular quantal 
finite size effects) 
and
are absent from semiclassical models.
We are well aware that our 
work represents only a starting point for a comprehensive understanding of
vortices in the inner crust of neutron stars and of their possible 
relation to the phenomenon of  glitches, because of various
limitations  which remain at the basis of our approach. 
In particular:

- We treat pairing only at the mean field level. It is well known that 
medium polarization effects have a deep influence on pairing correlations
in homogeneous matter \cite{refgaps}. This is even more so in the case of the
inhomogeneous systems associated with each of the 
Wigner-Seitz cells studied by Negele and Vautherin \cite{Gori04}.

- We restrict ourselves to the region of spherical nuclei, 
while the largest contribution to the moment of inertia of the inner crust come
from the deepest layers, where the assumption of 
spherical nuclei breaks down, and other phases  (like the pasta phase) become important.

- We consider the interaction between a single nucleus and the vortex,
neglecting the influence of other nuclei building the Coulomb lattice.
The soundness of this approximation depends on the vortex radius - 
becoming unacceptable when the   vortex radius is of the order, 
or larger than, the radius of the Wigner-Seitz cell.

- We assume axial
symmetry for the nucleus-vortex system. 
As a consequence  we can only consider special spatial configurations,
and we are not able to calculate the pinning energy as a function of the distance
between the center of the nucleus and the vortex axis.

-  In our calculations 
we assume
a simple isotopic composition for the nuclei, using always
the same number of protons $Z=40$, corresponding to a proton shell closure, as we neglect
the spin-orbit term in the single-particle potential. 
A better calculation should consider the spin-orbit term of the mean field
and  use an updated version of the original Negele and Vautherin analysis
\cite{Baldo}, in which  the effects of pairing correlations are 
considered in determining the favoured number of protons and the 
lattice step, as a function of density.

 Finally, there are limitations which are intrinsic to the present status of    
the theory of nuclear interactions and nuclear structure, in particular concerning the 
choice of the  effective interaction used to produce the mean field. In fact, we  shall see that 
some of our results depend, sometimes even qualitatively, on 
particular features of the selected  nuclear interaction.

\section{Hartree-Fock-Bogoliubov equations}\label{chap:hfb}
We want to study the properties of nuclei embedded in a sea of superfluid 
free neutrons, as well
as the properties of particular excitations of such a system, namely 
vortices. The best available tool to carry out  such a study at the mean field 
level
is provided by the Hartree-Fock-Bogoliubov (HFB) theory. The associated 
equations are widely used in both nuclear
and solid state physics, where they are better known as 
the Bogoliubov-De Gennes equations
\cite{Ringschuck, Degennes}. 
The HFB equations allow, starting from two-body forces $V_{ph}$ and 
$V_{pair}$ (assumed to be of zero range) in the particle-hole channel and  
in the particle-particle channel respectively, to
determine in a self-consistent way the single-particle 
Hartree-Fock Hamiltonian $H = T + U^{HF}$ and
the pairing field $\Delta(\mathbf{x})$, as well as the amplitudes 
$u_{i},v_{i}$ 
and the energies $E_{i}$ of the quasi particles. For a given particle species
(neutrons or protons), the HFB equations can be written as  
\begin{equation}\label{HFB2}
\left\{
\begin{array}{l}
 ~~(H(\mathbf{x})-\lambda) u_{i}(\mathbf{x})+ \Delta(\mathbf{x}) 
 v_{i}(\mathbf{x}) = 
 E_{i} u_{i}(\mathbf{x})\\
~\Delta^*(\mathbf{x}) u_{i}(\mathbf{x})-(H(\mathbf{x})-\lambda)
v_{i}(\mathbf{x}) = 
E_{i} v_{i}(\mathbf{x})\,
\end{array}
\right.
\end{equation}
where $\lambda$ denotes the chemical potential, which
is fixed so as to yield the desired average number of particles.

\begin{figure}[h!]
\centering
\includegraphics[height=6.5cm,width=8cm]{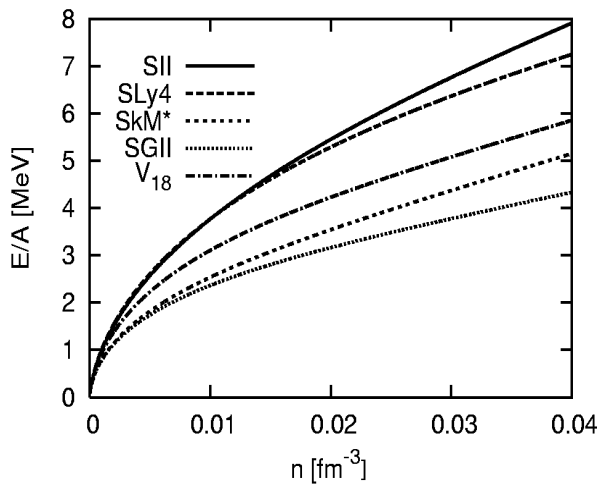}
\caption{(left) The energy per baryon calculated in neutron matter
with Br\"uckner theory
with the bare nucleon-nucleon Argonne interaction $V_{18}$  
(dot-dashed curve) \cite{Baldo04} is compared to the results obtained   
with the  Skyrme interactions SII  (solid curve), SLy4 (dashed curve), 
SkM* (short dashed curve ) and SGII (dotted curve). 
}
\label{fig:equation-state}
\end{figure}

\begin{figure}[h!]
\centering
\includegraphics[width=8cm]{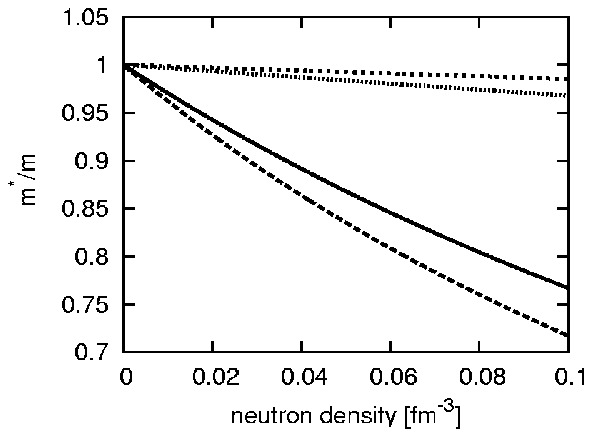}
\includegraphics[width=8cm]{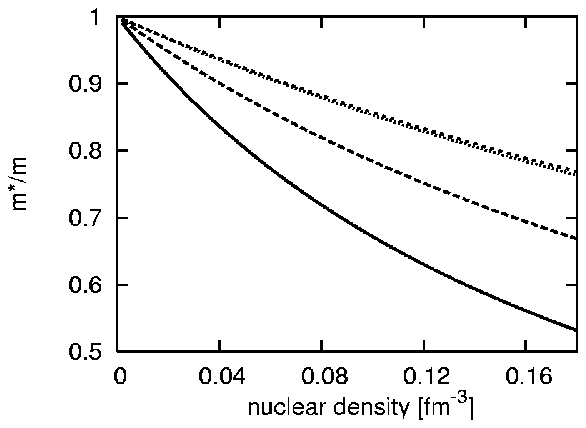}
\vsize=0.3cm
\caption{The effective mass in neutron matter (left panel) 
and in symmetric nuclear matter (right panel) for the different Skyrme interactions
used in this paper:  SII (solid curve), SLy4 (dashed curve), 
SkM* (short dashed curve),   and SGII (dotted curve). } 
\label{fig:effmass_Skyrme}
\end{figure}

While this framework is quite solid, the situation is very different 
concerning the choice of residual interactions acting in the particle-hole and 
in the particle-particle channel.
The situation 
is made particularly trying, because in the Wigner-Seitz cell we 
must deal both  with strongly asymmetric matter
(in the interior of the nuclei) and with low-density neutron matter. 
The parameters of the commonly
adopted  effective forces are usually derived fitting binding energies and other properties of
atomic nuclei  (mostly magic nuclei) in their ground state, as well as the binding energy of nuclear
matter. In some cases also results in neutron matter 
have been taken into account, but at  higher densities than those relevant  here.
On the other hand, many-body calculations should give reliable results for the equation of state of 
low-density neutron matter,
because the uncertainties related to three-body force play a minor role. The density dependence of 
neutron matter  energy resulting from such calculations has in fact been compared to  that of current
Skyrme forces (cf. e.g. \cite{Pasta1,Baldo04}).
We shall make use of the two  Skyrme forces SkM$^*$ and SLy4, 
which display a number of
attractive properties. 
The value of the energy per nucleon $E/A$, calculated with Br\"uckner theory 
based on a $G$-matrix derived by the bare nucleon-potential, and  
shown in Fig. \ref{fig:effmass_Skyrme},  lies in between the
results obtained with these two forces.   
A very relevant quantity  in our studies appears to be the effective mass 
associated with the effective forces. To investigate this point,
we have decided to show also  results obtained with two other forces:
SII and SGII, whose effective mass is close to that of SLy4 and of SkM$^*$ 
respectively (cf. Fig. \ref{fig:equation-state}).

\begin{figure}
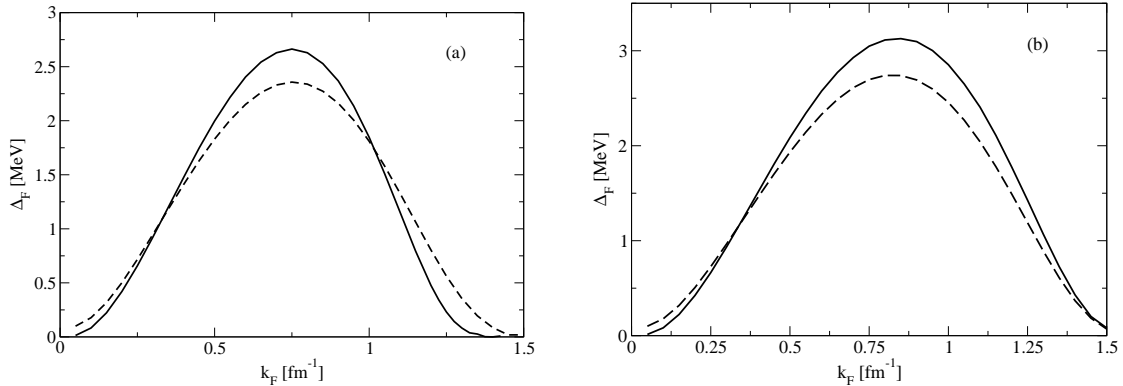

\begin{center}
\includegraphics[width=7cm]{delta_sly4_check.eps}
\hspace{5mm}
\includegraphics[width=7cm]{delta_skm_check.eps}
\end{center}
\caption{(a) The BCS pairing gap $\Delta_F$ in uniform neutron matter 
at the Fermi energy, 
is  shown as a function  of the Fermi momentum $k_F= (3 \pi^2 n)^{1/3}$,
calculated  with the zero-range,
density dependent pairing interaction (\ref{interpair}) (dashed curve) 
or with the Argonne  interaction (solid curve). In both cases    
the single-particle levels have been calculated with  the SLy4 interaction. 
(b) The same, but with single-particle levels calculated with the SkM*
interaction.}
\label{fig:delta_NM}
\end{figure}

Concerning the particle-particle channel, we shall use the density dependent 
contact force   
\begin{equation}\label{interpair}
V_{pair}(\mathbf{x}, \mathbf{x}')= V_0\cdot \left( 1- \eta 
\cdot \left( \frac{n(\textbf{x})}{0.08}\right) ^{\alpha}\right)\delta(\mathbf{x}- \mathbf{x}'),
\end{equation}
whose parameters $V_0 = $ -481 MeV fm$^{3}$,
$\eta= 0.7$ and $\alpha= 0.45$ have been chosen so as to reproduce the pairing
gap of neutron matter  obtained making use of a realistic nucleon-nucleon
interaction  \cite{Garrido}.
Because of the zero-range nature of the interaction, all values of  the 
momentum transferred in the nucleon-nucleon interaction processes are allowed. As a
consequence one needs  to introduce  a cutoff parameter $E_{cut}$. We have
implemented this condition in the basis used to calculate the pairing 
matrix elements (cf. Appendix A), taking only single-particle states
with energies  lower than $ E_{cut}=$ 60 MeV. In Fig. \ref{fig:delta_NM}  
we compare the pairing gaps at the Fermi energy  calculated in neutron matter 
with this force and with the bare Argonne nucleon-nucleon potential.
In Fig. \ref{fig:delta_NM}(a) and (b)  we use the single-particle levels calculated 
with the SLy4 and the SkM$^*$ interaction respectively: the latter yields higher values of the gap,
because of the higher effective mass and thus 
of the higher single-particle level density.  
The contact force reproduces rather well the density-dependence of the gap obtained  with the 
bare force,  although it underestimates the value at the maximum by 10-20\%.

The single-particle 
potential is obtained self-consistently
from the calculated density $n(\mathbf{x}) = \sum_i v_i(\mathbf{x})^2$.
We shall follow ref. \cite{Chabanat}, but we shall neglect the
spin-orbit $H_{so}$ and spin coupling
$H_{sg}$ terms in the Hamiltonian (cf. the comments at the end of Section IV B
nd Fig. \ref{fig:spin-orbit}).
Furthermore, we shall neglect the terms associated with the time-odd momentum density
(see for example \cite{Engel}),
which are associated with the presence of the vortex. In fact,
we have evaluated  their  contribution to the energy perturbatively, finding
that their contribution is negligible for the scope of this paper.

The 
pairing gap is obtained self-consistently
from the so-called abnormal density $ \kappa (\mathbf{x}) = \frac{1}{2} \sum_i 
u_i(\mathbf{x}) 
v_i^*(\mathbf{x})$, according to 
\begin{equation}
\Delta(\mathbf{x}) = - V_{pair}(\mathbf{x}) \kappa(\mathbf{x}).
\label{self2}
\end{equation}

\begin{figure}
\includegraphics[width=10cm]{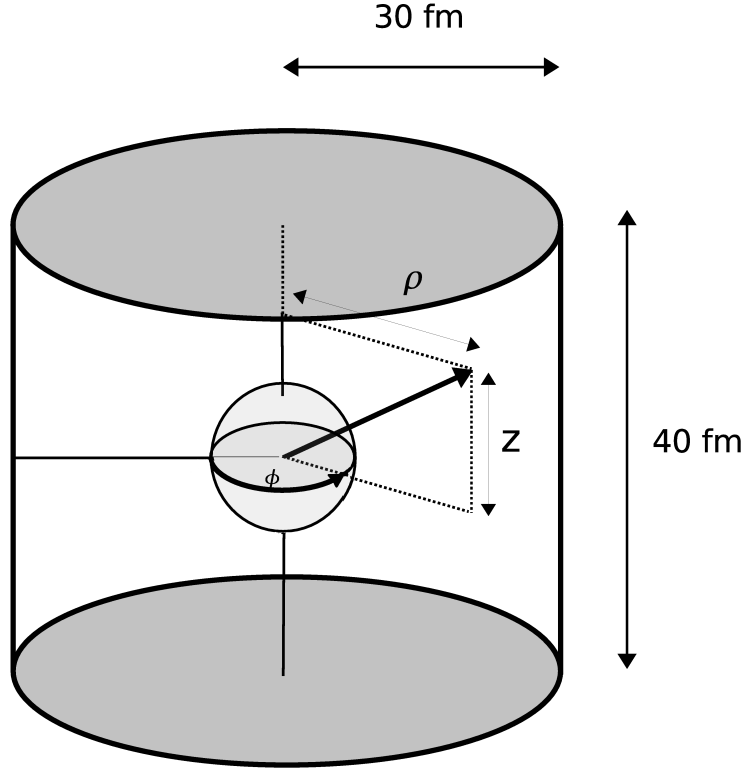}
	\caption{Geometrical features of the cylindrical cell used
	in most calculations of this paper.} 
\label{fig:cilindro}
\end{figure}

The HFB equations are solved enclosing the system  in a cylindrical box
of radius $\rho_{box}$ = 30 fm and height $h_{box}$  = 40 fm (cf. Fig. \ref{fig:cilindro}). 
We assume perfectly
reflecting walls, with single-particle wavefunctions which vanish
at the edges of the box (see the remarks on this issue in Appendix A).

In the calculations discussed in the following, we shall only consider axial symmetry, in which case 
the pairing field can be written as \cite{Degennes,Bohrmott62,Gigi}
\begin{equation}
 \Delta(\rho,z,\phi)=\Delta(\rho,z)e^{i\nu\phi},
\end{equation}
$\nu$ being the vortex number, which indicates the number of quanta of angular momentum carried by each 
Cooper pair.
In particular, if we set $\nu=0$ we obtain the usual  HFB equations (no vortex). Setting instead $\nu=1$, 
we can describe a vortex excitation, in which each Cooper pair carries one unit of angular momentum
along the $z-$axis. 
Experiments on superfluids indicate that it is  energetically more 
favorable to develop an array of $\nu=1$ vortices rather than a  few vortices 
carrying many quanta of angular momentum and in the following we shall apply this condition also in neutron 
stars.
When $\nu \neq 0$, one obtains a current given by  
\begin{equation}\label{velocita}
J(r,z,\phi) = -\frac{i\hbar}{m\rho }\sum_{i}v
_{i}^{*}(\rho,z,\phi)\frac{\partial v_{i}(\rho,z,\phi)}{\partial \phi},
\end{equation}
and an associated  velocity field $J/n$.
Aside from axial symmetry, the system has mirror 
symmetry respect to the $x-y$ plane (see Fig. \ref{fig:cilindro}), so that $\Delta(\rho,z)=\Delta(\rho,-z)$. 
Applying the parity operator ($\phi\rightarrow\phi+\pi$ and $z\rightarrow -z$) 
to the pairing field we obtain
\begin{equation}
  \hat P \Delta(\rho,z)e^{i\nu\phi}= e^{i\nu\pi} \Delta(\rho,z)e^{i\nu\phi}. 
\end{equation}

Therefore the pairing field is an eigenstate of the parity operator with eigenvalue
$ (-1)^{\nu}$.
This condition implies that the Cooper pairs involved in the $\nu=1$ excitation 
are constructed out  of single particle  levels of opposite parity.

Rather than solving the HFB equations directly in coordinate space, we expand them on a single-particle basis.
For details concerning the numerical procedure, we refer to Appendix A.

\section{The pinning energy}

Models that relate the glitch phenomenon to  the vortex dynamics in the 
inner crust, usually assume that the vortex lines which thread the neutron superfluid 
pin to the crust, being attracted or  repelled by the nuclei forming the Coulomb lattice.
If the vortex lines remain fixed, the velocity of the superfluid remains constant, 
while the velocity of the crust decreases due to the magnetic braking. At some critical
velocity difference, the Magnus force unpins the vortex lines, and angular momentum
is given back to the star, causing the glitch.

The determination of an 'optimal' vortex line depends, besides the lattice properties and the
vortex tension, on the vortex-nucleus interaction, or, more precisely, on the free energy
as a function of the distance between the nucleus and the vortex axis. 
As it was mentioned in the Introduction,  
although in the present paper we shall improve  in different
ways on previous models,
we shall only be able to 
compare  the two limiting configurations: a vortex whose axis passes through the center of the nucleus, 
and a vortex  placed far from the nucleus. 
\begin{figure}[h!]
\centering
\includegraphics[width=6cm,height=7cm]{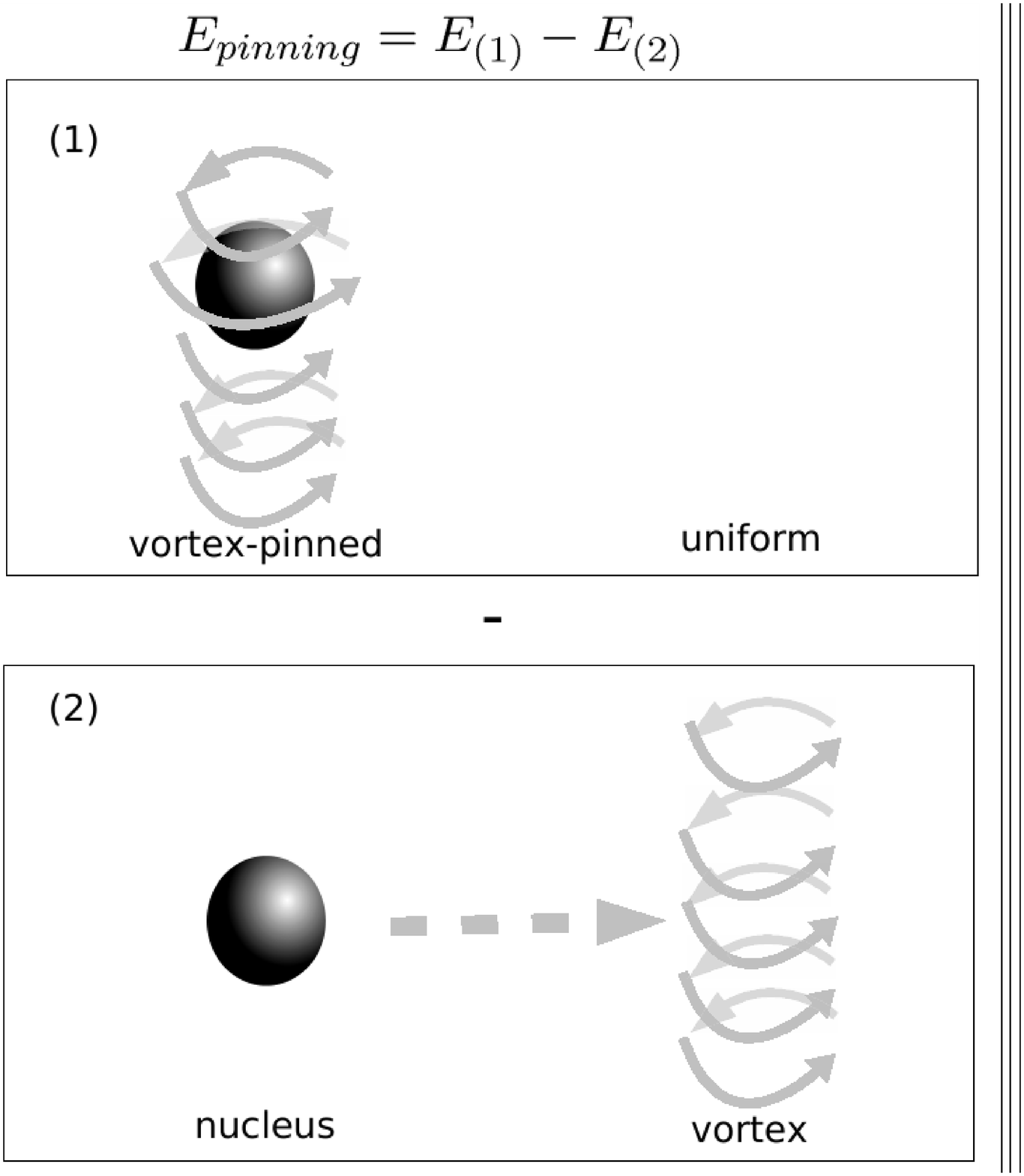}
\includegraphics[width=8.5cm,height=7cm]{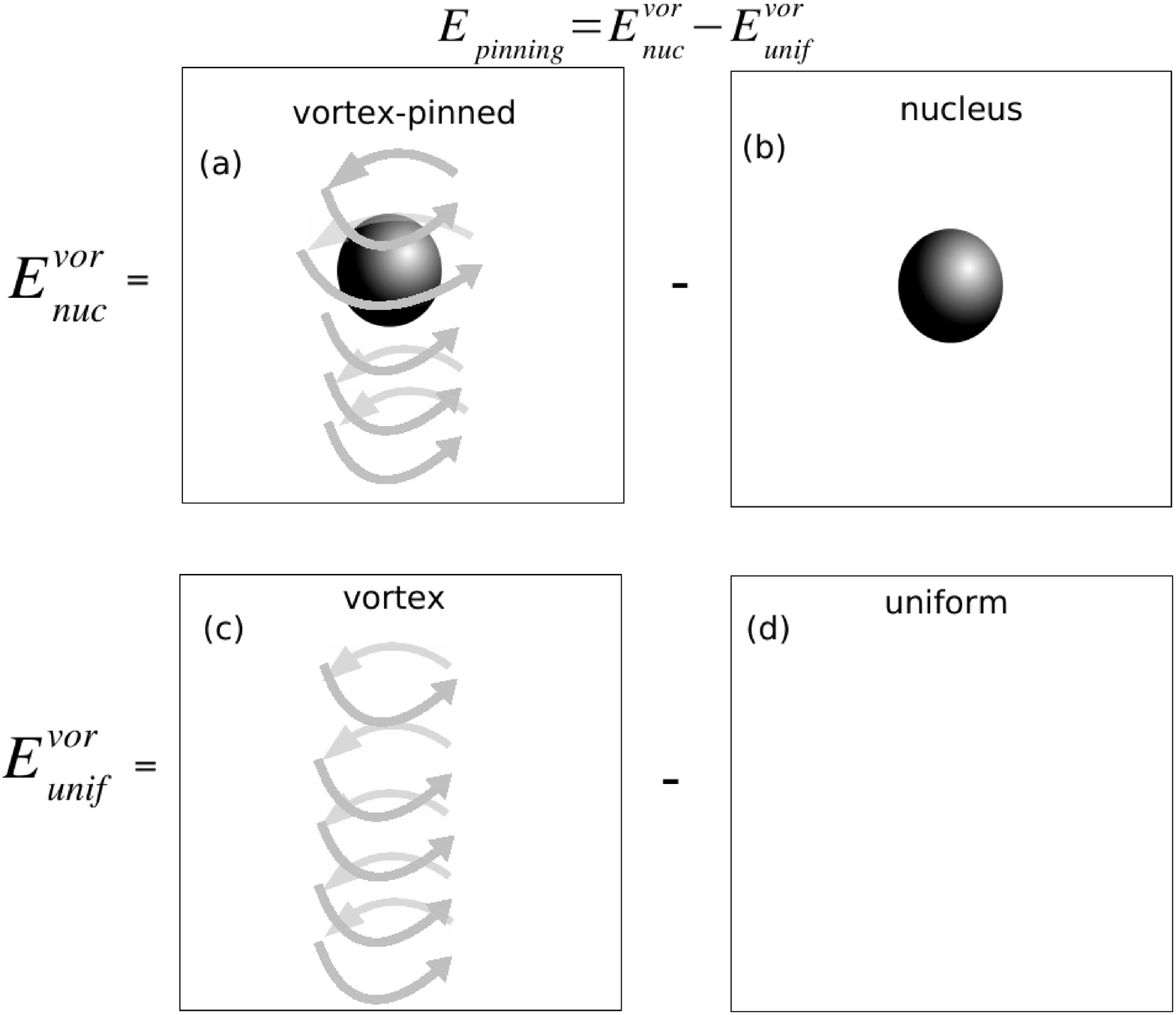}
\caption{(left) The pinning energy $E_{pinning}$ is obtained taking the difference
between  the energy of the pinned
configuration (1), in which the vortex axis passes through the center of the 
nucleus,  
and the energy of the interstitial configuration (2), 
in which the vortex is far from the nucleus. 
(right) We rearrange the configurations on the left,
indicating that the pinning energy is equivalent to the 
difference between the energy cost to build the vortex on a nucleus
($E^{vor}_{nuc}$) 
and in uniform matter ($E^{vor}_{unif}$), so that 
$E_{pinning}=E_{nuc}^{vor}-E_{unif}^{vor}$.}
\label{pinn_schem}
\end{figure}


We shall now discuss more precisely how we compute the pinning energy, 
$E_{pinning}$, at a given 
density in the crust, following (except for a sign)  the definition by  Epstein and 
Baym \cite{Epstein88}. At  a given density,  they calculated, within the Ginzburg-Landau approximation,
the energy of the vortex-nucleus system as a function of the distance
between the vortex axis and the center of the nucleus, determining the 
configuration of minimum energy, which often
turned out to correspond to zero distance.  
Then they calculated the
pinning energy as the difference between this minimum energy
and the energy  of a configuration
where the vortex and the nucleus are sufficiently far apart, 
so that their mutual  interaction 
can be neglected. Due to the axial symmetry condition, 
we shall only compare the energies of the zero-distance 
configuration (vortex axis passing through the center of the nucleus) and of the 
non-interacting one.  
In this framework we only deal with one nucleus at a time, 
neglecting the interaction of the vortex with distant nuclei. 
The situation can then be  schematically depicted 
as in the left part of Fig. \ref{pinn_schem}. We have to compare 
the energies of the {\it pinned} configuration (1) 
(vortex axis passing through the center of the nucleus) and 
of the {\it interstitial} configuration (2) (vortex axis far from the  nucleus). 
We can then 
consider $E_{pinning}$
as  the difference between two excitation energies: 
\begin{equation}
 E_{pinning}=E_{nuc}^{vor}-E_{unif}^{vor}
 \label{eq:epinn}, 
\end{equation}
where $E_{unif}^{vor}$ is the energy of a vortex in the uniform matter configuration, 
compared to the energy  of the same configuration
in its ground state, while 
$E_{nuc}^{vor}$  is the energy of the vortex pinned on a nucleus (immersed in the neutron sea),
compared to the energy of the isolated nucleus immersed in the neutron sea; that is, 
\begin{equation}
\begin{split} 
 E_{nuc}^{vor}=E_{pinned}-E_{nucleus}\\
 E_{unif}^{vor}=E_{vortex}-E_{uniform},
\end{split}
\label{eq:ecost}
\end{equation}
where the energy associated with each of the four configurations
is calculated as described in Appendix A.
Pinning the vortex on the  nucleus is  energetically convenient,
if the pinning energy is negative (note that Epstein and Baym define the pinning energy with the  
opposite sign). 
We note that the total and pairing energy of the various
configurations, as well as the energy cost to create the vortex, depend
on the size of the box. On the contrary,  the value of the 
pinning energy, which is the result of a phenomenon localized on the nucleus,
converges for a sufficiently large box.

\begin{figure}[h!]
\centering
\includegraphics[width=6cm]{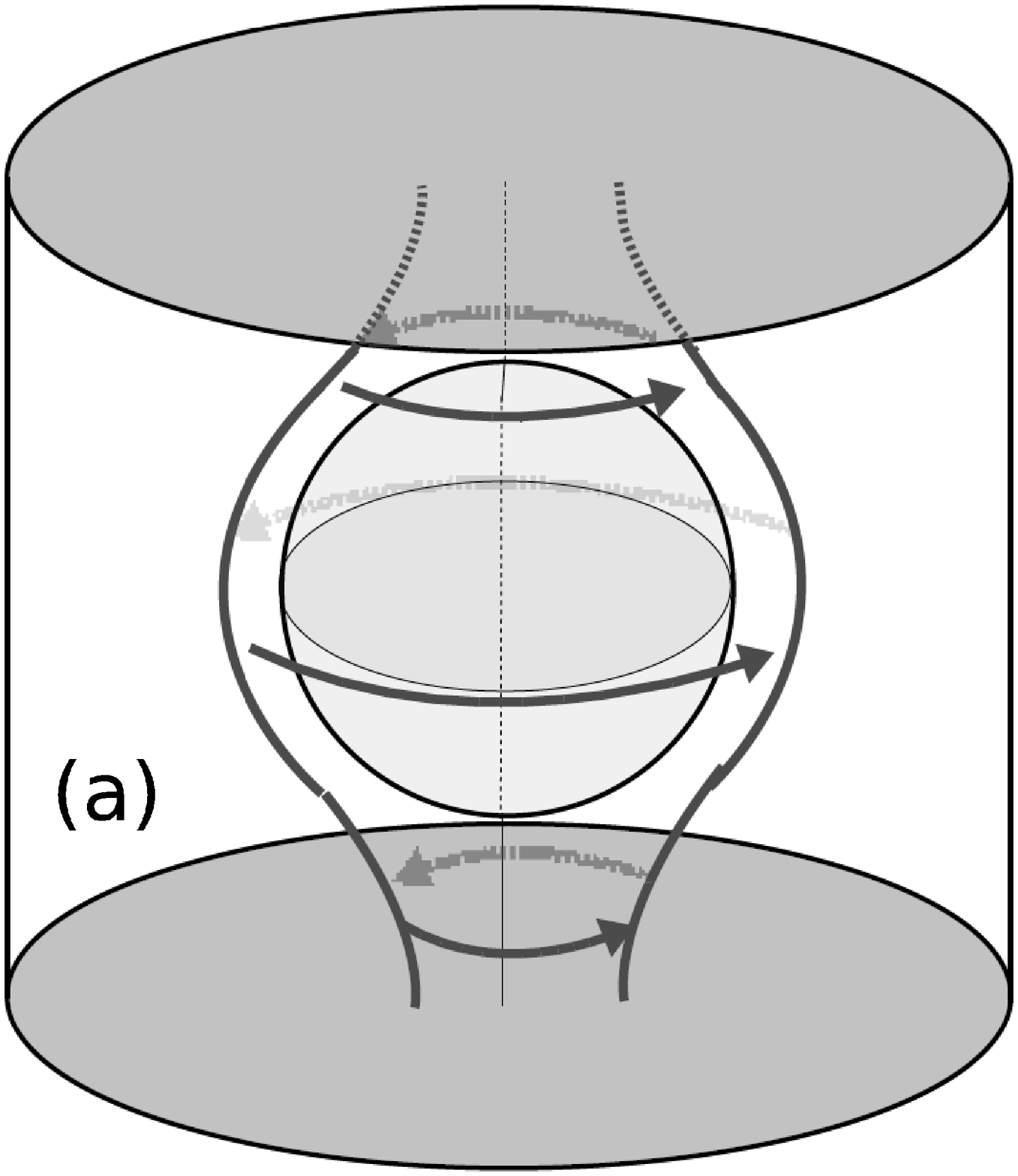}
\includegraphics[width=6cm]{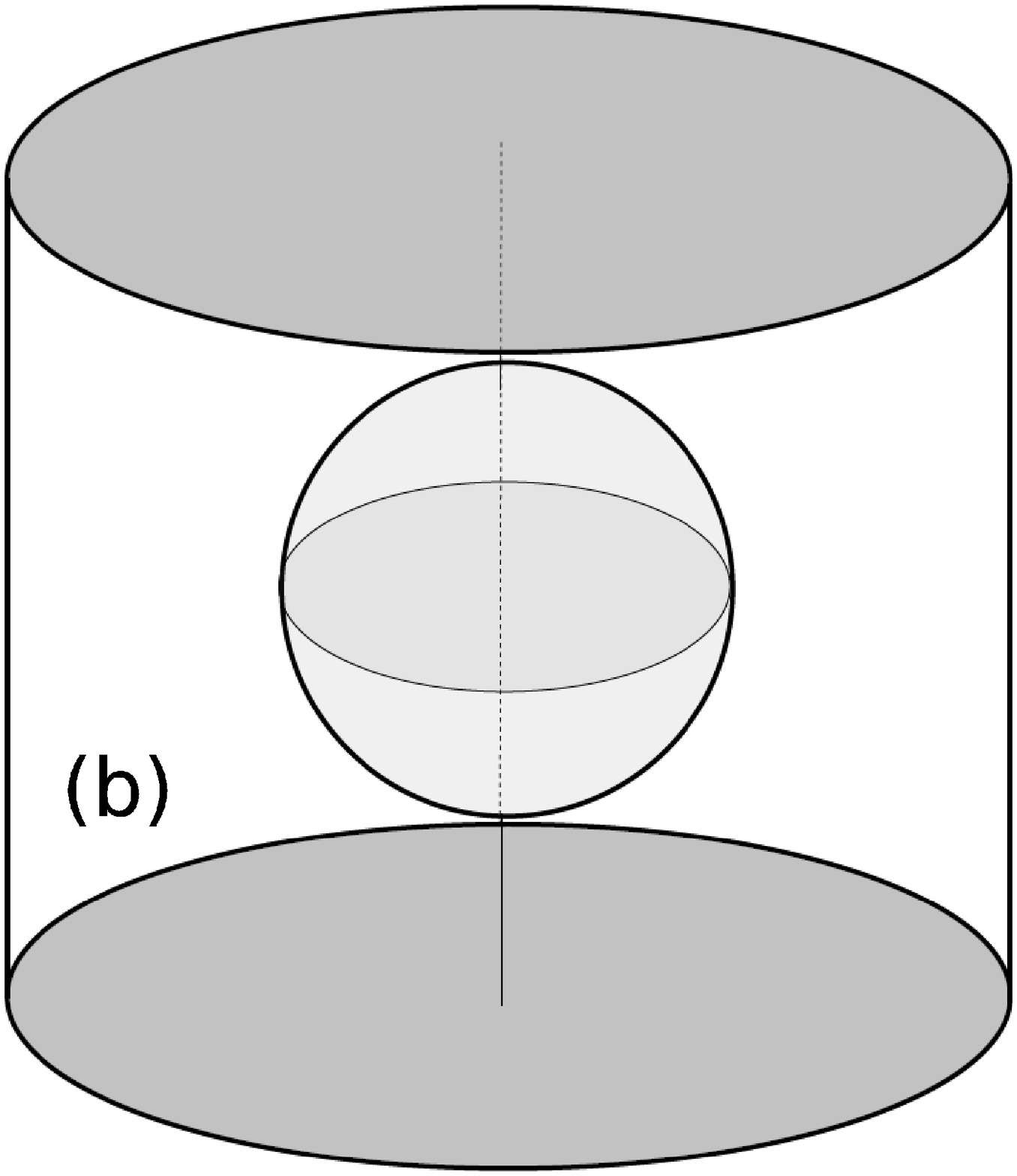}
\includegraphics[width=6cm]{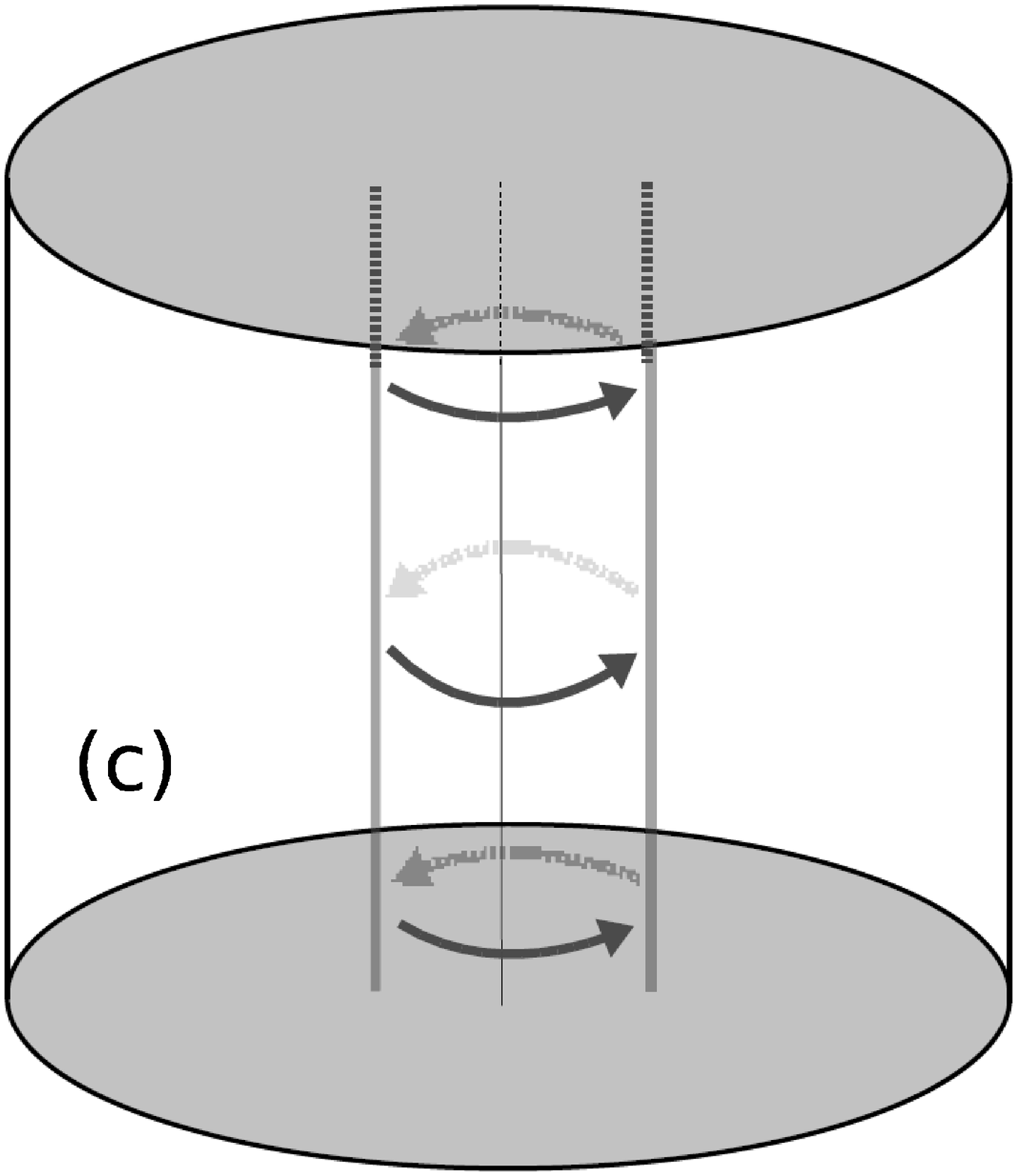}
\includegraphics[width=6cm]{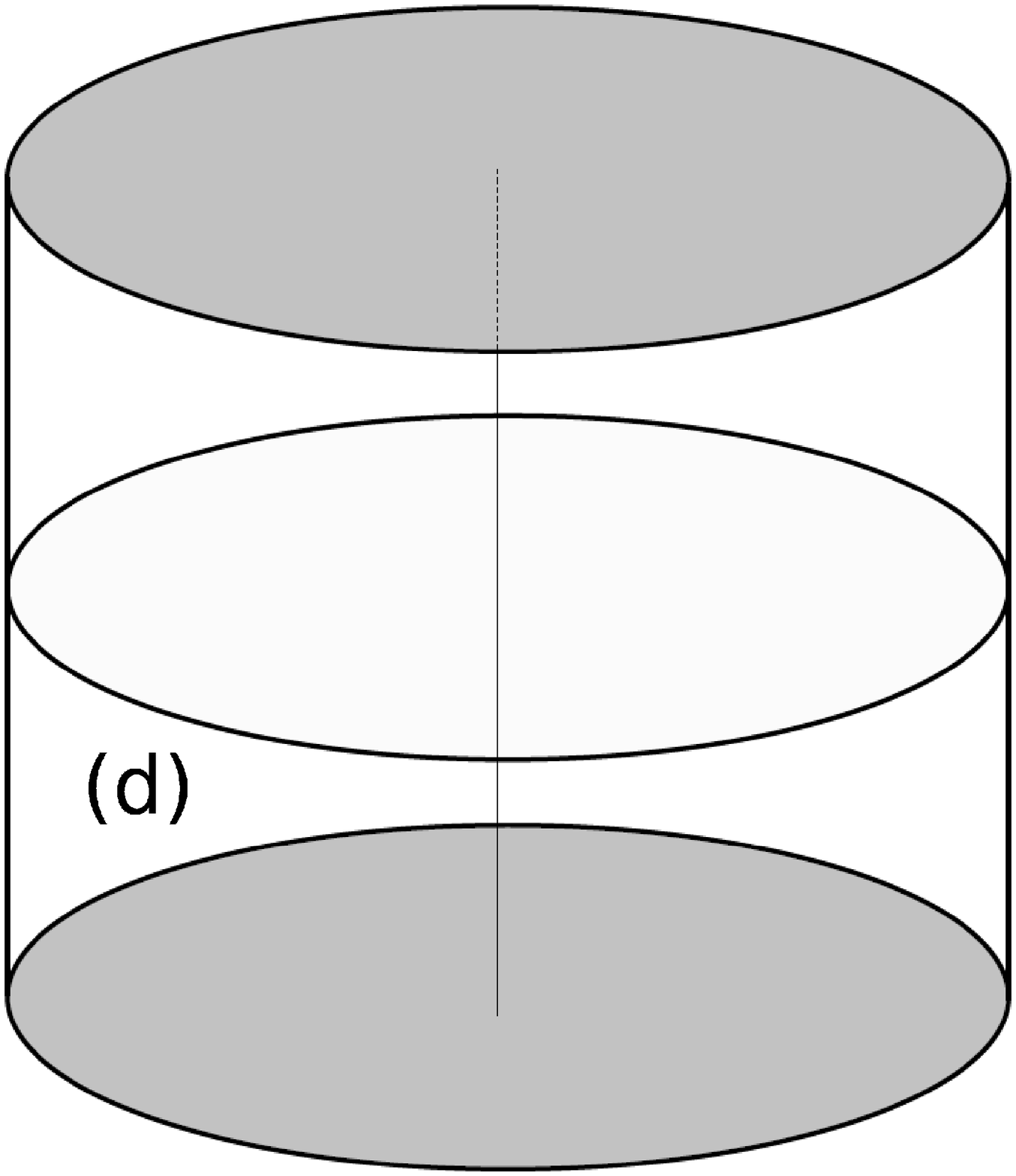}
\caption{Schematic picture of the four different physical configurations 
needed to calculate the pinning energy: (a) cell with a vortex 
pinned on a nucleus ; (b) cell with a nucleus in the neutron sea ; 
(c) cell with a vortex in the neutron sea; 
(d) cell in the uniform neutron sea.} 
\label{fig:configurazioni}
\end{figure}

In practice, four different HFB calculations have to be performed, using the same cylindrical cell
(cf. Fig. \ref{fig:configurazioni}): one with $Z=40$, $\nu=1$ (Fig. \ref{fig:configurazioni}(a)),
yielding $E_{pinned}$;
one with $Z=40$, $\nu=0$ (Fig. \ref{fig:configurazioni}(b)), 
yielding $E_{nucleus}$;  one with $Z=0$, $\nu=1$ 
(Fig. \ref{fig:configurazioni}(c)), yielding $E_{vortex}$; and finally, one with $Z=0$, $\nu=0$ 
(Fig. \ref{fig:configurazioni}(d)), yielding $E_{uniform}$.
To obtain the correct pinning energy it is essential that the calculations 
refer to 
the same asymptotic neutron density.
In the calculations (b) and (d), without the  vortex, we use the same 
value of the chemical potential $\lambda$,
yielding $N_{nucleus}$ and $N_{uniform}$ neutrons in the cylindrical cell.
The presence of the vortex in calculations (a) and (c) 
leads in each case to a small decrease 
of the number of particles. 
We compensate this reduction by a slight increase in the value of $\lambda$.
In practice, rather than attempting a very fine tuning of $\lambda$
we prefer  to account for the residual difference in the number of particles
subtracting  the  terms 
\begin{equation}
\Delta E_{nuc} = \lambda (N_{pinned} - N_{nucleus}) \quad ;   \quad
\Delta E_{unif} = \lambda (N_{vortex} - N_{uniform})  
\label{DeltaE}   
\end{equation}
respectively to $E^{vor}_{nuc}$ and to $E^{vor}_{unif}$.
Even if the pinning energies represent only a small fraction of the total energy, 
of the order of $10^{-3}-10^{-4}$, 
the subtraction scheme we have just outlined produces numerically reliable 
results (cf. Appendix B), which will be presented 
in the next Section. One should notice that the size 
of the cylindrical cell 
does not have to coincide with
that of the physical Wigner-Seitz cell; it must only be large enough, 
so as to obtain convergent
results for $E_{pinning}$. It is clear, however, that neglecting neighbouring nuclei 
can be inconsistent, if the radius of the box becomes of the order of 
the lattice constant. This point will be further discussed below in Section IV C.

\section{Results}
We have performed calculations at different densities in the inner crust,
ranging between $n \approx 0.001 $ fm$^{-3}$ and 
$n \approx 0.04 $ fm$^{-3}$. Our discussion will be mostly based on the
results obtained at eight densities with the interaction SLy4 and SkM$^*$, but we shall also 
present results obtained with the interactions SII and SGII 
(cf. Tables I and II).


In the following we first present  the results obtained  
for a nucleus embedded in the neutron sea, in the absence of vortex 
(HFB calculations with $\nu=0$).
We then discuss the structure of the vortex ($\nu=1$), without or with the nucleus.
Finally, we present the pinning energies obtained as a function of density.

\subsection{Calculations with $\nu = 0$ }

 \begin{table}[h]
     \begin{center}
        \begin{tabular}{|c|c|c|c|c|}
        \hline
        \hline
\;\; Interaction \;\; & $n_{\infty}$ [$fm^{-3}$] &$a$ [$fm$]& $n_0$ [$fm^{-3}$]& $R_0$[$fm$] \\
    \hline
    \hline
 SLy4       & 0.001 & 0.82    & 0.098 & 6.06  \\
    \hline      
 SLy4       & 0.002 & 0.82   & 0.098 & 6.04  \\
    \hline
 SLy4        & 0.004 & 0.82   & 0.090 & 6.81  \\
    \hline
 SLy4        & 0.008 &  0.86  & 0.091 & 6.75  \\
    \hline
 SLy4        & 0.011 & 0.91   & 0.089 & 6.76  \\
    \hline                    
 SLy4        & 0.017 & 0.96   & 0.081 & 7.33  \\
    \hline
 SLy4        & 0.026 & 1.05   & 0.070 & 7.83  \\
    \hline
 SLy4        & 0.037 & 1.13   & 0.057 & 8.43  \\
    \hline
    \hline
  SII        & 0.001 & 0.73   & 0.093 & 6.15 \\
    \hline
  SII        & 0.002 & 0.73   & 0.093 & 6.14 \\
    \hline
  SII        & 0.004 & 0.74   & 0.092 & 6.10 \\
    \hline
  SII        & 0.008 & 0.77   & 0.086 & 6.74 \\
    \hline      
  SII        & 0.011 & 0.78   & 0.085 & 6.71 \\
    \hline                      
  SII        & 0.016 & 0.81   & 0.082 & 6.75 \\
    \hline
  SII        & 0.024 & 0.87   & 0.073 & 7.18 \\
    \hline
  SII        & 0.035 & 0.94   & 0.062 & 7.60 \\
        \hline

         \end{tabular}
     \end{center}
\caption{Parameters of the Fermi function (\ref{fermifun}),
fitting the  calculated  neutron densities
in the Wigner Seitz cells with the SLy4 and SII interactions.
The values of the density have been rounded off at the third digit.}
\end{table}\label{tab:fit-nuc}

 \begin{table}[h]
     \begin{center}
        \begin{tabular}{|c|c|c|c|c|}
        \hline
        \hline
\;\; Interaction \;\; & $n_{\infty}$ [$fm^{-3}$] &$a$ [$fm$]& $n_0$ [$fm^{-3}$]& $R_0$[$fm$] \\
    \hline
    \hline
 SkM*       & 0.001 &  0.88   & 0.097 & 6.12  \\      
   \hline   
 SkM*       & 0.002  &  0.90  & 0.097 & 6.17  \\
   \hline
  SkM*      & 0.004  &  0.96  & 0.091 & 6.76  \\
   \hline 
  SkM*      & 0.008  &  0.97  & 0.088 & 6.73 \\
    \hline  
  SkM*      & 0.012  &  1.00  & 0.086 & 6.78  \\ 
    \hline 
  SkM*      & 0.016  &  1.05  & 0.080 & 7.03  \\
    \hline
  SkM*      & 0.025  &  1.09  & 0.070 & 7.33  \\  
    \hline   
  SkM*      & 0.038  &  1.17  & 0.054 & 7.95  \\
    \hline 
    \hline                   
 SGII      & 0.001  &  0.86  & 0.097 & 6.91 \\
    \hline
 SGII      & 0.002  &  0.88  & 0.097 & 6.91 \\ 
    \hline
 SGII      & 0.004  &  0.90  & 0.096   & 6.82  \\
    \hline
 SGII      & 0.008  &  0.94    & 0.092 & 6.86 \\
    \hline
 SGII      & 0.011  &  1.00    & 0.086 & 7.34 \\
    \hline 
 SGII      & 0.016  &  1.00    & 0.082 & 7.38 \\
    \hline
 SGII      & 0.026  &  1.07    & 0.070 & 7.77 \\
    \hline       
 SGII      & 0.037   &  1.13   & 0.057 & 8.27 \\
    \hline                            \end{tabular}
     \end{center}
\caption{Parameters of the Fermi function (\ref{fermifun}),
fitting the  calculated  neutron densities
in the Wigner Seitz cells with the SkM* and SGII interactions.
The values of the density have been rounded off at the third digit.}
\end{table}\label{tab:fit-nuc2}

We first discuss the density and the pairing gaps  obtained solving the HFB equations 
for $\nu=0$ in the cylindrical box,
for the nuclei immersed in the neutron sea at the
various densities.
The pairing field  in the Wigner-Seitz cell has been studied 
by various groups, either  based on Woods-Saxon potentials \cite{Pizzo_APJ} 
or on selfconsistent HFB calculations
with Skyrme forces \cite{Montani, Sandu, Baldo}. 
The effects of temperature, and the specific heat,
have also been discussed \cite{Pizzo_APJ,Sandu}, but we shall limit
ourselves to zero temperature. Our calculations are performed assuming 
axial symmetry, and they should respect  the spherical symmetry of 
isolated nuclei. 
In our calculations the neutron density is a function of the coordinates $\rho$ and $z$. 
To test the sphericity of our calculations we evaluate the mean value of the density 
within a spherical shell 0.05 fm wide and centered around a mean distance $r$ 
from the center of the nucleus and then evaluate 
the fluctuations at every point. 
The relative error (rms/average density) is everywhere lower than 0.03.

\begin{figure}
\centering
\includegraphics[width=6cm]{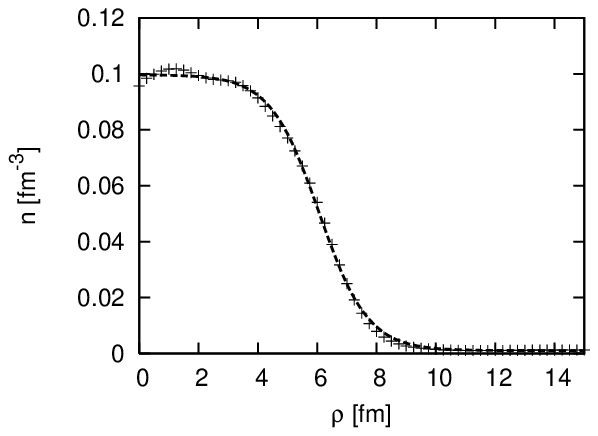}
\includegraphics[width=6cm]{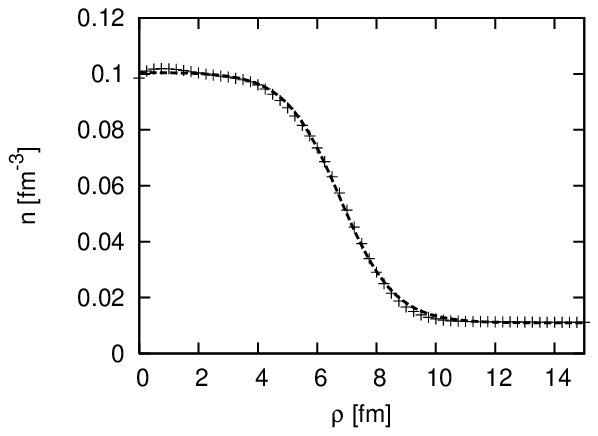}
\includegraphics[width=6cm]{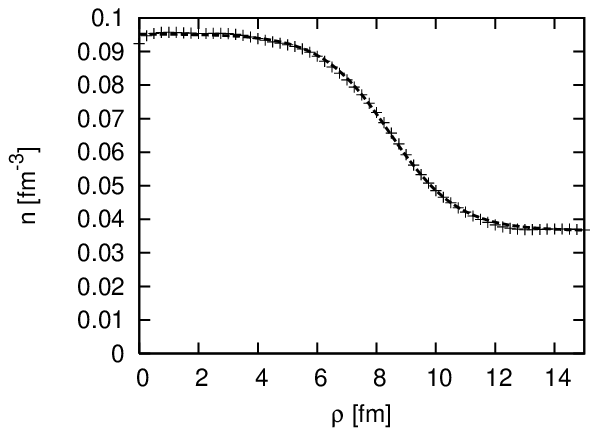}
	\caption{The neutron density calculated in the cylindrical box with the nucleus on
the $z=0$ plane (crosses) 
is compared with the fit obtained with the Fermi function
(cf. Eq. \ref{fermifun}). As we discuss in the text, the calculated 
density is spherically symmetric to a good approximation.
The figures refer to the asymptotic densities $n_{\infty}$= 0.001, 0.011 and 0.037 fm$^{-3}$,
and have been computed with the SLy4 interaction.} 
\label{fig:fits}
\end{figure}

As first remarked by Negele and Vautherin
\cite{NegeleVaut},
nuclei in the inner crust turn out to be fatter than atomic nuclei, due to their interaction with the neutron sea.
We have fitted the neutron density  
resulting from our calculations with a Fermi function:
\begin{equation}
f(r)=n_{\infty}+ \frac{n_0}{\left(1+{\rm exp}\left(\frac{(r-R_0)}{a}\right)
\right)}, 
\label{fermifun}
\end{equation} where 
$n_{\infty}$ gives the asymptotic value of the neutron density 
far from the nucleus, and we have determined
the values of  $n_0$, $R_0$ and $a$  
which best reproduce the computed density profile and are reported in  
Tables I and II. In the calculations we have adjusted 
the value of $\lambda$ 
so as to  obtain the same set of values of 
$n_{\infty}$  for  the various interactions, but there remain slight
differences in a few  values (cf. also Tables III and IV).
We compare the  calculated and fitted densities for three cases 
in Fig. \ref{fig:fits}.
The diffusivity $a$ increases from about 0.7  to about 1.1 fm with increasing density, 
to be  compared with the typical value of 0.65 fm 
for finite nuclei,
and the radius $R_0$ (up to 8 fm 
at the highest densities) also  turns out  to be  
rather large compared to ordinary nuclei.

\begin{figure}[!h]
\centering
\includegraphics[width=6cm]{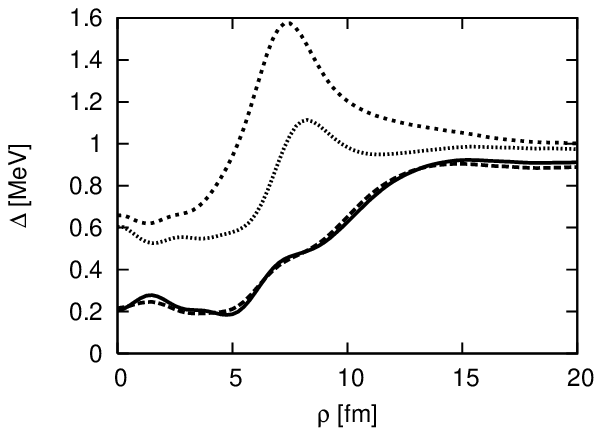}
\includegraphics[width=6cm]{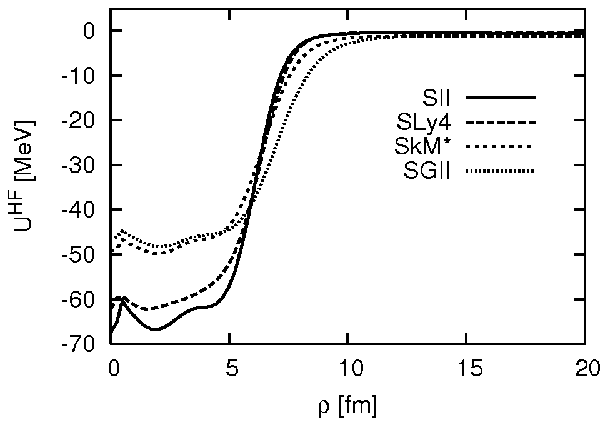}
	\caption{Pairing gap (left) and  selfconsistent potential (right) 
in the cylindrical cell without the vortex ($Z=40, \nu=0$) for $n_{\infty}\approx $ 0.001 {\rm fm}$^{-3}$ ($k_F \approx 0.3 $ fm$^{-1}$). 
The different interactions are: 
SLy4 (dashed curve), 
SkM* (short dashed curve),
SII (solid curve), 
and SGII (dotted curve).} 
\label{fig:delta1}
\end{figure}

\begin{figure}[!h]
\centering
\includegraphics[width=6cm]{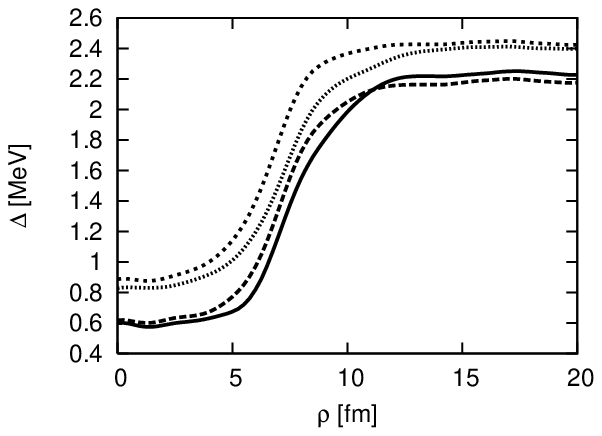}
\includegraphics[width=6cm]{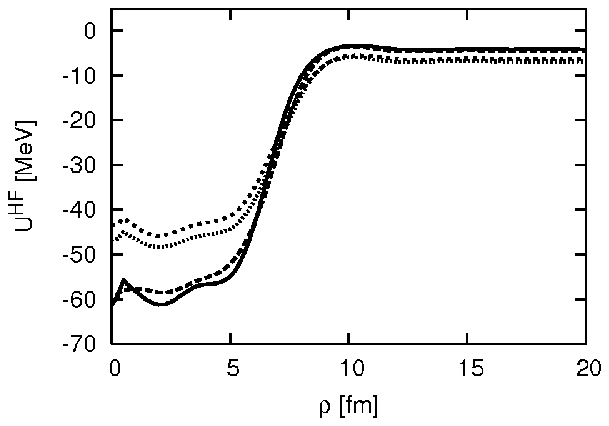}
\caption{Pairing gap (left) and  selfconsistent potential (right) 
in the cylindrical cell ($Z=40, \nu=0$) for $n_{\infty}\approx $ 0.011 {\rm fm}$^{-3}$
($k_F \approx 0.7 $ fm$^{-1}$). 
The different interactions are: 
SLy4 (dashed curve), 
SkM* (short dashed curve),
SII (solid curve) and SGII (dotted curve).}
\label{fig:delta2}
\end{figure}

\begin{figure}[!h]
\centering
\includegraphics[width=6cm]{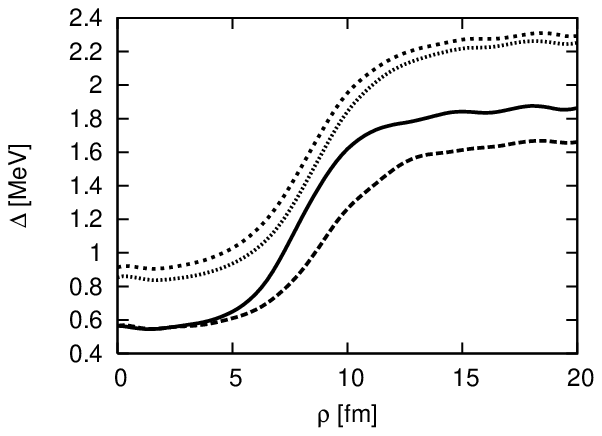}
\includegraphics[width=6cm]{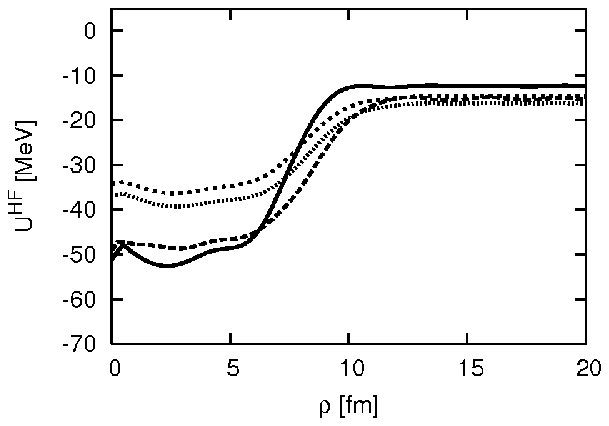}
\caption{Pairing gap (left) and  selfconsistent potential (right) 
in the cylindrical cell ($Z=40, \nu=0$) for $n_{\infty}\approx $ 0.035 {\rm fm}$^{-3}$
($k_F \approx 1 $ fm$^{-1}$). 
The different interactions are: 
SLy4 (dashed curve), 
SkM* (short dashed curve),
SII (solid curve), 
and SGII (dotted curve).}
\label{fig:delta3}
\end{figure}

\begin{figure}[!h]
\centering
\includegraphics[width=6cm]{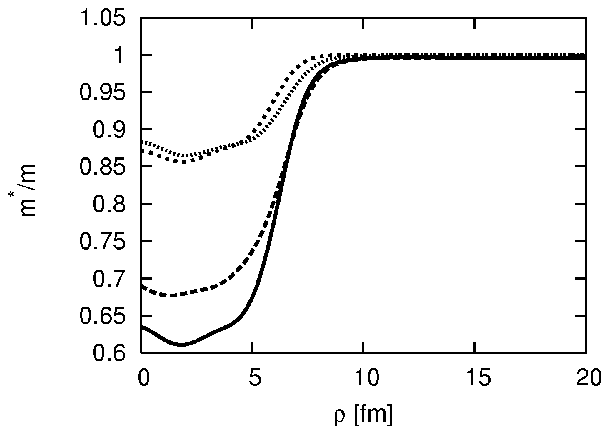}
\includegraphics[width=6cm]{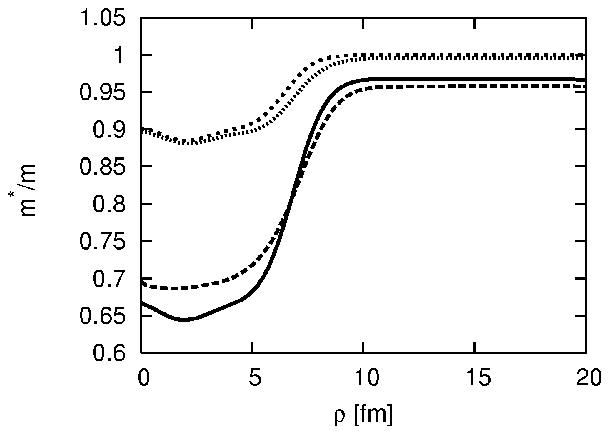}
\vsize=1cm
\includegraphics[width=6cm]{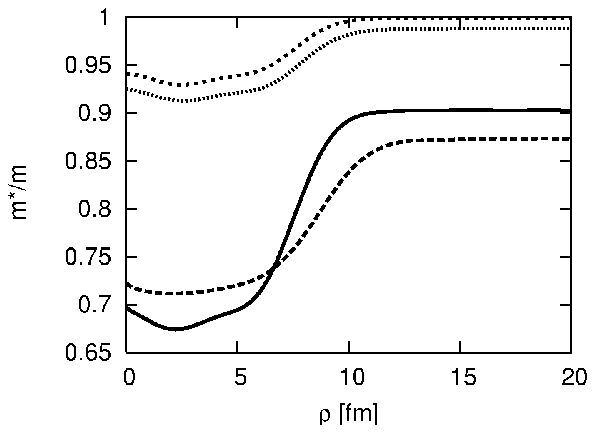}
\caption{Effective masses associated with the calculations presented 
in Fig. \ref{fig:delta1}  
($n_{\infty}\approx $ 0.001 {\rm fm}$^{-3}$, top left), 
in Fig. \ref{fig:delta2}
($n_{\infty}\approx $  0.01  {\rm fm}$^{-3}$ , top right),
and 
in Fig. \ref{fig:delta3}
($n_{\infty} \approx $  0.035  {\rm fm}$^{-3}$ , bottom)
for the four interactions SII (solid curve), 
SLy4 (dashed curve), SkM* (short dashed curve) and SGII (dotted curve).} 
\label{fig:mas1}
\end{figure}

In Figs. \ref{fig:delta1}-\ref{fig:delta3}  we  show 
the pairing gaps and selfconsistent 
potentials as a function of the distance from the nucleus,
calculated for 
$n_{\infty} \approx $ 0.001, 0.01 and 0.035 {\rm fm}$^{-3}$; 
the associated effective masses 
are shown in  Fig. \ref{fig:mas1}.

As we anticipated (cf. Fig \ref{fig:effmass_Skyrme}), 
the behaviour of the forces can be divided into two groups: the effective masses 
associated with SGII and SkM* are quite similar ( $m^{*}/m\approx 0.9$ in the interior of the nucleus), 
as are those associated with SLy4 and SII  ($m^{*}/m\approx0.7$).
In order to reproduce the same asymptotic density the SII and SLy4 interactions produce 
a deeper selfconsistent potential 
than the SkM* and SGII. 
In all cases the pairing gap increases, going from the nuclear region to the sea of superfluid neutrons. 
This is in keeping with the density dependence of the pairing gap in 
uniform matter, previously shown in Fig. \ref{fig:delta_NM}: the  interior of the nucleus corresponds
to large values of the local Fermi momentum where the pairing gap tends to
vanish, while  even at the smallest densities considered here, the gap 
in neutron matter is at least close to 1 MeV. 
In the case of the SkM* and SGII interactions the pairing gaps 
show a peak in the nuclear surface region
for $n_{\infty}\approx $ 0.001 {\rm fm}$^{-3}$ (cf. Fig. \ref{fig:delta1}). 
The pairing gaps are smaller with SII and SLy4, due to their lower effective mass and 
level density.


\subsection{Calculations with $\nu=1$}

\begin{figure}[!h]
\centering
\includegraphics[width=7cm]{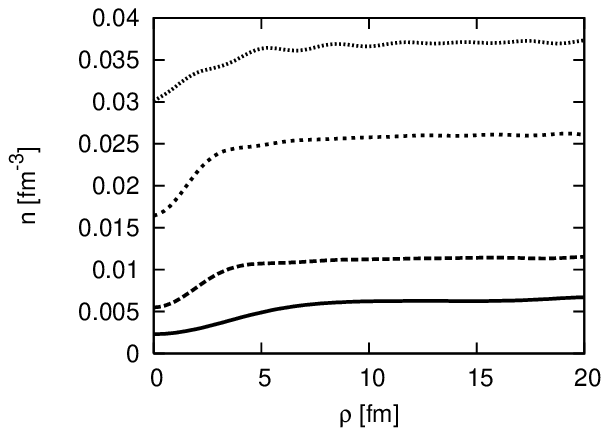}
\includegraphics[width=7cm]{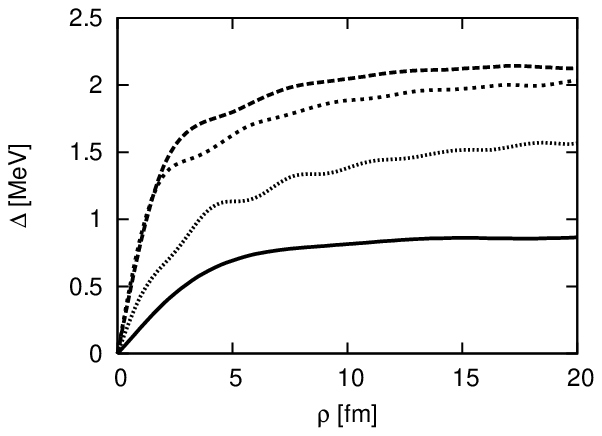}
\caption{ (left)
Neutron density profiles for the vortex solution, without the nucleus
at the center of the cell ($Z=0,\nu=1$),  
calculated with the SLy4 interaction at different 
densities: $n_{\infty}= $ 0.0013 {\rm fm}$^{-3}$;  (solid curve, the density shown has been multiplied 
by 5), $n_{\infty}= $0.011  {\rm fm}$^{-3}$; (dashed curve), 
$n_{\infty}= $ 0.026 {\rm fm}$^{-3}$ (short dashed curve) and 
$n_{\infty}= $ 0.037 {\rm fm}$^{-3}$; 
(dotted curve).  
(Right) The corresponding pairing gaps.}
\label{fig:dens-vortice1}
\end{figure}

\begin{figure}[h!]
\centering
\includegraphics[width=7cm]{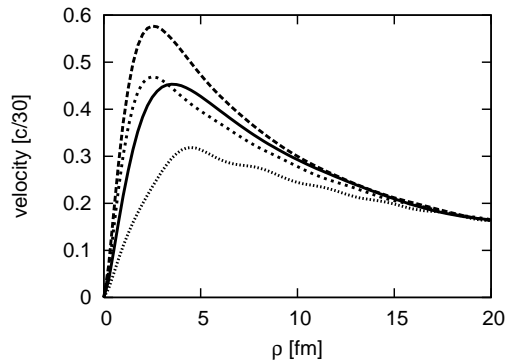}
\caption{
Velocity field of a vortex   calculated in the cell without the 
nucleus ($Z=0,\nu=1$), with 
the SLy4 interaction calculated with 
$n_{\infty}= $ 0.0013 {\rm fm} $^{-3}$; (solid curve), 
$n_{\infty}= $ 0.011 {\rm fm} $^{-3}$; (dashed curve), 
0.026 {\rm fm}$^{-3}$; (short dashed curve),  
and 0.037  {\rm fm}$^{-3}$;   (dotted curve). 
For large values of $\rho$ the
velocity tends to the Onsager limit $ \hbar /2 m \rho$.
}
\label{fig:vort-vel}
\end{figure}

We now turn our attention to the  HFB calculations of the vortex ($\nu=1$). 
We first discuss
the results we obtain for the uniform neutron sea ($Z=0$). 
In this case  the  vortex configuration keeps cylindrical symmetry, and a cut at a fixed  value of $z$  
provides complete information. 
The density profiles and the pairing gap obtained with the SLy4 interaction 
at four different densities are reported 
in Fig. \ref{fig:dens-vortice1}. 
The pairing gap vanishes on the vortex axis, and grows linearly for about 3-5 fm; then it slowly increases
towards the   value in  $\nu=0$ neutron matter. 
In keeping with Fig. \ref{fig:delta_NM},
the asymptotic value of $\Delta$  reaches its maximum 
for $k_F \sim 0.8$ fm$^{-1}$, that is, for $n \sim $ 0.02 fm$^{-3}$.
Near the vortex axis the pairing gap displays oscillations that have been interpreted as 
Friedel-like oscillations, due to 
the presence of bound states in the vortex core  \cite{elgaroy}.
Our results are very similar to those  obtained by Bulgac
and Yu, who have solved the HFB equations for a vortex  in
uniform, infinite   neutron matter for the first time \cite{Bulgac2003}.
In particular, the depletion of the density close to the vortex axis 
(cf. Fig. \ref{fig:dens-vortice1})  and the behaviour of the velocity field  (cf. Fig. \ref{fig:vort-vel})
are in good agreement with their results. Note however that we use a different 
pairing interaction and that we consider densities
larger than those calculated in ref.  \cite{Bulgac2003}.

\begin{figure}[h]
\centering
\includegraphics[width=8cm]{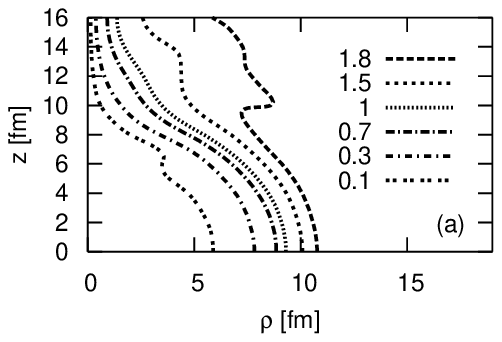}
\includegraphics[width=8cm]{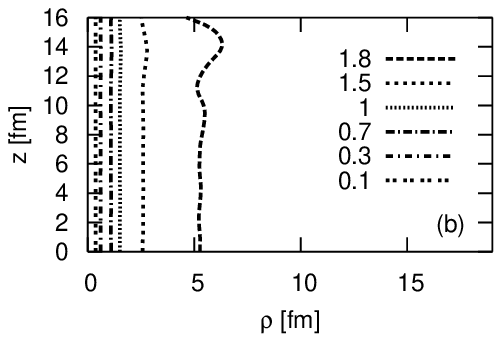}
\includegraphics[width=8cm]{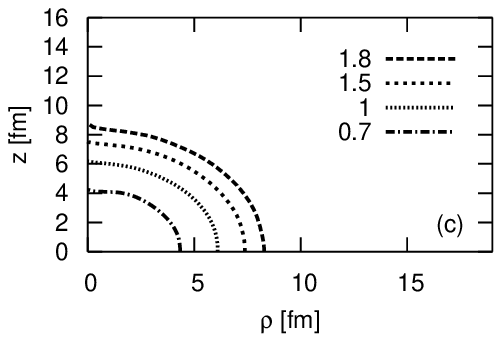}
\caption{
The level curves (values in MeV) of the pairing gap associated with  
a vortex pinned on a nucleus 
($Z=40,\nu=1$, top left),  with a vortex in uniform matter ($Z=0,\nu=1$, top right) 
and with a nucleus in the neutron sea ($Z=40,\nu=0$, bottom) obtained with a SLy4 force for  
$n_{\infty}= $ 0.011  {\rm fm} $^{-3}$. 
The nucleus is at $\rho=z=0$ and the vortex is along the $z$-axis. 
}
\label{fig:pair3d5.8}
\end{figure}

\begin{figure}
\centering
\includegraphics[width=6cm]{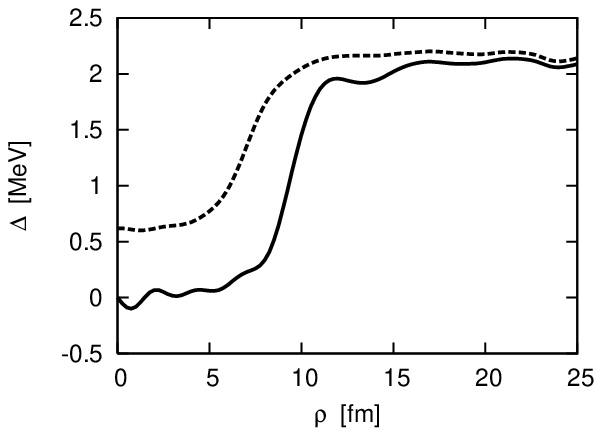}
\includegraphics[width=6cm]{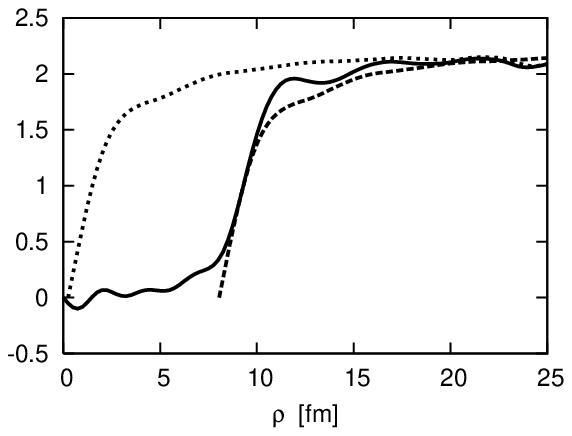}
\caption{
(left) The pairing gap associated with   the pinned  vortex  
on the $z=0$ plane ($Z=40,\nu=1$, solid line) 
is compared with the  gap associated with the nucleus 
in the absence of the vortex 
($Z=40,\nu=0$, dashed line). 
(right) The pairing gap associated with the pinned vortex
($Z=40,\nu=1$, solid line) is now compared to the gap associated with the   
vortex in uniform matter ($Z=0,\nu=1$, short dashed line); the latter is 
also shown  translated by $7.8$  fm (dashed line). 
The calculation is for 
the SLy4 interaction at $n_{\infty}= $ 0.011 {\rm fm}$^{-3}$.}
\label{fig:pair-2d-ent-nuc}
\end{figure}

\begin{figure}
\centering
\begin{center}
\includegraphics[width=6cm,angle=0]{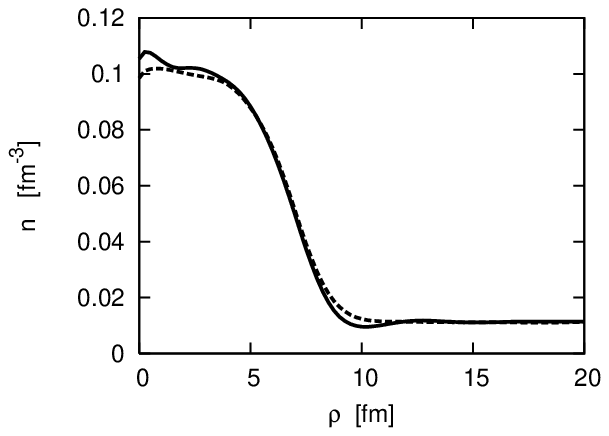}
\includegraphics[width=6cm,angle=0]{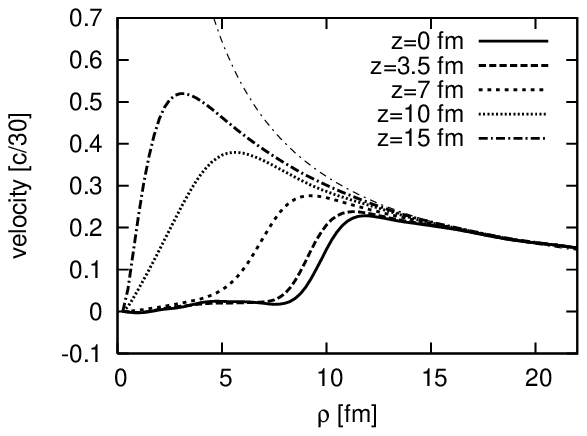}
\end{center}
\caption{
(left) Density with ($Z=40,\nu=1$, solid line) and without ($Z=40,\nu=0$, dashed line) the vortex, 
calculated  at $n_{\infty}= $ 0.011 {\rm fm}$^{-3}$ on the $z=0$ plane 
with the SLy4 force.
(right) Velocity field of the pinned vortex.
The different curves refer to different cuts;
the dot-dashed line is the $\hbar/2m\rho$ function. }
\label{fig:dens-ent-nuc}
\end{figure}

\newpage
We now discuss the calculation of the vortex pinned on the nucleus ($Z=40, \nu=1$).
The main results are illustrated in Figs. \ref{fig:pair3d5.8} and \ref{fig:pair-2d-ent-nuc},
which refer to a calculation performed at
$n_{\infty}= $ 0.011 {\rm fm}$^{-3}$ with the SLy4 interaction.
In Fig. \ref{fig:pair3d5.8} we show the gap in the $\rho, z$ plane. In   
Fig. \ref{fig:pair-2d-ent-nuc} the gap is shown as a function 
of $\rho$ in the $z=0$ equatorial plane.
In Fig. \ref{fig:pair-2d-ent-nuc} 
we also make a comparison with the pairing gaps
obtained for a vortex in uniform matter 
and for a nucleus in the neutron sea, previously discussed.
The neutron density and the velocity field are shown in 
Fig. \ref{fig:dens-ent-nuc}.

For large values of $z$ the properties of the vortex are the same as those calculated 
for uniform matter (cf. Fig. \ref{fig:pair3d5.8}). Approaching the nuclear surface 
the  vortex is strongly modified, and the associated pairing gap 
is suppressed 
within its volume and also on its surface
(for example $\Delta \approx $ 0.3 MeV at $\rho \approx 7$ fm and $z=$0 ),
both compared to the case of an isolated nucleus in the absence of the vortex 
($\Delta \approx$ 1 MeV)
and compared to the case of a vortex in uniform neutron matter 
($\Delta \approx $ 2 MeV). 
The typical linear rise of the gap 
away from the vortex axis is delayed by about 8 fm, which is about the 
radius of the nucleus,  as compared to the uniform case 
(cf. Fig. \ref{fig:pair-2d-ent-nuc}).
In other words, the vortex opens  up and surrounds the nuclear volume,
being unable to penetrate into it.
The suppression of the $\nu=1$ pairing field inside the nucleus was already
found (but not thoroughly discussed)  in a quantum calculation by Elgar\o y
and De Blasio \cite{elgaroy}, although their study was not self-consistent
and assumed an unrealistic  cylindrical shape for the nucleus.
The same effect appears concerning the velocity of the vortex flow. 
The velocity vanishes inside the nuclear volume, and
the typical rise of the velocity field from zero to the asymptotic
Onsager dependence $v = \hbar/ 2m \rho$ is delayed, again  by about 8 fm at $z=0$.  
The vortex induces some changes on the nuclear density, which becomes  
somewhat elongated along the vortex axis, increases close to the
center of the nucleus and decreases close to the surface, mostly 
around the  equatorial plane (cf. Fig. \ref{fig:dens-ent-nuc}). 
This behaviour can be compared with the case
of the vortex in uniform matter, where the depletion takes place close to the
axis (cf. Fig. \ref{fig:dens-vortice1} above).

\begin{figure}[h]
\centering
\includegraphics[width=6cm]{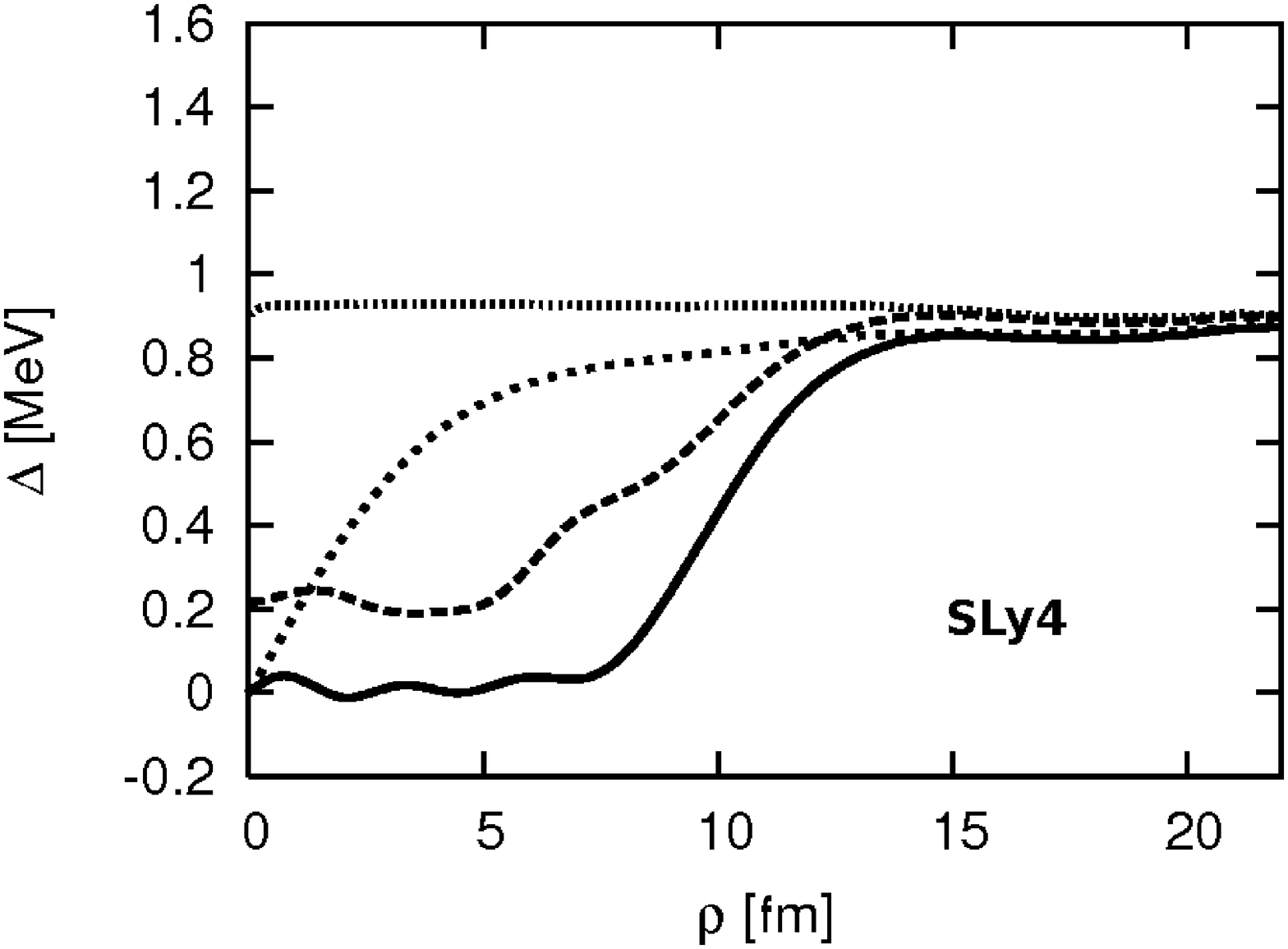}
\includegraphics[width=6cm]{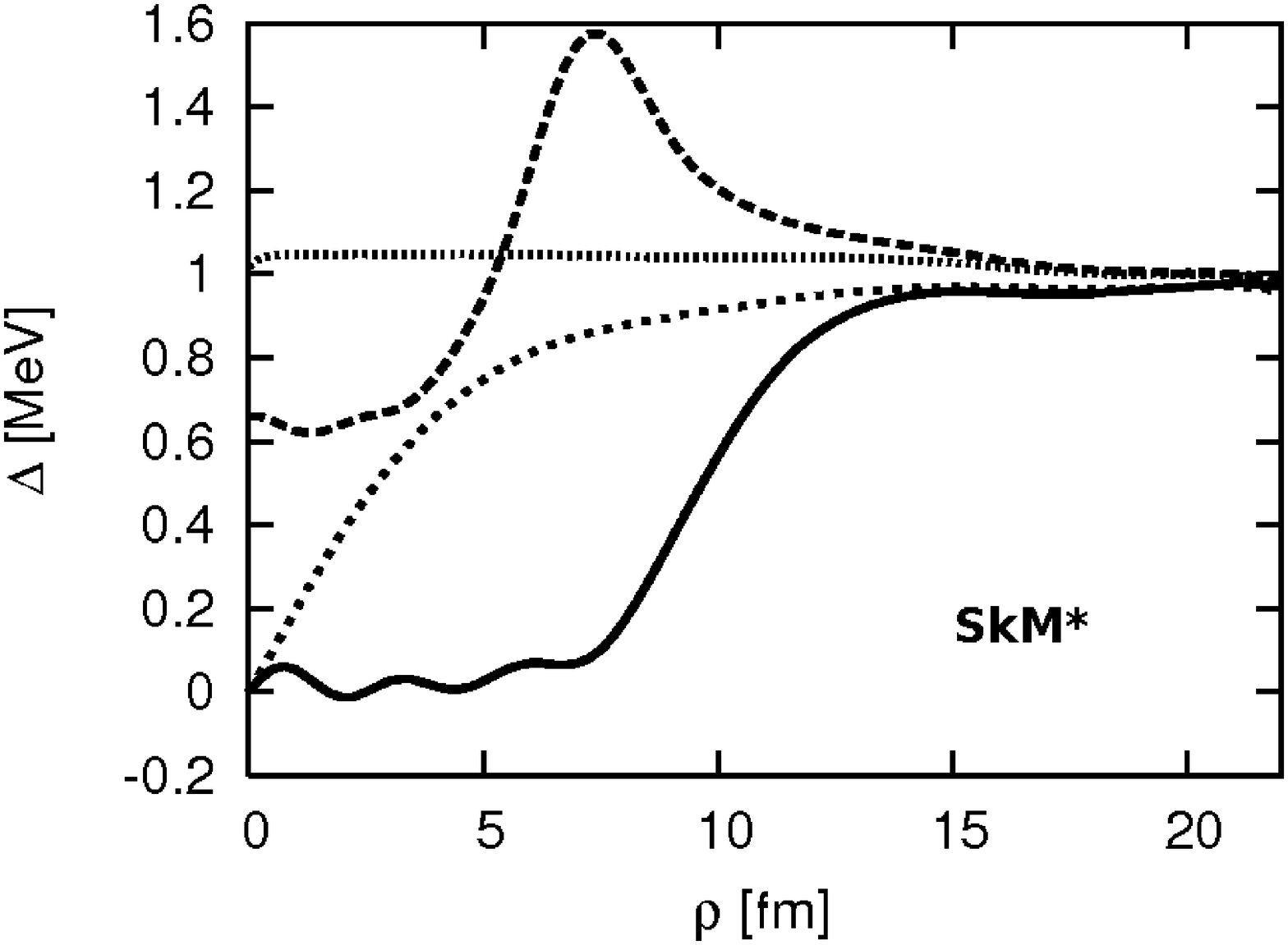}
\caption{The pairing gaps associated with the four different
configurations (pinned vortex ($Z=40,\nu=1$): solid curve; 
nucleus in the neutron sea  ($Z=40,\nu=0$): dashed curve; vortex the neutron sea ($Z=0,\nu=1$): short dashed curve; 
uniform neutron sea ($Z=0,\nu=0$: dotted curve) 
calculated with the SLy4 (left) 
and the SkM* (right) interactions 
at $n_{\infty} \sim $ 0.001 {\rm fm}$^{-3}$.}
\label{fig:tutti-gap-16-skm-sly}
\end{figure}

\begin{figure}[h]
\centering
\includegraphics[width=6cm]{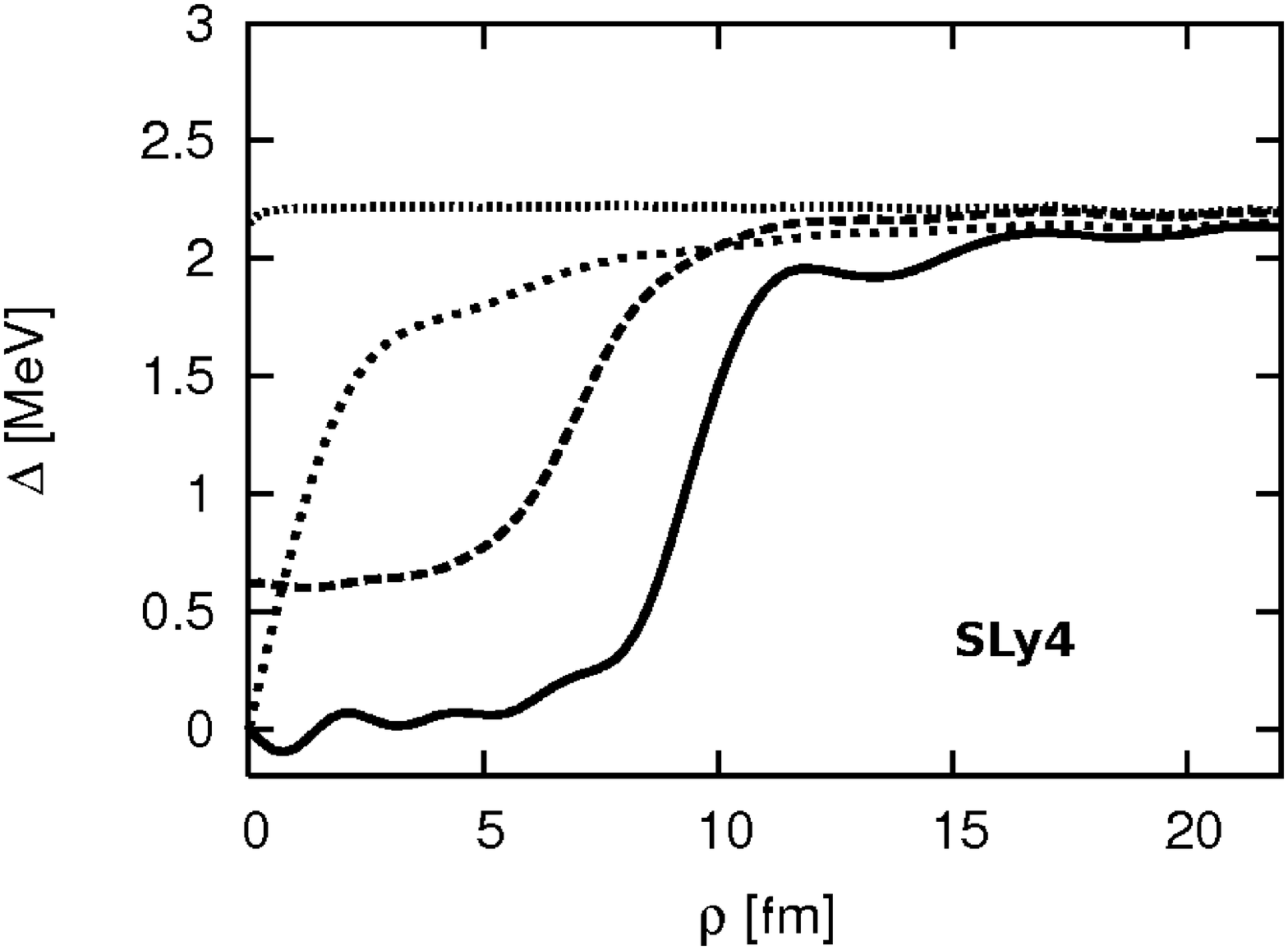}
\includegraphics[width=6cm]{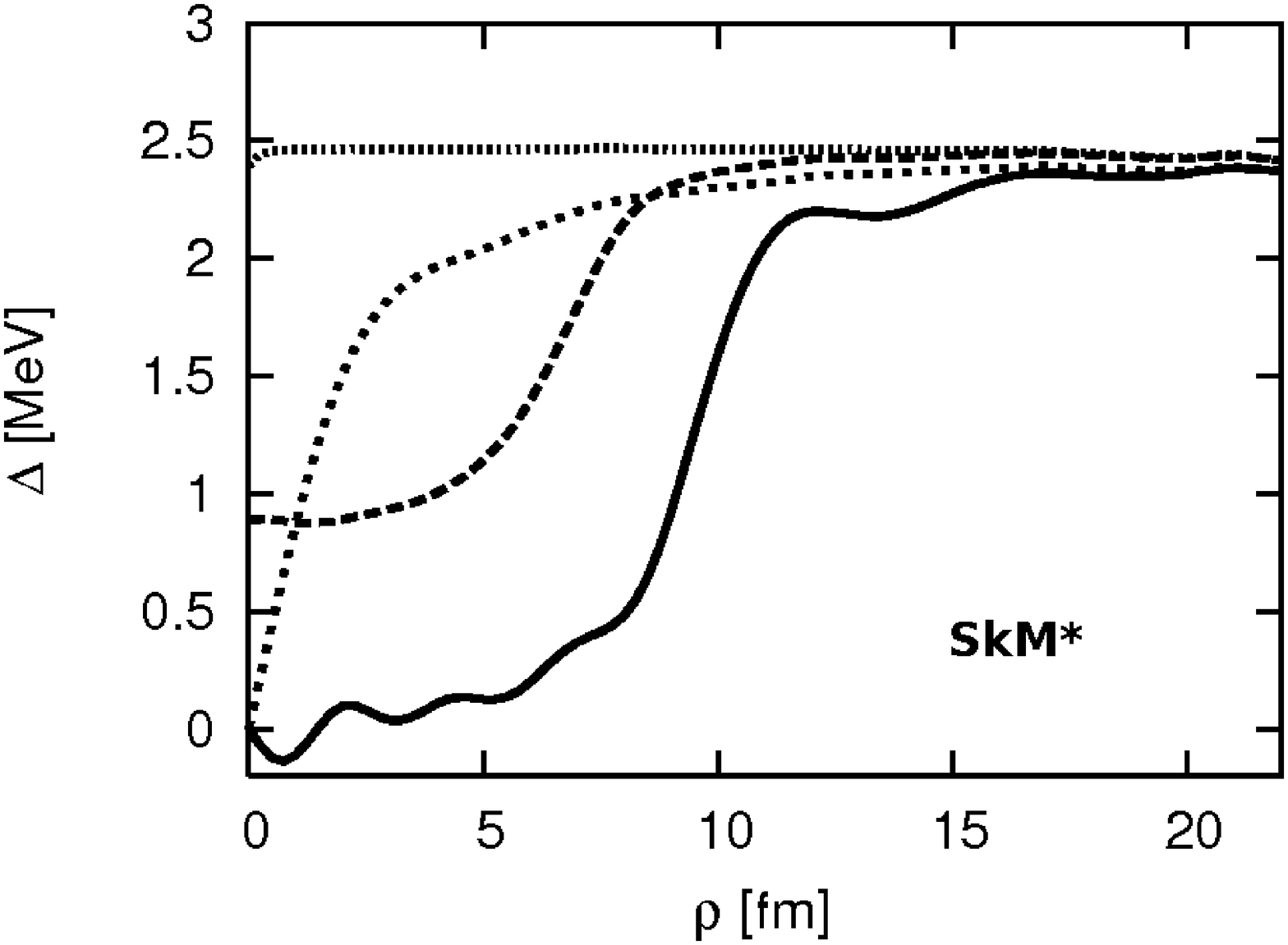}
\caption{
The same as Fig. \ref{fig:tutti-gap-16-skm-sly}, but 
for the density $n_{\infty} \sim $ 0.01 {\rm fm}$^{-3}$.}
\label{fig:tutti-gap-58-skm-sly}
\end{figure}

\begin{figure}[b!]
\centering
\includegraphics[width=6cm]{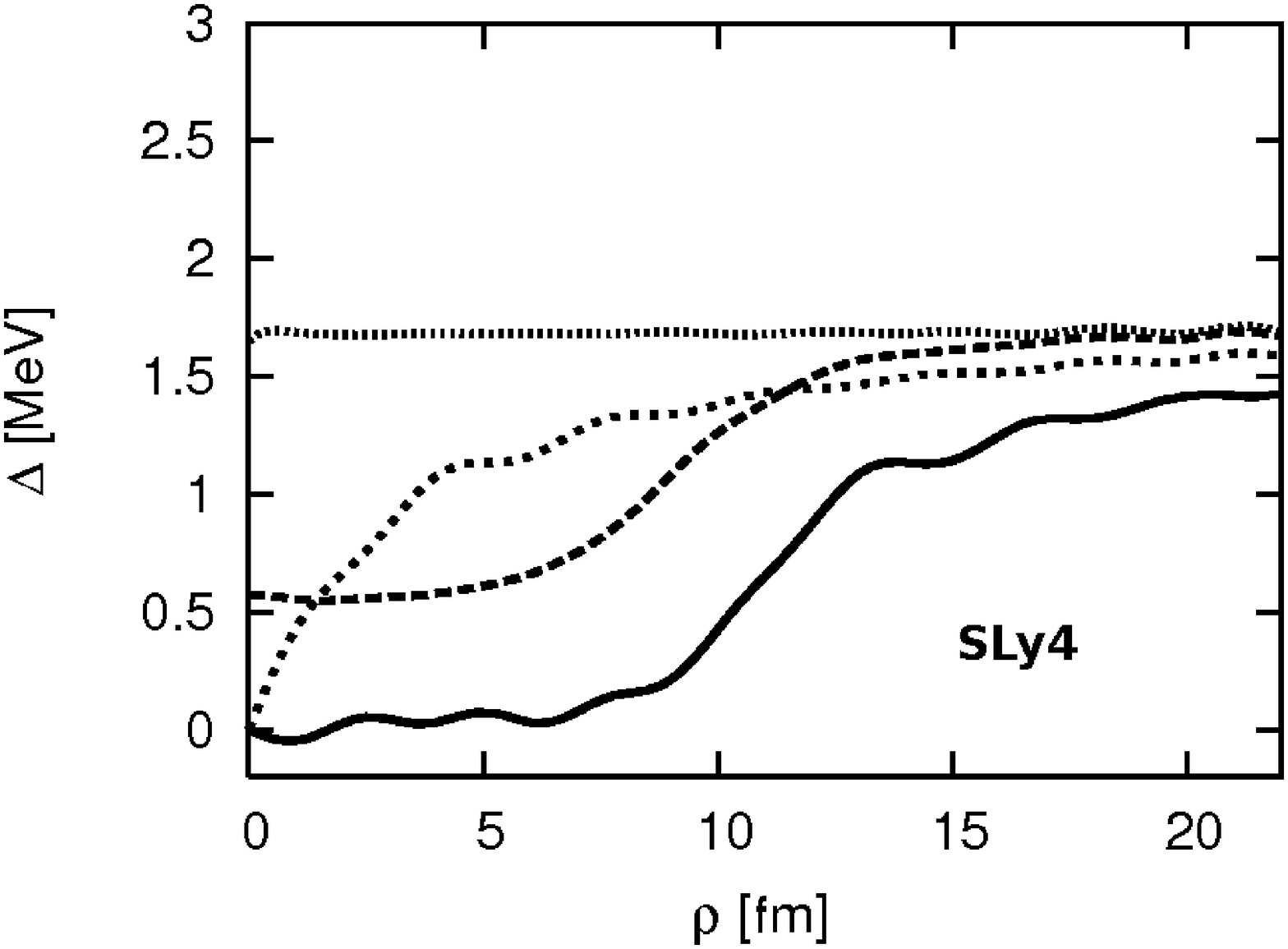}
\includegraphics[width=6cm]{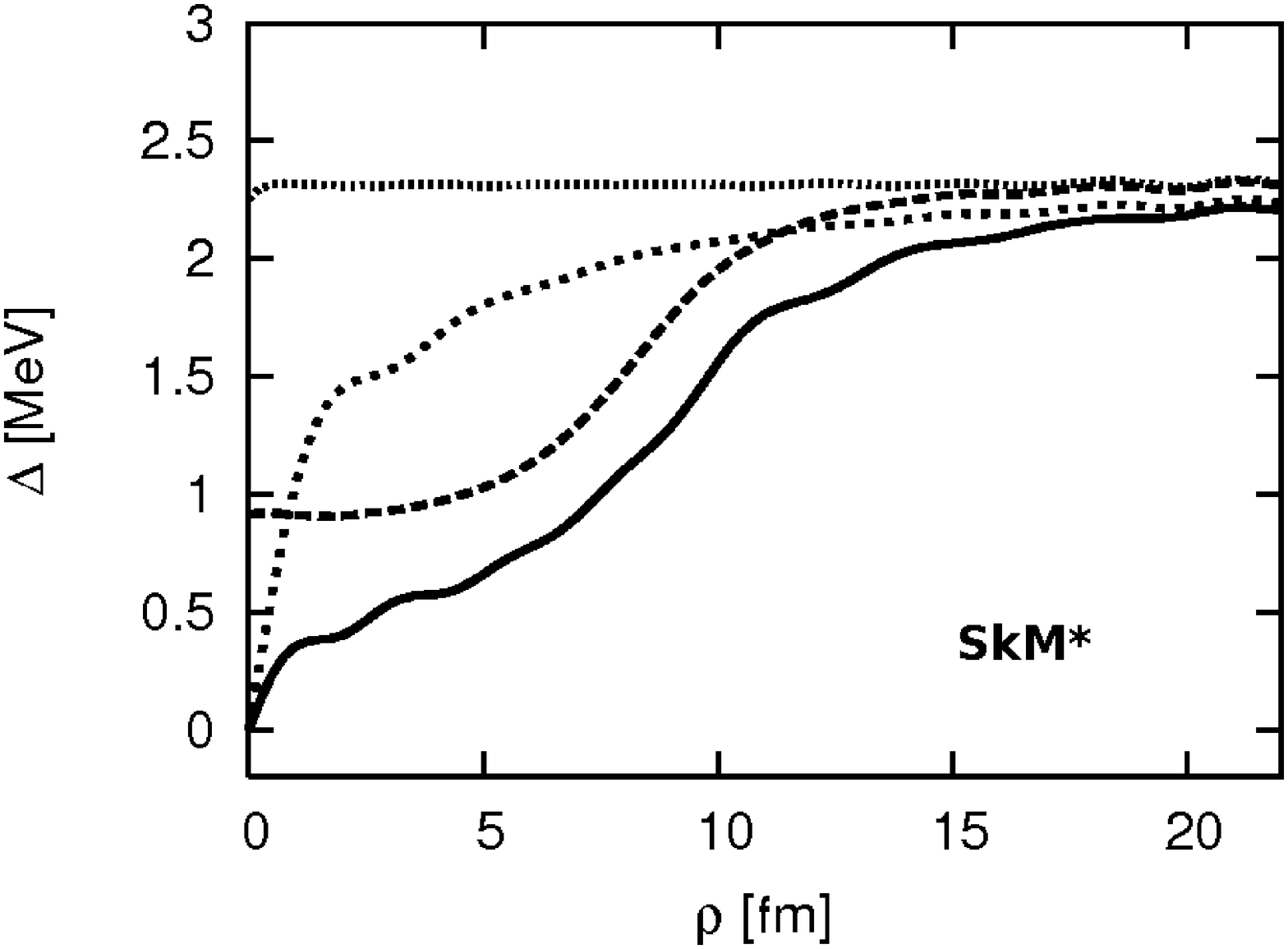}
\caption{
The same as Fig. \ref{fig:tutti-gap-16-skm-sly}, but 
for the density 
$n_{\infty}= $ 0.035 {\rm fm}$^{-3}$.}
\label{fig:tutti-gap-113-skm-sly}
\end{figure}

\begin{figure}[h]
\centering
\begin{center}
\includegraphics[width=6cm,angle=0]{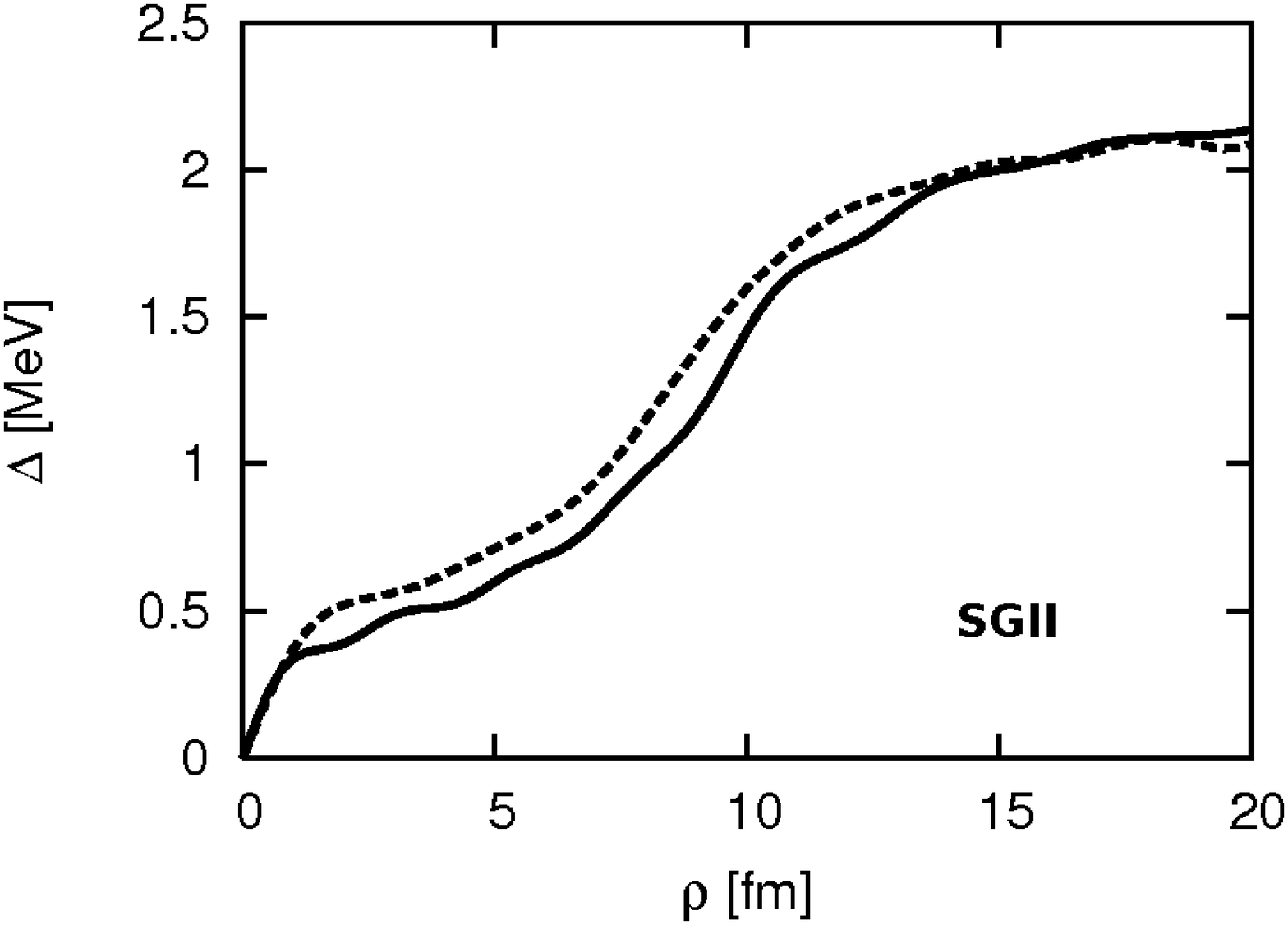}
\includegraphics[width=6cm,angle=0]{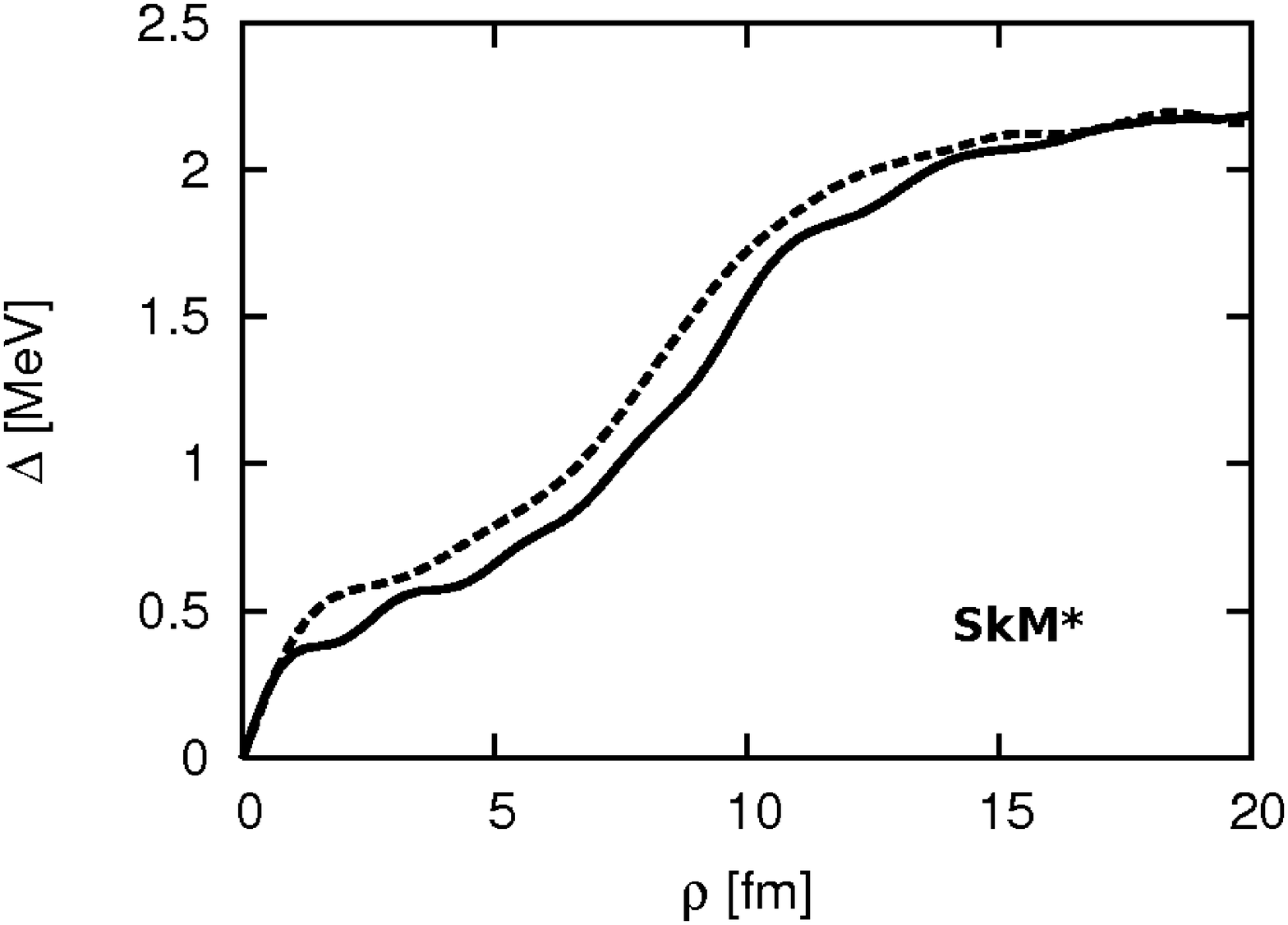}
\end{center}
\caption{The pairing gaps calculated with the  SGII (left) and the SkM* (right)
interactions 
at $n_{\infty}\sim $ 0.035 {\rm fm}$^{-3}$ (solid lines) are compared with
the expression  (\ref{eq:superp}) (dashed lines).}
\label{fig:semi-skm}
\end{figure}

The pairing gaps calculated for the four configurations at 
$n_{\infty}\sim $ 0.001, 0.01 and 0.035 {\rm fm}$^{-3}$
with the interactions SLy4 and SkM* are collected in 
Figs. \ref{fig:tutti-gap-16-skm-sly}-\ref{fig:tutti-gap-113-skm-sly}.
The results obtained with SLy4 at all densities reflect  the main features discussed above
(the results obtained with the SII interaction are very similar).
One can observe that at the highest density, the results obtained with the SkM* 
interaction show  a different behaviour. 
In fact, in this case the pairing gap associated with the vortex 
is not suppressed in the interior of the nucleus, but it shows a constant 
increase: the vortex can penetrate the nucleus.
Differences are present also in the density (which presents a 
slight depletion close to the vortex
axis) and in the velocity (which does not go to zero in the interior of the nucleus), which 
are  closer 
to the  case of the vortex in uniform matter.
The pairing gap associated with the pinned vortex obtained with the  SkM*  
(and with the  SGII)
interaction can be approximated 
rather well assuming that   the gap suppressions produced by the nuclear field  
($\Delta_{nucleus}(\rho)/\Delta_{\infty}$) 
and by the vortex in uniform matter ($\Delta_{vortex}(\rho)/\Delta_{\infty}$) 
respect to the value in uniform matter
$\Delta_{\infty}$ 
act 
simultaneously and independently (cf. Fig. \ref{fig:semi-skm}). 
The superposition of the effects is then the product of the two suppressions: 
\begin{equation}
\Delta (\rho)  = (\Delta_{vortex} (\rho)/\Delta_{\infty}) \cdot (\Delta_{nucl} (\rho)/\Delta_{\infty}) 
\cdot \Delta_{\infty}.
\label{eq:superp}
\end{equation}
This approximation would fail badly in the case of SLy4 interaction, where the pairing of the 
pinned vortex is suppressed much more strongly in the interior, as discussed above.

\newpage

\begin{figure}[!t]
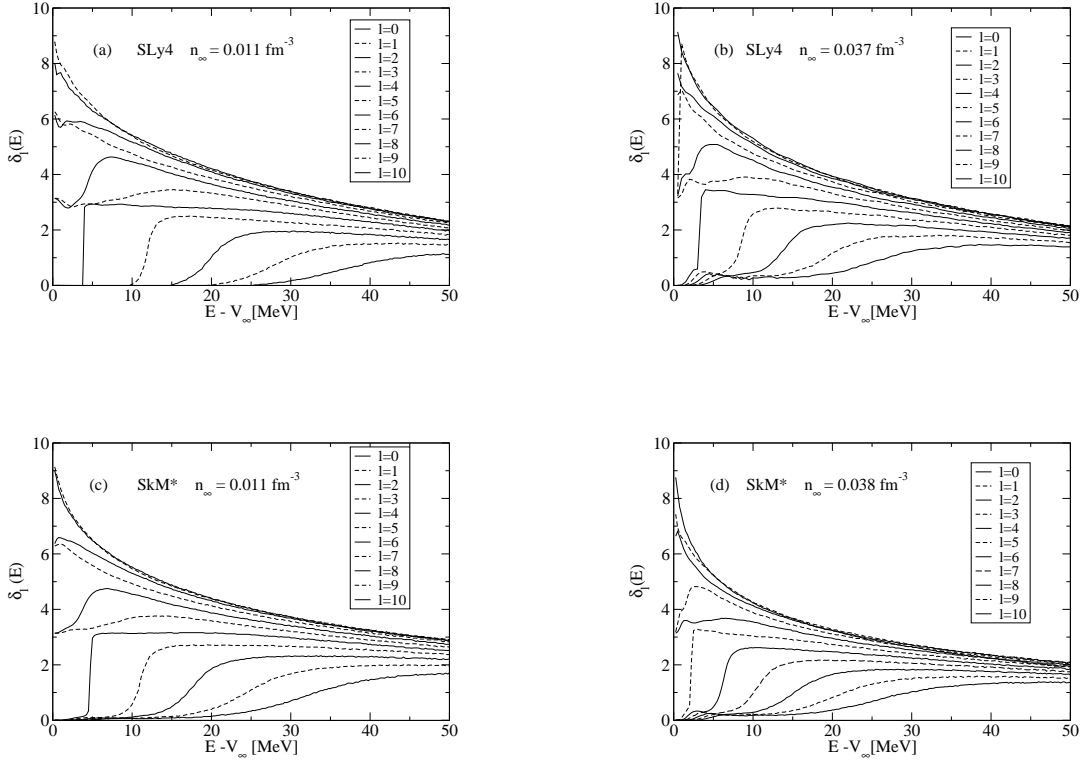

\centering
\begin{center}
\includegraphics[width=6cm,angle=0]{phase_sly4_58.eps}
\hspace{2cm}
\includegraphics[width=6cm,angle=0]{phase_sly4_113.eps}
\end{center}
\vspace{1cm}
\includegraphics[width=6cm,angle=0]{phase_skm_58.eps}
\hspace{2cm}
\includegraphics[width=6cm,angle=0]{phase_skm_113.eps}
\caption{Phase shifts associated with the Hartree-Fock potential
for a spherical nucleus,  obtained with the SLy4 interaction
for n$_{\infty}$ = 0.011 fm$^{-3}$ ( corresponding  to  $\lambda = 5.8 $ MeV (a)) and
for n$_{\infty}$ = 0.037 fm$^{-3}$ ( corresponding  to  $\lambda = 11.3 $ MeV (b)), and 
 with the SkM$^*$ interaction
for n$_{\infty}$ = 0.011 fm$^{-3}$ ( corresponding  to  $\lambda = 5.8 $ MeV (c)), and 
for n$_{\infty}$ = 0.038 fm$^{-3}$ ( corresponding  to  $\lambda = 11.3 $ MeV (d)).  
The energy is referred to the asymptotic value of the mean field potential far
from the nucleus.}    
\label{fig:resonances}
\end{figure}

The different behaviours discussed above can be understood qualitatively in terms of nuclear shell
effects. As we remarked previously, the $\nu=1$ Cooper pairs are formed out of single-particle
levels of opposite parity. 
Shell effects strongly hinder the formation of such Cooper pairs inside the nuclear
volume, as they
separate  the levels of opposite parity  lying  within the nuclear potential
(i.e. the resonant states)
by several MeV, even if these levels lie in the continuum.
This can be seen 
in Fig. \ref{fig:resonances}(a) and (b), where 
we show a calculation of the phase shifts $\delta_l(E)$, for $l=0-10$, 
in the mean field (without the vortex)
associated with the interaction
SLy4  at the densities $n_{\infty}$ = 0.011 fm$^{-3}$ (cf. Fig. \ref{fig:delta2})
and $n_{\infty}$ = 0.037 fm$^{-3}$  (cf. Fig. \ref{fig:delta3}).
In the figures 
the energy $E$ is 
referred  to the asymptotic
value $V_{\infty}$ of the self-consistent potential $U^{HF}$ far from the nucleus
(cf. Fig. \ref{fig:delta2} and Fig. \ref{fig:delta3}).
In Fig. \ref{fig:resonances}(c) and (d) we show the same calculation for
the SkM$^*$ interaction.
We recall that the resonant states and the resonant energy 
are characterized by the conditions  $\delta_l(E_{res}) = (2n+1) \pi/2, n=0,1,..$, 
and 
$d \delta_l(E_{res}) /dE > 0$, which in the present cases are fulfilled by states 
with energies up to about 30-40 MeV.
At low density (cf. Fig. \ref{fig:resonances}(a) and (c)) the
phase shifts are similar for the two interactions. Close to the Fermi energy 
$\lambda - V_{\infty}$, which is about 10 MeV,   
one finds  a rather narrow resonance for $l=7$ ($\Gamma = \pi/(d \delta_l(E_{res}) /dE) \approx 3 $MeV ); 
one also finds narrow  resonances for $l=6$ ($\Gamma \approx 0.5 $MeV ) and
rather well defined resonances for 
$l=8$ ($\Gamma \approx 15$ MeV ), but they lie about 7 MeV below or above the Fermi energy.

In Fig. \ref{fig:resonances}(b) and (d)  
we show a calculation at  higher density, corresponding to 
$\lambda - V_{\infty}$ about 25 MeV. One still finds resonant states
in the region around the Fermi energy, associated with $l=8-9$,
but they are much broader for the SkM$^*$ (the width becoming larger than the centroid energy) 
than for the SLy4 interaction ($\Gamma \approx 9 $ MeV compared to $ E_{res}-V_{\infty} \approx 15 $ MeV
for $l=8$, and $\Gamma $ is still of the order of $E_{res}-V_{\infty}$ for $l=9$), in 
keeping with the fact that the mean-field potential is much shallower.
The role of shell effects in the SkM$^*$ case is correspondingly much 
reduced, and the pairing interaction  is able to mix levels of opposite parity
and to create a pairing field of $\nu=1$ character inside the nuclear volume.

\begin{figure}[!h]
\centering
\begin{center}
\includegraphics[width=6cm,angle=0]{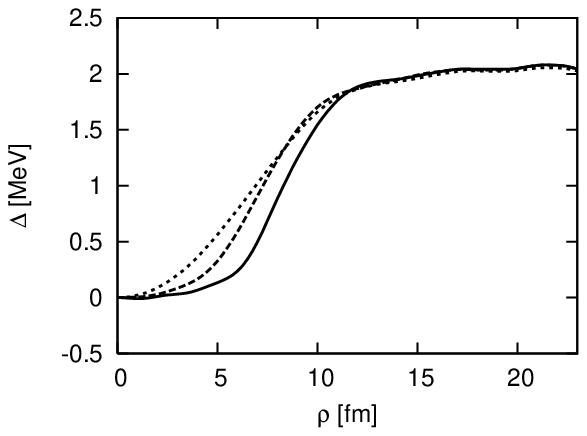}
\includegraphics[width=6cm,angle=0]{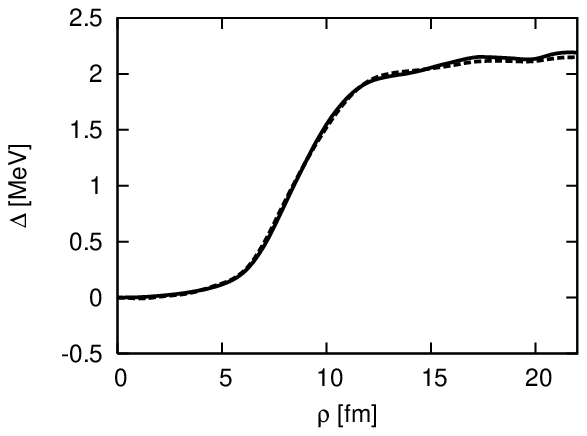}
\end{center}
\caption{
(left): The pairing gap associated with a $\nu=2$ vortex pinned on the nucleus
($Z=40,\nu=1$), calculated with 
the SLy4 interaction at $n_{\infty}= $ 0.011 {\rm fm}$^{-3}$. The cuts refer to fixed value of $z=$ 10 fm 
(short dashed curve),
$z=5$ fm (dashed line) and $z=0 $  fm (solid line). 
(right) The pairing gap at $z=0$ (dashed line) 
is compared to the approximation (\ref{eq:superp}) (solid line) .}    
\label{fig: nu2}
\end{figure}

This interpretation is reinforced by a calculation of a $\nu=2$ vortex. In this case, 
Cooper pairs are formed out of single-particle states of the same parity, and one can see
in Fig. \ref{fig: nu2} that already for 
$n_{\infty}= $ 0.011 {\rm fm}$^{-3}$
 the pairing gap obtained with the SLy4 interaction
does not vanish within the nucleus: its value can be reproduced 
by the simple approximation of Eq. (\ref{eq:superp}),
similar to the case of the SkM* interaction at high density 
(cf. Fig. \ref{fig:semi-skm}).

\begin{figure}
\centering
\begin{center}
\includegraphics[width=10cm,angle=0]{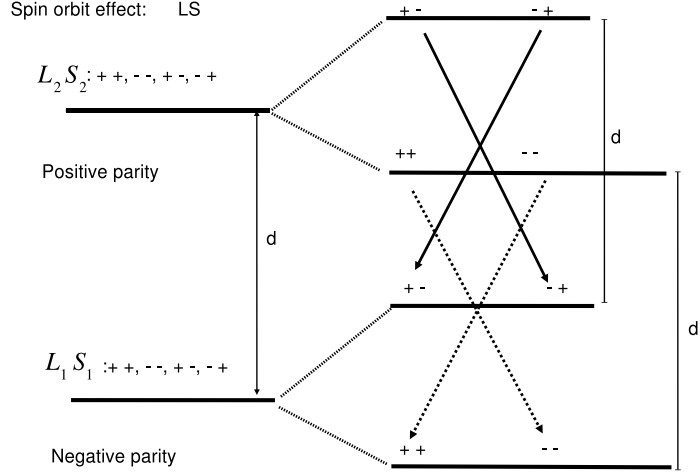}
\end{center}
\caption{Schematic figure showing that the spin-orbit interaction
tends to shift the energy of single-particle pairs involved  in the formation of $S=0, \nu=1$ Cooper pairs 
by the same amount (cf. text).}  
 \label{fig:spin-orbit}
\end{figure}

It could be argued that the spin-orbit interaction, not considered in the present calculation, 
will shift the relative energies of levels of different parity around the Fermi energy, eventually modifying 
the pairing gap we obtain. However the $\nu=1 $
pairing field connects single-particle levels with 
opposite spin projections, and with $L_z$-values such that $L_z(1)+L_z(2)=\nu$.
This implies (except for $L_z=0-1$) that the products  $L_z(1)*S_z(1)$ and $L_z(2)*S_z(2)$ have
the same sign, so that the two levels will be  shifted by about the same amount by the spin-orbit interaction,
as illustrated in Fig. \ref{fig:spin-orbit}.     
One thus does not expect a significant change of the phase space density relevant 
for the vortex formation.

\subsection{Pinning energies}

We now turn to the calculation of the pinning energy, defined above 
in Section III. 
We recall that we compare the energy of the 'pinned' configuration (vortex axis 
passing through the center of the nucleus) 
with the 'interstitial' configuration, in which we assume that the vortex is  
placed far from the nuclei, so that its properties are the same as in uniform matter.
In both cases, we assume that the interaction of the vortex with  'distant' nuclei can
be neglected.
In order to address the validity 
of our approximation in 
a quantitative way, one would need to perform systematic calculations taking 
into account the band structure associated with the lattice \cite{Chamel}.
One expects, however, that our approximation should be reasonable if the 
vortex radial dimension is smaller than the distance between nuclei, as given by
the diameter of the Wigner-Seitz cell calculated by Negele and Vautherin.
To obtain some insight about this point, we 
need an estimate of the extension of the vortex. 
Different definitions
of the vortex radius in uniform matter 
have been introduced. 
\begin{figure}[h!]
\centering
\includegraphics[width=6cm]{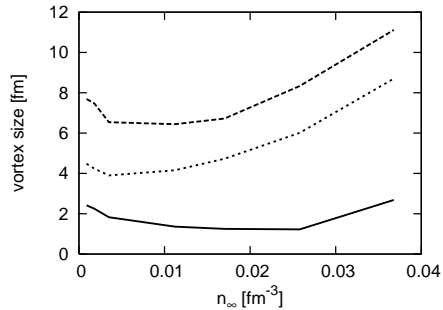}
\caption{ 
Comparison between different definitions of the vortex radius:
$R_{50\%}$ (solid curve), $R_{90\%}$ (dashed curve) and  
coherence length (short dashed curve), calculated with the SLy4 interaction
for a vortex in uniform neutron matter   at various neutron densities.}
\label{fig:vort-rad}
\end{figure}
One often employs  the 
coherence length, $\xi=\hbar^2 k_{F}/m \pi \Delta$, shown 
in Fig. \ref{fig:vort-rad} by the short-dashed curve.
We also show two different estimates of the vortex radius \cite{elgaroy}, 
namely  the distances $R_{90\%}$ or $R_{50\%}$
from the axis where the pairing gap reaches 90\% or  50\% 
of its asymptotic value.
According to this figure,
the coherence length and $R_{90\%}$   display a similar dependence  on density,
while  the quantity $R_{50\%}$ is small and almost independent of 
the density. 
We shall take $R_{90\%}$ as a measure of the vortex radius, 
as this is the most conservative estimate.

Let us then consider the interstitial configuration first.

\begin{figure}[!h]
\centering
\includegraphics[height=6.5cm,width=8cm]{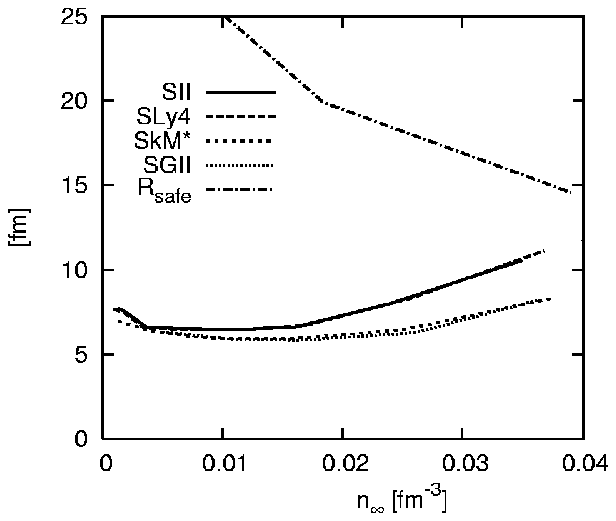}
\includegraphics[width=8cm]{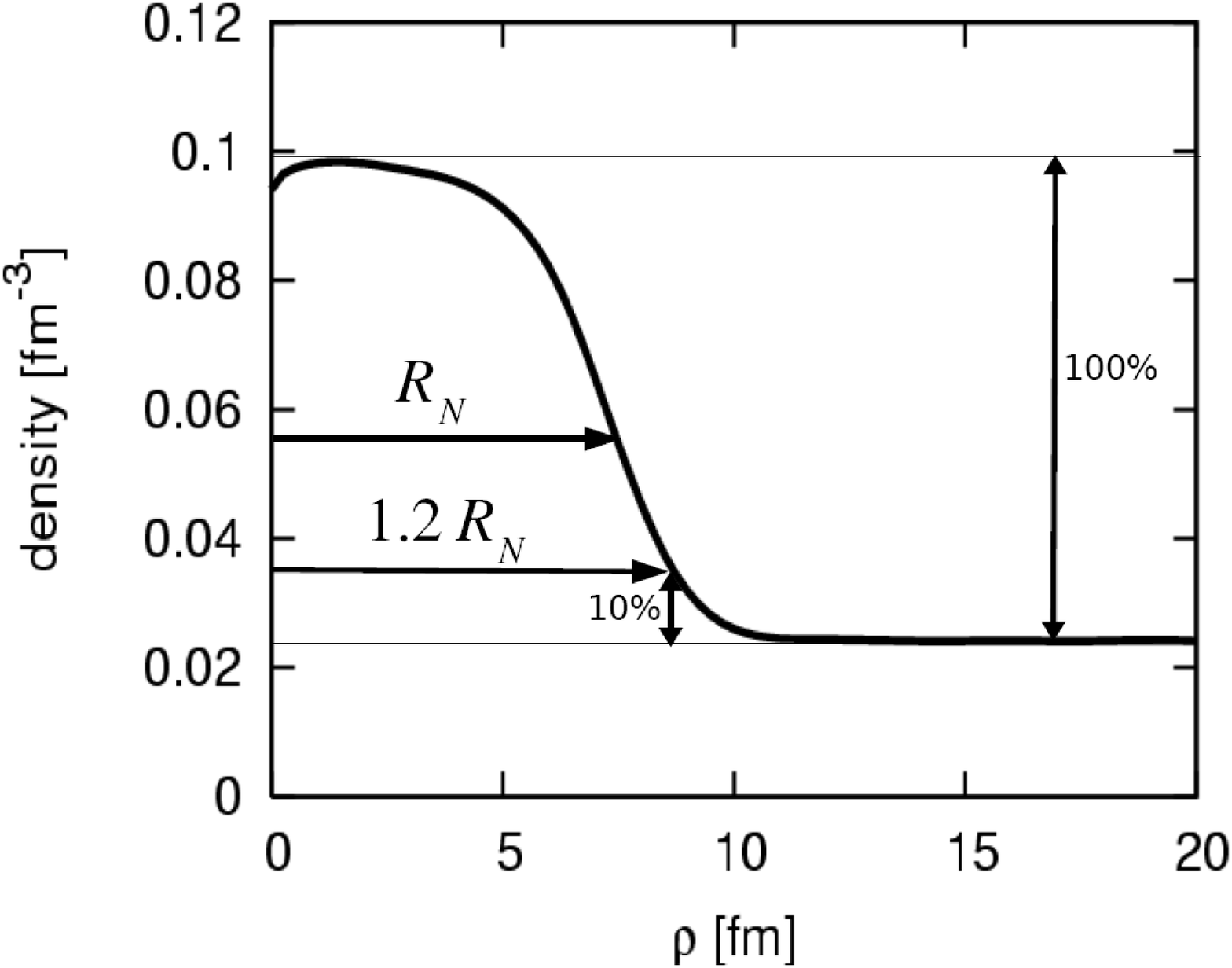}
\caption{(left) Comparison between the vortex radius ($R_{90\%}$) 
and the distance $R_{safe}$. The interactions SII and SLy4 
have very similar radii  at a fixed neutron density  
and their curves are almost superimposed; the same is true 
for SkM* and SGII. 
(right) Illustration of the density profile obtained with the parametrization 
(\ref{fermifun}). At a distance 1.2 $R_0$ from the center of the nucleus, 
the difference between the neutron density and its asymptotic 
value has reduced to 10\% of its maximum value.}
\label{fig:schema1}
\end{figure}

In this case we require that when the vortex has reached its asymptotic gap,
the neutron density is still close to its asymptotic $n_{infty}$, that is, it is not appreciably
affected by the neighbouring nuclei. We then introduce the distance 
$R_{safe} \equiv R_{WS}-1.2 R_N$.
According to  the parametrization 
(\ref {fermifun}),
the quantity  $R_{safe}$ represents the distance 
where the difference between the neutron density and its asymptotic 
value has reduced to 10\% of its maximum value. 
It turns out (see Fig. \ref{fig:schema1}) that the dimension of the vortex  is
always smaller than  $R_{safe}$, although it approaches  this value 
at the largest densities we have considered in our calculations.

\begin{figure}[h!]
\centering
\begin{center}
\includegraphics[width=6cm]{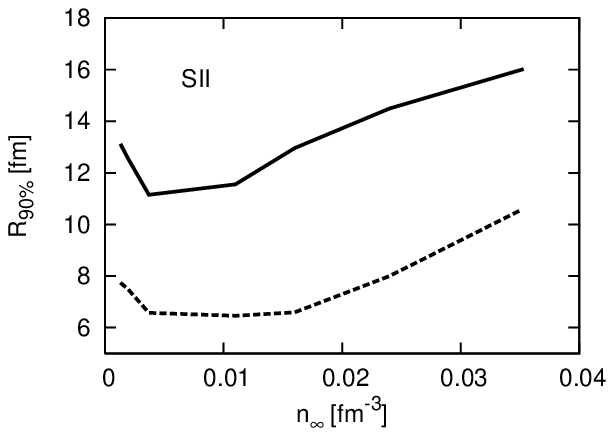}
\includegraphics[width=6cm]{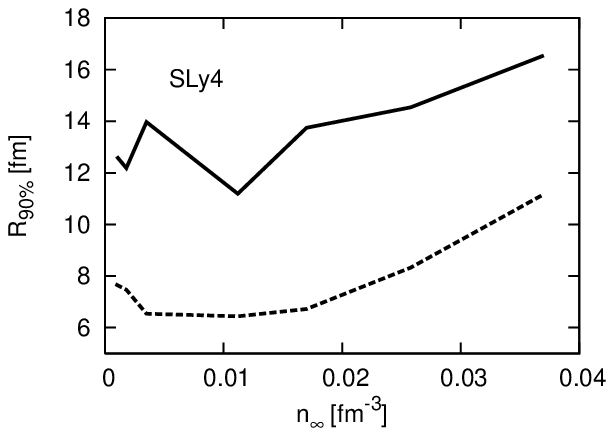}
\includegraphics[width=6cm]{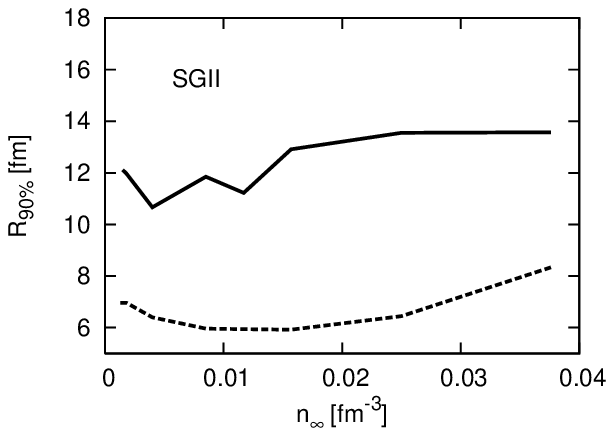}
\includegraphics[width=6cm]{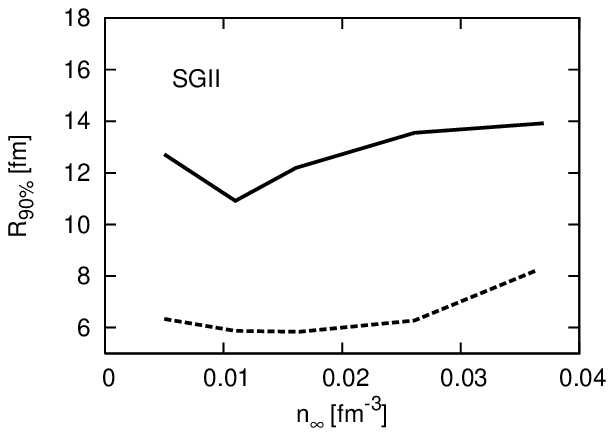}
\end{center}
\caption{The vortex radius, 
$R_{90\%}$ is shown for the pinned vortex (solid curves) 
and for the vortex in uniform matter (dashed curves), 
for the four Skyrme interactions.}
\label{fig:raggi-parag}
\end{figure}

Concerning the pinned configuration,   
the radii of the pinned vortices, measured through $R_{90\%}$, are 4-6 fm larger than in  
uniform matter, depending on the interaction (cf. Fig. \ref{fig:raggi-parag}).
However, the neglect of neighbouring nuclei appears to be even better justified  
for the pinned than for the interstitial configuration discussed previously.
In fact, in the pinned case one should compare $R_{90\%}$ with  the distance between
neighbouring nuclei 2$R_{WS}$ (or better with 2$R_{safe}$).
In all cases the calculated radii of the pinned vortices  
are considerably smaller than this value. 

Consequently we can safely describe the structure of the vortex in the inner crust, both in the
interstitial and in the pinned configuration, within cylindrical boxes
of the order of $R_{90\%}$.
However, in the calculation of the pinning energy we need a common box size value for the two vortex
configurations. This implies using a box of radius close to   the value 
$R_{90\%}$  for the pinned configuration, 
which is the largest one: this quantity remains smaller than $R_{safe}$, as can be seen
comparing Fig. \ref{fig:raggi-parag} with Fig. \ref{fig:schema1}.

In practice we have used  somewhat larger radii in the calculation of the pinning energy 
in order to check that our boundary conditions 
do not produce any relevant effect (see appendix A). 

The fact that the presence of neighbouring nuclei is not relevant for the calculation of the 
pinning energy for the densities considered in this paper, is confirmed by explicit estimates
carried out in refs. \cite{Epstein88,DonatiPizzo} (see Appendix C).

\newpage

\begin{figure}[h]
\centering
\includegraphics[width=6cm]{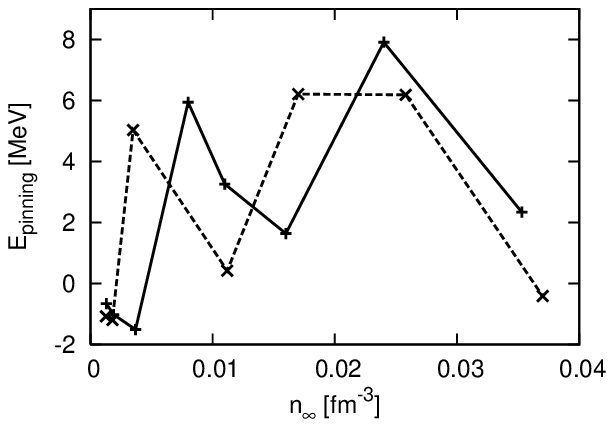}
\includegraphics[width=6cm]{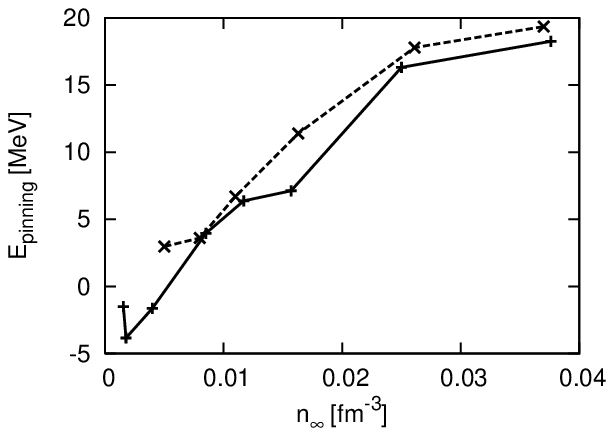}
\caption{((left) Pinning energy calculated 
as a function of the asymptotic neutron density with the interactions SLy4 (dashed line)
and SII (solid line).
The lines connect the
calculated points.
 (right) The same for the interactions SkM* (solid line)  and SGII (dashed line). }
\label{fig:tutti-pinn}
\end{figure}

After calculating the energies of the four different configurations of Fig. \ref{fig:configurazioni}, and
using Eqs.(\ref{eq:epinn}) and  (\ref{eq:ecost})  we compute the pinning energies 
shown in Fig.  \ref{fig:tutti-pinn} for different values of $\lambda$ and $n_{\infty}$ (cf. Tables I and II).
An estimate of the numerical errors associated with  the calculations
is given in Appendix B. 
Some details of the energy calculations performed with the interactions SLy4 and SkM* are reported
in Tables III and IV, in which we list the values of the chemical potential,
of the number of neutrons, of  the total energy $E_{tot}$ and of  the
pairing energy $E_{pair}$ (cf. Appendix A). The total energy associated with the 
two calculations with a  vortex  (labeled 'Pinned' and 'Vortex'), 
has been corrected by the term $\Delta E$ introduced at the end of Section III,
in order to account for the  difference in 
number of neutrons compared to the corresponding calculations 
without vortex (labeled 'Nucleus' and 'Uniform').

All the interactions  yield weak pinning ($E_{pinning} \lesssim$ 0) at the smallest densities
we have calculated
($n_{\infty} \sim 0.001$ fm$^{-3}$), and  
yield a few MeV antipinning for $n_{\infty} = 0.01-0.02$ fm$^{-3}$.  
At higher density,
the SkM* and SGII interaction lead to a strong antipinning ($E_{pinning} >$ 0) , in marked contrast with the
other two interactions, which instead again produce weak pinning at the highest
density points. This different density dependence can be related to 
the qualitative differences in the vortex structure, discussed in the previous section.  



\newpage

 \begin{table}[h!]
    \begin{center}
       \begin{tabular}{|c|c|c|c|c|c|c|c|}
       \hline
\multicolumn{8}{|c|}{SLy4 \;\;\; $n_{\infty}$=0.001 fm$^{-3}$}\\
       \hline
       & $\lambda$  & N & $E_{pair}$ & $E_{tot}$ & $E_{tot}$+$\Delta E$ &\; $E^{vor}$ \; &  \; $E_{pinning}$ \; \\
       \hline
     Pinned  & 1.65  &  \; 188.52 \; &  \; -103.99 \; & \; -904.05 \;  & \;-902.68 \; &  \multirow{2}{*}{3.37} &  \\         \cline{1-6}
     Nucleus  & 1.6  &  189.35 &  -120.61 & -906.05   &       & & \multirow{2}{*}{-1.08} \\         \cline{1-7}
     Vortex  & 1.65  &  82.09  &  -112.32 & 93.60   &   96.02   &  \multirow{2}{*}{4.45} & \\         \cline{1-6}
     Uniform  & 1.6  &  83.56  &  -130.98 & 91.57    &            &   & \\         \hline
       \end{tabular}
    \end{center}
\label{Tab_sly4_1.6}
\end{table}

\begin{table}[h!]
    \begin{center}
       \begin{tabular}{|c|c|c|c|c|c|c|c|}
       \hline
\multicolumn{8}{|c|}{SLy4 \;\;\; $n_{\infty}$=0.002 fm$^{-3}$}\\
       \hline
       & $\lambda$  & N & $E_{pair}$ & $E_{tot}$ & $E_{tot}$+$\Delta E$ &\; $E^{vor}$ \; &  \; $E_{pinning}$ \; \\
       \hline
     Pinned   & 2.05    &   \;229.87   \; &  \;-167.31  \;   & \;-825.98   \;  & \; -821.42     \; &  \multirow{2}{*}{5.69  } &  \\         \cline{1-6}
     Nucleus  & 2.00    &     232.09      &    -195.74       &   -827.11      &            &                         & \multirow{2}{*}{ -1.20 } \\         \cline{1-7}
     Vortex   & 2.05    &     125.93      &    -181.19       &   177.47       &    181.11        &  \multirow{2}{*}{ 6.89 } & \\         \cline{1-6}
     Uniform  & 2.00    &     127.71      &    -208.13       &   174.22       &            &                         & \\         \hline
       \end{tabular}
    \end{center}
\end{table}

 \begin{table}[h!]
    \begin{center}
       \begin{tabular}{|c|c|c|c|c|c|c|c|}
       \hline
\multicolumn{8}{|c|}{SLy4 \;\;\; $n_{\infty}$=0.004 fm$^{-3}$}\\
       \hline
       & $\lambda$  & N & $E_{pair}$ & $E_{tot}$ & $E_{tot}$+$\Delta E$ &\; $E^{vor}$ \; &  \; $E_{pinning}$ \; \\
       \hline
     Pinned  & 3.05  &  \; 413.36 \; &  \; -354.79 \; & \; -288.67 \;  & \;-288.11 \; &  \multirow{2}{*}{19.93} &  \\         \cline{1-6}
     Nucleus  & 3.0  &  413.55 & -447.36 &  -308.04   &  & & \multirow{2}{*}{5.03} \\         \cline{1-7}
     Vortex  & 3.05 &  272.14 &   -407.00   &   567.28   & 576.10 &  \multirow{2}{*}{14.90} & \\         \cline{1-6}
     Uniform  & 3.0  &  275.03  &  -456.72    &   561.20  &     &   & \\         \hline
       \end{tabular}
    \end{center}
\label{Tab_sly4_4.3}
\end{table}

\begin{table}[h!]
    \begin{center}
       \begin{tabular}{|c|c|c|c|c|c|c|c|}
       \hline
\multicolumn{8}{|c|}{SLy4 \;\;\; $n_{\infty}$=0.008 fm$^{-3}$}\\
       \hline
       & $\lambda$  & N & $E_{pair}$ & $E_{tot}$ & $E_{tot}$+$\Delta E$ &\; $E^{vor}$ \; &  \; $E_{pinning}$ \; \\
       \hline
     Pinned   & 4.75   &  783.54 \;       & -816.18     & \; 1220.08  \;  & \;1252.77      \; &  \multirow{2}{*}{ 34.82 } &  \\         \cline{1-6}
     Nucleus  & 4.70    &  790.42         & -937.12        &   1218.06        &            &                         & \multirow{2}{*}{ 0.40 } \\         \cline{1-7}
     Vortex   & 4.75    & 653.65          &  -884.15         & 2128.650         &  2148.74          &  \multirow{2}{*}{ 34.42 } & \\         \cline{1-6}
     Uniform  & 4.70    &  657.88         & -977.32         & 2114.32         &            &                         & \\         \hline
       \end{tabular}
    \end{center}
\end{table}

 \begin{table}[h!]
    \begin{center}
       \begin{tabular}{|c|c|c|c|c|c|c|c|}
       \hline
\multicolumn{8}{|c|}{SLy4 \;\;\; $n_{\infty}$=0.011 fm$^{-3}$}\\
       \hline
       & $\lambda$  & N & $E_{pair}$ & $E_{tot}$ & $E_{tot}$+$\Delta E$ &\; $E^{vor}$ \; &  \; $E_{pinning}$ \; \\
       \hline
     Pinned  & 5.73  &  \; 1086.7 \; &  \; -1068.50\; & \; 2858.24 \;  & \;2839.90 \; &  \multirow{2}{*}{50.09} &  \\         \cline{1-6}
     Nucleus  & 5.63 &  1083.5 &  -1215.42 & 2789.81   &       & & \multirow{2}{*}{0.42} \\         \cline{1-7}
     Vortex  & 5.74  &  965.4 &  -1146.52 & 3816.77  &   3816.77  &  \multirow{2}{*}{49.67} & \\         \cline{1-6}
     Uniform  & 5.67  & 962.6  &  -1258.76 & 3751.11   &            &   & \\         \hline
       \end{tabular}
    \end{center}
\label{Tab_sly4_5.8}
\end{table}

 \begin{table}[h!]
    \begin{center}

       \begin{tabular}{|c|c|c|c|c|c|c|c|}
       \hline
\multicolumn{8}{|c|}{SLy4 \;\;\; $n_{\infty}$=0.017 fm$^{-3}$}\\
       \hline
       & $\lambda$  & N & $E_{pair}$ & $E_{tot}$ & $E_{tot}$+$\Delta E$ &\; $E^{vor}$ \; &  \; $E_{pinning}$ \; \\
       \hline
     Pinned  & 7.05  &  \; 1647.33 \; &  \; -1245.87 \; & \; 6544.28 \;  & \;6595.52 \; &  \multirow{2}{*}{77.09} &  \\         \cline{1-6}
     Nucleus  & 7.0  &  1654.60 &  -1460.55 & 6518.42   &       & & \multirow{2}{*}{6.21} \\         \cline{1-7}
     Vortex  & 7.05  &  1498.32 &  -1375.65 &  7322.81  &   7346.86   &  \multirow{2}{*}{70.88} & \\         \cline{1-6}
     Uniform  & 7.0  &  1501.73  &  -1523.55 & 7275.98    &            &   & \\         \hline
       \end{tabular}
    \end{center}
\label{Tab_sly4_7}
\end{table}

\vsize=10cm
\begin{table}[h!]
    \begin{center}
       \begin{tabular}{|c|c|c|c|c|c|c|c|}
       \hline
\multicolumn{8}{|c|}{SLy4 \;\;\; $n_{\infty}$=0.026 fm$^{-3}$}\\
       \hline
       & $\lambda$  & N & $E_{pair}$ & $E_{tot}$ & $E_{tot}$+$\Delta E$ &\; $E^{vor}$ \; &  \; $E_{pinning}$ \; \\
       \hline
     Pinned   & 8.55    &   \;2452.99   \; &  \; -1174.10 \;   & \;  12956.57 \;  & \;  13058.04    \; &  \multirow{2}{*}{ 104.02 } &  \\         \cline{1-6}
     Nucleus  & 8.50    &    2464.86       &    -1467.41       &     12954.02     &            &                         & \multirow{2}{*}{ 6.19 } \\         \cline{1-7}
     Vortex   & 8.55    &    2301.11       &    -1362.87       &     13712.91     &     13714.88        &  \multirow{2}{*}{ 97.83 } & \\         \cline{1-6}
     Uniform  & 8.50    &    2301.34       &    -1541.69       &     13617.05     &            &                         & \\         \hline
       \end{tabular}
    \end{center}
\end{table}


 \begin{table}[h!]
    \begin{center}
       \begin{tabular}{|c|c|c|c|c|c|c|c|}
       \hline
\multicolumn{8}{|c|}{SLy4 \;\;\; $n_{\infty}$=0.037 fm$^{-3}$}\\
       \hline
       & $\lambda$  & N & $E_{pair}$ & $E_{tot}$ & $E_{tot}$+$\Delta E$ &\; $E^{vor}$ \; &  \; $E_{pinning}$ \; \\
       \hline
     Pinned  & 10.05  &  \; 3332.59 \; &  \; -762.77 \; & \; 22821.65 \;  & \;22814.52 \; &  \multirow{2}{*}{122.01} &  \\         \cline{1-6}
     Nucleus  & 10.0 &  3332.53 &  -1159.46   & 22692.50 & & & \multirow{2}{*}{-0.41} \\         \cline{1-7}
     Vortex  & 10.05&  3500.58 &  -1012.38   & 23418.71 &  23419.28 & \multirow{2}{*}{122.42} & \\         \cline{1-6}
     Uniform  & 10.0 &  3501.29  &  -1230.38   & 23296.86 &   &   & \\         \hline
       \end{tabular}
    \end{center}
    \caption{Calculation of the pinning energy with the interaction SLy4 for the various  
neutron densities we have considered.
(cf. Table I).
In each case, we give the values  of the chemical potential $\lambda$,
(in MeV), the number of neutrons $N$, the pairing energy $E_{pair}$ (in MeV), 
the  total energy $E_{tot}$ (in MeV) (cf. Appendix A),
associated with the four configurations labeled {\it Pinned} (vortex on a nucleus), {\it Nucleus}
(nucleus in the neutron sea), {\it Vortex} (vortex in uniform matter), {\it Uniform} (uniform matter).
For the configurations  {\it Pinned} and {\it Vortex} we give the total energy $E_{tot}$+$\Delta E$ (in MeV),
corrected in order to obtain the same number of neutrons as in the configurations {\it Nucleus} and {\it Uniform}.
We also give the cost $E^{vor}$ (in MeV) to create the vortex pinned on the nucleus and the vortex in uniform
matter, and finally the pinning energy $E_{pinning}$ (in MeV). }
\label{Tab_Sly4_11.3}
\end{table}

\newpage

 \begin{table}[h!]
    \begin{center}
       \begin{tabular}{|c|c|c|c|c|c|c|c|}
       \hline
\multicolumn{8}{|c|}{SkM* \;\;\; $n_{\infty}$=0.001 fm$^{-3}$}\\
       \hline
       & $\lambda$  & N & $E_{pair}$ & $E_{tot}$ & $E_{tot}$+$\Delta E$ &\; $E^{vor}$ \; &  \; $E_{pinning}$ \; \\
       \hline
     Pinned  & 1.27  &  \; 199.36 \; &  \; -120.64 \; & \; -946.97 \;  & \;-938.50 \; &  \multirow{2}{*}{3.49} &  \\         \cline{1-6}
     Nucleus  & 1.2  &  206.14 &  -169.11 & -941.99   &       & & \multirow{2}{*}{-1.51} \\         \cline{1-7}
     Vortex  & 1.25  &  89.99  &  -127.51 & 82.77    &   86.23   &  \multirow{2}{*}{5.00} & \\         \cline{1-6}
     Uniform  & 1.2  &  92.74  &  -149.53 & 81.23    &            &   & \\         \hline
       \end{tabular}
    \end{center}
\label{Tab_SkM*_1.6}
\end{table}

\begin{table}[h!]
    \begin{center}
       \begin{tabular}{|c|c|c|c|c|c|c|c|}
       \hline
\multicolumn{8}{|c|}{SkM* \;\;\; $n_{\infty}$=0.002 fm$^{-3}$}\\
       \hline
       & $\lambda$  & N & $E_{pair}$ & $E_{tot}$ & $E_{tot}$+$\Delta E$ &\; $E^{vor}$ \; &  \; $E_{pinning}$ \; \\
       \hline
     Pinned   &  1.40  & \;221.48 \;  &   \;-156.20   \; &  \; -917.12 \;   & \; -894.54  \;  &  \multirow{2}{*}{ 6.45 } &  \\         \cline{1-6}
     Nucleus  &  1.35  & 237.61  &   -229.53        &    -897.14       &          &                       & \multirow{2}{*}{-3.85  } \\         \cline{1-7} 
     Vortex   & 1.40   &  117.20 &   -172.05        &  120.66         & 124.78         & \multirow{2}{*}{ 2.61 } & \\         \cline{1-6}
     Uniform  & 1.35   & 120.15  &   -198.97        &   118.33        &             &                         & \\         \hline
       \end{tabular}
    \end{center}
\end{table}

\begin{table}[h!]
    \begin{center}
       \begin{tabular}{|c|c|c|c|c|c|c|c|}
       \hline
\multicolumn{8}{|c|}{SkM* \;\;\; $n_{\infty}$=0.004 fm$^{-3}$}\\
       \hline
       & $\lambda$  & N & $E_{pair}$ & $E_{tot}$ & $E_{tot}$+$\Delta E$ &\; $E^{vor}$ \; &  \; $E_{pinning}$ \; \\
       \hline
     Pinned   & 2.05    &   \; 378.80  \; &  \;  -413.59 \;   & \; -626.93  \;  & \;  -499.45    \; &  \multirow{2}{*}{13.70  } &  \\         \cline{1-6}
     Nucleus  & 2.00    &      440.98     &      -524.10      &    -513.16      &            &                         & \multirow{2}{*}{-1.63  } \\         \cline{1-7}
     Vortex   & 2.05    &      278.08    &      -440.18      &     417.50      &      429.03      &  \multirow{2}{*}{ 15.33 } & \\         \cline{1-6}
     Uniform  & 2.00    &      283.70     &      -495.26      &     413.70      &            &                         & \\         \hline
       \end{tabular}
    \end{center}
\end{table}

\begin{table}[h!]
    \begin{center}
       \begin{tabular}{|c|c|c|c|c|c|c|c|}
       \hline
\multicolumn{8}{|c|}{SkM* \;\;\; $n_{\infty}$=0.008 fm$^{-3}$}\\
       \hline
       & $\lambda$  & N & $E_{pair}$ & $E_{tot}$ & $E_{tot}$+$\Delta E$ &\; $E^{vor}$ \; &  \; $E_{pinning}$ \; \\
       \hline
     Pinned   &  3.05  &  \;785.03 \; &  \; -919.51 \; & \; 459.09 \; &  \; 512.59 \; &  \multirow{2}{*}{39.67  } &  \\         \cline{1-6}
     Nucleus  & 3.00 & 802.57    &     -1085.98      &   472.92        &               &                         & \multirow{2}{*}{ 3.95 } \\         \cline{1-7}
     Vortex   & 3.05 &  659.10   &     -1002.84      &   1459.46        &  1486.87     &  \multirow{2}{*}{35.72 } & \\         \cline{1-6}
     Uniform  & 3.00 &  668.09    &     -1106.10      &   1451.15        &              &                         & \\         \hline
       \end{tabular}
    \end{center}
\end{table}

 \begin{table}[h!]
    \begin{center}
       \begin{tabular}{|c|c|c|c|c|c|c|c|}
       \hline
\multicolumn{8}{|c|}{SkM* \;\;\; $n_{\infty}$=0.012 fm$^{-3}$}\\
       \hline
       & $\lambda$  & N & $E_{pair}$ & $E_{tot}$ & $E_{tot}$+$\Delta E$ &\; $E^{vor}$ \; &  \; $E_{pinning}$ \; \\
       \hline
     Pinned  & 3.62  &  \; 1057.68 \; &  \; -1235.92 \; & \; 1420.66 \;  & \;1436.74 \; &  \multirow{2}{*}{57.07} &  \\         \cline{1-6}
     Nucleus  & 3.53  &  1062.12 & -1412.41   & 1379.67    &  & & \multirow{2}{*}{6.38} \\         \cline{1-7}
     Vortex  & 3.60 &  926.77 &   -1317.69    &   2401.44   & 2398.33 &  \multirow{2}{*}{50.69} & \\         \cline{1-6}
     Uniform  & 3.53  &  925.91  & -1433.06    &     2347.63      & &   & \\         \hline
       \end{tabular}
    \end{center}
\label{Tab_SkM*_5.8}
\end{table}

\begin{table}[h!]
    \begin{center}
       \begin{tabular}{|c|c|c|c|c|c|c|c|}
       \hline
\multicolumn{8}{|c|}{SkM* \;\;\; $n_{\infty}$=0.016 fm$^{-3}$}\\
       \hline
       & $\lambda$  & N & $E_{pair}$ & $E_{tot}$ & $E_{tot}$+$\Delta E$ &\; $E^{vor}$ \; &  \; $E_{pinning}$ \; \\
       \hline
     Pinned   & 4.25    &  \;1397.33 \;   &  \; -1534.87 \;   & \; 2827.25  \;  & \; 2970.22     \; &   \multirow{2}{*}{73.78  } &  \\         \cline{1-6}
     Nucleus  & 4.20    &   1430.97        &   -1749.45        &   2896.34       &            &                         & \multirow{2}{*}{ 7.97 } \\         \cline{1-7}
     Vortex   & 4.25    &  1280.31         &    -1636.09       &   3864.68       &    3901.98       &  \multirow{2}{*}{ 65.81 } & \\        \cline{1-6}
     Uniform  & 4.20    &   1289.09        &    -1789.18       &    3837.70      &            &                         & \\         \hline
       \end{tabular}
    \end{center}
\end{table}

\begin{table}[h!]
    \begin{center}
       \begin{tabular}{|c|c|c|c|c|c|c|c|}
       \hline
\multicolumn{8}{|c|}{SkM* \;\;\; $n_{\infty}$=0.025 fm$^{-3}$}\\
       \hline
       & $\lambda$  & N & $E_{pair}$ & $E_{tot}$ & $E_{tot}$+$\Delta E$ &\; $E^{vor}$ \; &  \; $E_{pinning}$ \; \\
       \hline
     Pinned   &  5.65   &   \; 2261.44  \; &  \; -1897.82 \;   & \; 7286.06  \;  & \;  7358.10    \; &  \multirow{2}{*}{116.90  } &  \\         \cline{1-6}
     Nucleus  &  5.60   &      2274.19     &     -2161.68      &    7241.19      &            &                         & \multirow{2}{*}{16.33  } \\         \cline{1-7}
     Vortex   &  5.65   &      2128.52     &     -2026.77      &    8249.63      &      8286.71       &  \multirow{2}{*}{100.57  } & \\         \cline{1-6}
     Uniform  &  5.60   &      2135.08     &     -2227.12      &    8186.13      &            &                         & \\         \hline
       \end{tabular}
    \end{center}
\end{table}

 \begin{table}[h!]
    \begin{center}
       \begin{tabular}{|c|c|c|c|c|c|c|c|}
       \hline
\multicolumn{8}{|c|}{SkM* \;\;\; $n_{\infty}$=0.038 fm$^{-3}$}\\
       \hline
       & $\lambda$  & N & $E_{pair}$ & $E_{tot}$ & $E_{tot}$+$\Delta E$ &\; $E^{vor}$ \; &  \; $E_{pinning}$ \; \\
       \hline
     Pinned  & 7.55  &  \; 3481.75 \; &  \; -1857.15 \; & \; 15563.57 \;  & \;15566.52 \; &  \multirow{2}{*}{154.43} &  \\         \cline{1-6}
     Nucleus  & 7.5  &  3482.14 &  -2130.98 & 15412.09   &       & & \multirow{2}{*}{18.27} \\         \cline{1-7}
     Vortex  & 7.55  &  3342.92 &  -1965.67 & 16476.96  &   16486.52  &  \multirow{2}{*}{136.16} & \\         \cline{1-6}
     Uniform  & 7.5  &  3344.18  &  -2210.59 & 16350.36   &            &   & \\         \hline
       \end{tabular}
    \end{center}
    \caption{
The same as in Table \ref{Tab_Sly4_11.3}, but for the SkM* interaction.}    
\label{Tab_SkM*_11.3}
\end{table}

\clearpage

\section{Comparison with previous models}

The density dependence of the pinning energy presented above is
remarkably different from that obtained in previous works.

Epstein and Baym \cite{Epstein88} calculated the pinning energy solving
the Ginzburg-Landau equation.  They 
assumed that in the pinned
configuration 'the order parameter in the nucleus would be similar to that for the 
unperturbed vortex line'.
They calculated the kinetic energy associated with  the vortex axis at a distance $s$ from
the center of the nucleus, assuming the same functional form 
and spatial dependence for the order
parameter associated with the vortex, with or without the presence of the nucleus.
Due to their basic assumption, 
the difference in energy between the pinned ($s=0$) and the interstitial ($s \to \infty$) configurations
only involved the condensation energy (cf. Eqs. (5.2), (A13) and (C3) of their paper).
As a consequence, the pinning configuration becomes favoured when less pairing energy 
is suppressed by the vortex, than in uniform matter.  This in turn implies, that the 
pairing gap inside the nucleus must be smaller than in the outer gas: this happens
in their calculation for all  densities larger than 10$^{12.98}$ g  cm$^{-3}$, or 0.0055 fm$^{-3}$.

\begin{figure}
\includegraphics{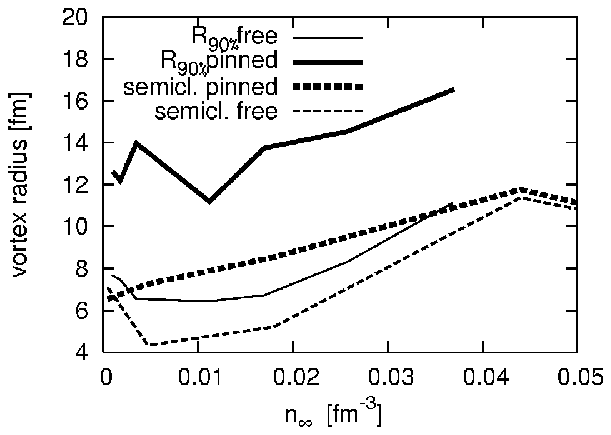}
\caption{The radii of the free and pinned vortex according to our quantal 
model (cf. Fig. \ref{fig:raggi-parag}) calculated with the SLy4 and with the SkM$^*$
mean fields are compared with the radii obtained in the semiclassical 
model of ref. \cite{DonatiPizzo}} 
\label{fig:raggi}
\end{figure}

The results obtained by Epstein and Baym are qualitatively similar to those obtained
in the semiclassical model  studied 
by  Donati and Pizzochero \cite{DonatiPizzo} 
using the Argonne force, who found antipinning at low densities and pinning
for densities larger than about 0.01 fm$^{-3}$ (cf. Fig. \ref{fig:eflowcond_hfb}(c)   
and Fig. \ref{figSemi}).
As we discuss in Appendix C, the differences between the quantal and the semiclassical
results are not due to differences
in the definition of the pinning energy.
In order to have some better insight,
we recall that the pinning energy is the difference between the
excitation energy $E_{unif}^{vor}$ associated with  a vortex created in uniform matter
 and the excitation
energy $E_{nuc}^{vor}$ associated with  a vortex built on the nuclear volume
(cf. Eq. (\ref{eq:ecost})).
In the model of ref. \cite{DonatiPizzo} these excitation energies result from two 
essential contributions. One of them is the condensation
energy $E_{cond}$, associated with the pairing fields calculated 
with and without the presence
of the vortex. The other is the kinetic energy $E_{flow}$ associated with the 
velocity field created by the vortex. 
For each of the four relevant  configuration (pinned, nucleus, vortex, uniform), 
$E_{flow}$ and $E_{cond}$ are  obtained 
integrating respectively  the 
kinetic energy density 1/2 $\Phi^2/n $, 
and the condensation energy density
-$3/8 \left [ n(\rho,z) \Delta^2(\rho,z)/(\lambda - U^{HF}(\rho,z)) \right]$.
The excitation energy associated with the pinned vortex  is estimated as 
$E_{nuc}^{vor} = [E_{flow} + E_{cond}]_{pinned} -[E_{flow} + E_{cond}]_{nucleus}$.
Similarly,
$E_{unif}^{vor} = [E_{flow} + E_{cond}]_{vortex} -[E_{flow} + E_{cond}]_{uniform}$.
Finally, the pinning energy is obtained as
$E_{pinning}= E_{nuc}^{vor} - E_{unif}^{vor}$.
The contribution of $E_{flow}$ to the pinning energy is then 
$E_{flow} \equiv [E_{flow}]_{pinned} -[E_{flow}]_{nucleus} -
[E_{flow}]_{vortex} +[E_{flow}]_{uniform}$, and similarly for $E_{cond}$.
The values of $E_{flow}$ and $E_{cond}$ obtained in
the semiclassical model of Ref.\cite{DonatiPizzo} 
are shown in Fig. \ref{fig:eflowcond_hfb}(c) as a function of density. 
It is seen that $E_{cond}$ is negative for densities larger than 
about 0.01 fm $^{-3}$. This is associated with the fact 
that the pairing
gap in the nucleus is smaller than in the external neutron gas.
Consequently, creating a vortex passing through the  nucleus destroys less
pairing energy than creating a vortex in uniform matter. 
The gain in condensation energy is largest, when the pairing gap in the 
neutron gas is maximum ($k_F \sim $ 0.8 fm$^{-1}$, or $n \sim 0.02$ fm$^{-3}$,
cf. Fig. \ref{fig:delta_NM}).
The radius of the vortex
core is 2-3 fm larger in the pinned than in the uniform case (cf. 
Fig. \ref{fig:raggi});  this leads to
a modest reduction of the kinetic energy in the pinned case. 
As a consequence the contribution of $E_{flow}$ is negative and 
rather small.  

In the semiclassical model, the radius of the vortex core is determined requiring that 
the absolute value of the condensation energy density be equal to the value of
the kinetic energy density associated with the vortex flow. 
In uniform matter the vortex core  is equal to the coherence length  $\xi$ which is of the order,
or larger than the nuclear radius $R_0$ . 
At high neutron densities in the crust $\xi > R_0$;  
the nucleus is included in the
vortex core and cannot influence its properties. The semiclassical model is
then very similar to Epstein and Baym's and the pinning energy is only due to
the contribution from the condensation energy (cf. Fig. \ref{fig:eflowcond_hfb}(c)). At lower neutron density,
 $\xi \sim R_0$. In this case the kinetic energy density 
increases  on the nuclear surface leading to an  
increase  of the core radius of 2-3 fm.   In this case the pinning energy receives a contribution
also from the kinetic energy.

In Fig. \ref{fig:eflowcond_hfb}(a) and (b)  we show the quantities  
$E_{flow}$ and $E_{cond}$
obtained from our calculated potentials, pairing gaps, currents and densities
associated with the SLy4 and SkM$^*$ mean fields. 
For densities lower than about  0.02 fm$^{-3}$,
the  results obtained with the two mean fields are rather similar, and are
substantially different from the semiclassical results:  
both  $E_{flow}$ and $E_{cond}$ are much larger in absolute value
and $E_{cond}$ takes positive values.
This difference is directly related to the much larger vortex radius 
we calculated in the pinned configuration (cf. Fig.\ref{fig:raggi-parag}). 
In turn, this is  associated with shell effects, which have been discussed
above in Sec. IV and which are not considered in the semiclassical model, 
As a consequence, in our calculations pairing is suppressed
not only inside the nuclear volume,
but also in a large layer (6-7 fm thick) 
in  the interface region between the nuclear surface and the neutron gas
(cf. Fig. \ref{fig:pair-2d-ent-nuc}). This
changes the energy balance as compared to the semiclassical model, 
and makes it less favourable to build the vortex on the nucleus. One should also
add that, due to  proximity effects, the pairing field does not vanish in the 
nuclear volume, as it happens in the Local Density  Approximation with the Argonne
potential (cf. Fig. \ref{fig:delta1}-\ref{fig:delta3}) \cite{Pizzo_APJ}.    
On the other hand, the large core radius implies the suppression of the 
velocity field in a large volume (cf. Fig. \ref{fig:dens-ent-nuc}), 
leading to large negative
values of  $E_{flow}$.
 
At densities larger than about $0.02$ fm$^{-3}$, one can notice 
a change in the density dependence of $E_{flow}$ and $E_{cond}$ 
in the case of the SkM$^*$ interaction: $E_{flow}$ increases and 
$E_{cond}$  decreases, reaching values closer to the semiclassical ones. 
This is related to the weakening  of shell effects for $\lambda >$ 10 MeV with
this interaction  (cf. Fig. \ref{fig:resonances}).

While in the semiclassical model the pinning energy is essentially  the sum of 
$E_{flow}$ and $E_{cond}$, in our self-consistent calculations, 
the changes in the mean field  induced by the vortex  play an
important  role.   This can be seen in Figs. \ref{fig:eflowcond_hfb}(a),(b), 
where we compare 
the pinning energy  with the sum  $E_{flow} + E_{cond}$.
The contributions of the changes in the mean field  to the pinning energy 
are of the  order of 5-10 MeV: they
are smaller than the range of values of $E_{flow}$ or 
$E_{cond}$ but are relevant for a quantitative assessment of the pinning
energy, leading in some cases to a change of its sign.


\begin{figure}
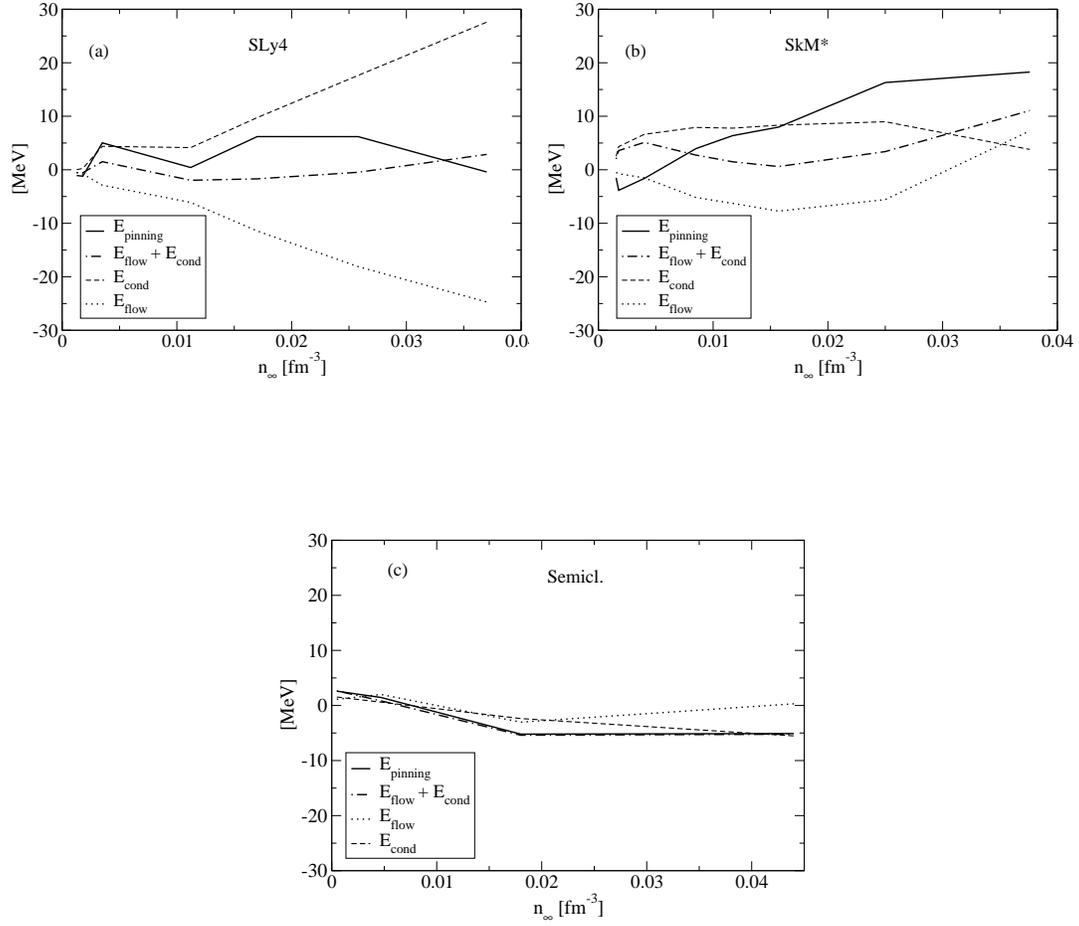

\includegraphics[width=7cm]{eflowecond_sly4.eps}
\vspace{2cm}
\includegraphics[width=7cm]{eflowecond_skm.eps}
\vspace{2cm}
\includegraphics[width=7cm]{eflowecond_semi.eps}
\caption{Pinning energy, flow energy and condensation
energy (in MeV) derived from the HFB calculations using 
the SLy4 mean field (a)
or the SkM$^*$ mean field (b). In (c) we show the same  quantities
calculated in the semiclassical model. }
\label{fig:eflowcond_hfb}
\end{figure}

\section{Conclusions}

Vortices are one of the clearest examples of the quantal nature
of superfluids. In particular of Fermi superfluids made out
of a large number  of overlapping Cooper pairs. Rotation can
only happen by giving one ($\nu=1$) or more ($\nu=2 ,...$) units of 
angular momentum to each Cooper pair. It is then not surprising that
the properties of these rotonic  excitations in a rather complex
many-body system like that of finite nuclei forming a Coulomb lattice
immersed in a sea of free neutrons display a complex dynamics, reflecting
simultaneously the variety of quantal behaviour associated with both 
finite and infinite systems. 
In particular spatial quantization typical of finite many-body systems 
which provides special bunchiness to the single-particle levels and 
is responsible not only for the special stability of certain nuclear species,
but also determines the structure of single-particle resonances in the continuum:
nuclear structure and nuclear reaction aspects of  the same phenomenon.

The effective mass ($m_k$) associated with the nucleon-nucleon interactions used
to describe the system makes these effects more or less important depending 
on whether $m_k$ is smaller than, or close to the bare mass $m$, in keeping with 
the fact that the associated mean fields are deeper or shallower, leading to 
larger or smaller  energy separations between resonances of different parities,
respectively.

While the associated  results for the pinning energy are quite different, 
in the first case ($m_k < m$) 
producing a bell shape  like pinning energy behaviour as a function of density,
and in the second case  ($m_k \sim m$) predicting a linearly increasing behaviour, both are
quite different from the semiclassical result. Consequently, while we cannot say which 
of the two behaviours is correct  due to the existing latitude in the parametrization 
of effective nuclear forces, 
one can conclude that there is likely no alternative to a full quantal treatment 
of the nucleus-vortex interaction in the determination of the pinning energies.
To become quantitative, one needs  to consider on par with
the bare nucleon-nucleon interaction, the induced interaction arising from 
the exchange of density and spin modes, taking into account both self-energy,
vertex and induced interaction type of processes.
There is a formidable task lying ahead, if one wants to become quantitative and realistic 
concerning the vortex-nucleus interaction in the situation typical of the inner crust of a neutron
star. We believe that one can attack this problem with good 
chances of success capitalizing on  the results obtained in the present work.

\section{Appendix A}\label{Appendice_A}

The HFB equations can be written in cylindrical coordinates as 
\begin{equation}\label{HFB2a}
\begin{array}{l}
 ~~(\hat T + U^{HF}(\rho, z)-\lambda) u_{qm}(\rho,z,\phi)+ 
 \Delta(\rho,z,\phi) v_{qm}(\rho,z,\phi) = 
 E_{qm} u_{qm}(\rho,z,\phi)\\
~\Delta^*(\rho,\phi,z) u_{qm}(\rho,z,\phi)
-(\hat T + U^{HF}(\rho,z)-\lambda)v_{qm}(\rho,z,\phi) = 
E_{qm} v_{qm}(\rho,z,\phi).
\end{array}
\end{equation}
The kinetic energy operator, $\hat T = - \frac{\hbar^2}{2} \nabla
\frac{1}{\mu(\rho,z)} \nabla$ contains the effective mass $\mu(\rho,z)$ associated
with the adopted Skyrme interaction,  while 
$\Delta$ is the pairing field
and $U^{HF}$ is the mean field
potential calculated using the Skyrme functional. We include all the 
terms of the Skyrme functional listed in Eqs.(2.1) of ref. \cite{Chabanat},
except for the spin-orbit and the spin gradient terms.
The index $q$ 
labels the different quasiparticle energies $E$ and amplitudes $u,v$
associated with  a projection $m$ of the orbital angular momentum along the $z-$axis.
Since we neglect the spin-orbit interaction, the spin degree of freedom
is taken into account simply by the degeneracy factor $g=2$.

For a system with axial symmetry respect to the $z$-axiz, the characteristic ansatz for the study
of the vortex is \cite{Degennes,Bohrmott62,Gigi}
\begin{equation}
\Delta(\rho,z,\phi)= \Delta(\rho,z) e^{i\nu \phi},
\label{ansatz}
\end{equation}
where $\nu=0,1,2 ...$ is the vortex index. 

The normal and abnormal densities are given by
\begin{equation}
n(\rho,z,\phi)= 2 \sum_{qm} v_{qm}(\rho,z,\phi)v_{qm}^*(\rho,z,\phi)
\end{equation}
and by
\begin{equation}
\kappa(\rho,z,\phi)= \sum_{qm} u_{qm}(\rho,z,\phi)v_{qm}^*(\rho,z,\phi),
\label{abnorm}
\end{equation}
and the gap is given by 
\begin{equation}
\Delta(\rho,z,\phi)= - V_{pair} \kappa(\rho,z,\phi).
\label{gapeq}
\end{equation}
The Hartree-Fock field $U^{HF}$
is calculated according to Eq. (2.8) of ref. \cite{Chabanat}.
We assume the following form for the dependence of the 
quasiparticle amplitudes on the angle $\phi$:
\begin{equation}
\begin{array}{l}
u_{qm}(\rho,z,\phi) \sim  e^{im\phi} \quad ; \quad \\
v_{qm}(\rho,z,\phi) \sim  e^{i(m-\nu)\phi}.
\end{array}
\end{equation}
In this way, Eqs.(\ref{abnorm}) and (\ref{gapeq}) lead to a pairing
field which is consistent with the ansatz (\ref{ansatz}).

We shall only solve the axially symmetric equation for the neutrons.  
According to the study of Negele and Vautherin \cite{NegeleVaut}, who neglected 
pairing in the minimization of their
energy functional,  the favoured number of protons is mostly 
determined by shell effects, and varies from 
$Z=40$ to $Z=50$. Because we neglect
the spin-orbit interaction, the natural choice for us is to assume $Z=40$, 
corresponding to a shell closure.
However, more recent studies including pairing  have shown that the value  
of $Z$ can depend on the 
details of the functional, being sometimes rather different 
from Negele and Vautherin's solution \cite{Baldo}. 
We have not investigated whether a different choice of $Z$ would lead to important 
changes in our results.  
The protons are deeply bound and  for this reason 
they are not expected to be influenced significantly by the vortex, and thus the proton field 
should essentially keep spherical symmetry. We then perform a simple spherical HF calculation  
for the 
protons, enclosing them in a spherical box of radius 15 fm, and   carry 
out a spherical average of the neutron part of the potential felt by the protons
(in particular at each iteration we replace  
the neutron density $n(\rho,z)$ 
by an isotropic $n_{av}(r)$ 
obtained averaging  $n(\rho,z)$  on the surface of a sphere of radius 
$r$).

We shall solve the HFB equations in a cylindrical cell of radius $\rho_{box}$ and
height $h_{box}$ (cf. Fig. \ref{fig:cilindro}), imposing the boundary
condition $u_{qm}(\rho,z,\phi)= v_{qm}(\rho,z,\phi)=0$ for $\rho= \rho_{box}$, or $z = \pm h_{box}/2$.
We shall expand the quasiparticle amplitudes
on a single-particle basis writing
\begin{eqnarray}
&u_{qm}(\rho,z,\phi)&= \sum_{ih} \; U_{hi;qm} \; \phi_{hm}(\rho) \;\psi_i(z) \; e^{im\phi} \\
\nonumber
&v_{qm}(\rho,z,\phi)& = \sum_{ih} \; V_{hi;qm} \;\phi_{h(m-\nu)}(\rho) \; \psi_i(z) \; e^{i(m-\nu)\phi}. 
\label{expans}
\end{eqnarray}
In the $z-$direction we have taken  the set of standing waves 
\begin{equation}
\psi_i(z)= \sqrt\frac{2}{h_{box}}{\rm sin}(k_i(z+h_{box}/2)), \quad k_i = \frac{\pi}{h_{box}}, \frac{2\pi}{h_{box}}, ... 
\end{equation}
In the $\rho$ direction we have taken the set of solutions associated with the
Schr\"odinger equation for free neutrons:
\begin{equation}
\left [ - \frac{\hbar^2}{2\mu_0} \frac{1}{\rho} \frac{d}{d\rho}\rho\frac{d}{d\rho} 
+ \frac{\hbar^2 m^2}{2\mu_0 \rho^2}\right] \phi_{hm}(\rho) = \epsilon_{hm}
\phi_{hm}(\rho),
\end{equation}
where $\mu_0$ indicates the bare nucleon mass.
Projecting the HFB equations on the chosen single-particle basis, we obtain two
algebraic equations for the expansion coefficients $U_{hi;qm},V_{hi;qm}$:
\begin{equation}
\begin{array}{l}
\sum_{hi} (K_{lhij,m} - \lambda) U_{hi;qm} + \sum_{hi} U^{HF}_{lhij,m} U_{hi;qm} + 
\sum_{hi} \Delta_{lhij,m} V_{hi;qm} = E_{qm} U_{lj;qm} \\
\sum_{hi} \Delta_{hlij,m} U_{hi;qm} - \sum_{hi} (K_{lhij,m-\nu} - \lambda) V_{hi,qm} 
- \sum_{hi} U^{HF}_{lhij,m-\nu} V_{hi;qm} = E_{qm} V_{lj;qm}.
\end{array}
\end{equation}
The matrix elements of the kinetic energy, of the mean field and of the pairing field
are given by
\begin{equation}
K_{lhij,m}= 2 \pi \int \; d\rho dz \rho  \; \phi_{lm}(\rho) \psi_j(z) \left[
\left( \frac{\mu_0}{\mu(\rho,z)} \epsilon_{lm} + 
\frac{\hbar^2 k_j^2}{2 \mu(\rho,z)} \right)
- \frac{\hbar^2}{2} ( \nabla \frac{1}{\mu(\rho,z)})\cdot \nabla \right] \phi_{hm}(\rho) \psi_i(z);
\end{equation}
\begin{equation}
U^{HF}_{lhij,m}= 2 \pi \int \;  d\rho dz \rho\; \phi_{lm}(\rho) \psi_j(z) U^{HF}(\rho,z) 
\phi_{hm}\psi_i(z);
\end{equation}
\begin{equation}
\Delta_{lhij,m}= 2 \pi \int \; d\rho dz \rho  \; \phi_{lm}(\rho) \psi_j(z) \Delta(\rho,z) 
\phi_{h(m-\nu)}\psi_i(z). 
\end{equation}

The total energy is given by $E_{tot} = E_{HF} + E_{pair}$, where is 
$E_{HF}$ calculated according to Eq. (2.2) of ref. \cite{Chabanat}
and $E_{pair}$ is obtained by
\begin{equation}
E_{pair} = - \frac{1}{2}  \int d\rho dz d\phi \; \rho \kappa^*(\rho,z,\phi)
\Delta(\rho,z,\phi) 
\end{equation}

\begin{figure}[h!]
\vspace{1cm}
\includegraphics[width=7.5cm]{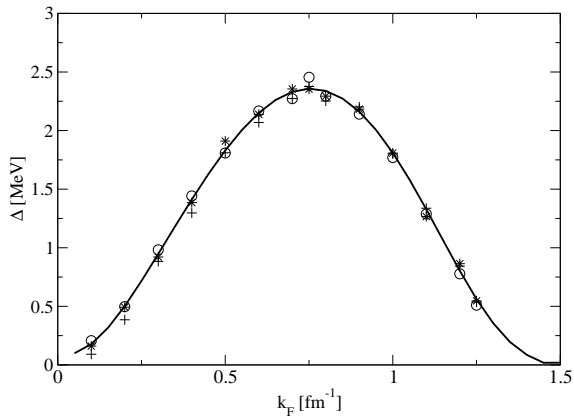}
\caption{ 
Comparison between the gap obtained requiring that the
radial single-particle wavefunctions $u_L(r)$ vanish 
at the edge of the spherical box for all 
orbital angular momenta $L$ (crosses), 
or that $u_L(R)$ (for $L$ even) and $du_L/dR (L$ odd) vanish at the edge of the box (stars),
or that $u_L(R)$ (for $L$ odd) and $du_L/dR (L$  even) 
vanish at the edge of the box (circles). The solid curve shows the pairing gap calculated in infinite
neutron matter.}
\label{fig:campane}
\end{figure}

In the literature, one finds different choices concerning the boundary conditions
applied to calculations in  Wigner-Seitz cells.
In Fig. \ref{fig:campane} we compare  the values of the gap at the Fermi energy
$\Delta_F$ obtained with the SLy4 mean field as a function of the  Fermi  momentum in a  spherical box
of radius $R$= 30 fm,  with those independently calculated in infinite neutron matter and already shown in
Fig. \ref{fig:delta_NM}. 
We have used  three different boundary conditions
for the single-particle wavefunctions: 
(a)  all the wavefunctions vanish at the edge of the cell (the boundary condition adopted in this paper)
(b)  the wavefunctions of even orbital angular momentum and the 
derivatives of the wavefunctions of odd orbital angular momentum vanish at the edge of the cell 
(the boundary condition adopted by Negele and Vautherin);
(c) the same as (b), but exchanging the role of odd and even angular momenta. 
It is seen that the agreement with the result obtained in infinite matter
is of the same quality in the three cases (cf. also the discussion in ref. \cite{Montani}). 

We have also checked that with our boundary conditions (a) we reproduce accurately the values of the
energy per particle in neutron matter 
calculated analytically  with the Skyrme interactions we have considered.
As we discussed in Section IVB, using boundary conditions (a) we have also been able to reproduce the 
pairing gap associated with the vortex ($\nu=1$),  calculated by  Bulgac and Yu in infinite neutron matter. 

For calculations with protons, that is, with a nucleus at the center of the box, we have found that 
in all the calculations reported in this paper, the pairing gap at large distances ($\rho > $ 20 fm)
resumes the value calculated in uniform neutron matter at the relevant density. Obviously 
with boundary conditions (a) the pairing gap drop quickly to zero at the edge of the box. 
We have checked that  this does not influence the calculations of the pinning energy, which is obtained subtracting the energies
of different configurations, as discussed in the next Section.

In all our calculations we have used boxes with radii equal to, or larger than, 30 fm.
In other works, where one has to calculate the pairing gap inside the actual Wigner-Seitz
cells, one finds some clear dependence on the boundary
condition for cells smaller than about 20 fm \cite{Baldo}. In this case, one should consider in detail 
the influence of the Coulomb lattice \cite{Chamel}.

\section{Appendix B}\label{Appendice_B}

In this Appendix we give some details concerning the numerical accuracy of the  calculation
of the pinning energy. This is important, because the pinning energy, which is of the order of a few MeV, 
results from the difference of the total 
energies of the various configurations, which are of the order of a  few GeV. 
The main parameter that influence the accuracy of our calculation 
is the mesh size. Furthermore, we have to check that the size of the box 
( that is, its height $h_{box}$ and its radius $\rho_{box}$)
is sufficiently  large, to obtain stable results. As we discussed in Section IVC, we expect that 
values like $h_{box} = $ 40 fm and $\rho_{box}$ = 30 fm are sufficiently large  to guarantee that the 
pinning process is not significantly influenced by the boundary. In fact, the typical 
vortex radius, as measured by $R_{90\%}$, is about 15 fm (cf. Fig. \ref{fig:raggi-parag}).
Moreover, the calculation of the pinning energy is more sensitive to the value of
$\rho_{box}$ than to $h_{box}$, because 
the distortion of the pinned vortex is much more relevant on the equator (cf. for example 
Fig. \ref{fig:pair3d5.8}). 
We have studied in a few cases the dependence of the computed values of the pinning energy 
on $\rho_{box}$, making calculations of the relevant configurations  with different boxes. 
In Tab. \ref{tab:s2-mesh} and in Tab. \ref{tab:SLy4-mesh} we show  
the values of the pinning energy calculated with different
box sizes for the SII and SLy4 interactions. 
The values show very modest fluctuations with the box size 
for the points associated with densities up to about $n < 0.017$ fm$^{-3}$, for which
we can estimate an uncertainty of the order of 200 keV. For the highest density
points, the fluctuations become larger, but do not exceed 1-2 MeV. This is related to
the fact that the vortex radius increases with density, particularly with the SII and SLy4
interactions (cf. Fig. \ref{fig:raggi-parag}). 

 \begin{table}[h!]
     \begin{center}
        \begin{tabular}{|c|c|c|c|c|}
        \hline 
         $\lambda$ [MeV]& $n_{\infty}$ [fm$^{-3}$] & $\rho_{box}$= 30 fm  & $\rho_{box}$= 25 fm \\ 
        \hline 
         3   & 0.004    &-1.51  &        -1.65 \\
        \hline
         5.8 & 0.011    &    3.26  &           3.46\\
        \hline 
         7   & 0.016    &          1.64 &             1.34\\
        \hline
        11.3 & 0.035    &        2.34  &       1.77  \\
        \hline 
        \end{tabular} 
     \end{center}
    \caption{Pinning energy (in MeV) calculated with the  SII 
interaction at various values of the chemical potential and of the asymptotic neutron densities, 
for two different radii of the cylindrical box.}
\label{tab:s2-mesh}
\end{table}

  \begin{table}[h!]
     \begin{center}
        \begin{tabular}{|c|c|c|c|c|c|c|c|}
        \hline 
 $\lambda$ [MeV] & $n_{\infty}$ [fm$^{-3}$]    &  $\rho_{box}$= 35 fm  &  $\rho_{box}$= 30 fm   &     $\rho_{box}$= 25 fm \\ 
        \hline 
 5.7  &  0.011   &   0.45       &   0.42    &        -0.3 \\
        \hline 
 7    &  0.017  &   6.07       &     6.21  &      5.54    \\
        \hline 
 10   &  0.037  &   1.64       &    -0.41  &         1.78   \\
        \hline 
        \end{tabular} 
     \end{center}
    \caption{
Pinning energy (in MeV) calculated with the  SLy4 
interaction at various values of the chemical potential and of the asymptotic neutron densities, 
for three different radii of the cylindrical box.}
\label{tab:SLy4-mesh}
\end{table}

 \begin{figure}[!h]
\centering
\includegraphics[width=6cm]{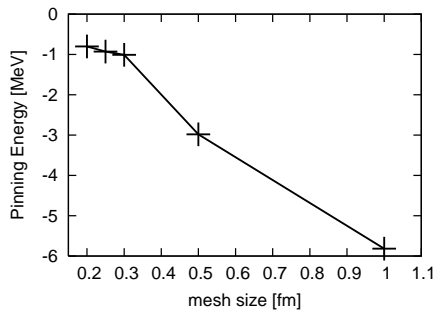}
	\caption{Pinning energy calculated with the SLy4 interaction at the density $n_{\infty} = 0.001$ fm$^{-3}$ as a function of the mesh size
used in the $\rho$-coordinate.  } 
\label{fig:mesh-pinning}
\end{figure}

The size of the mesh in $\rho$ and $z$ used in the code that solves the HFB equations and calculates the energy
is crucial for the accuracy of the calculations. 
We have studied the influence of the mesh (for the $\rho$ coordinate) in the case of the SLy4 interaction 
at the density $n_{\infty}= $ 0.0013  {\rm fm}$^{-3}$
and the results are shown in Fig. \ref{fig:mesh-pinning}. The fluctuations of the pinning energy for
a mesh smaller than 0.3 MeV  are of the order of hundreds of keV ($\lesssim$ 0.2 MeV).

\section{Appendix C}\label{Appendice_C}

In a semiclassical study of the problem under discussion \cite{DonatiPizzo} 
the pinning energy was defined as the difference  
of the energies associated with the pinned and the interstitial 
configurations, taking explicitly into account 
the presence of one of the  neighbouring nuclei
(cf. the left part of Fig. \ref{pinn_schem_catania}).
This definition of pinning energy reduces to ours,  
if the interaction of the vortex with the second nucleus is negligible. 
We have  tested the importance of 
the neighboring nucleus within the semiclassical model, for the range of densities
we have considered. 
We first computed the 
energy of each of the relevant configurations using  the semiclassical theory, 
the two pairing interactions (Argonne and Gogny) 
and the parameters presented  in Ref.~\cite{DonatiPizzo} , 
instead of solving the HFB equations.
We then calculated  the pinning energy using the scheme proposed in
Ref.~\cite{DonatiPizzo}: the resulting values are
connected by the dashed curves in Fig. \ref{figSemi}, 
and they  reproduce quite accurately
the  values reported in ref.~\cite{DonatiPizzo}, already shown in Fig. \ref{fig:eflowcond_hfb}(c)
in the case of the Argonne interaction.
In Fig. \ref{figSemi} we also show the pinning energies calculated 
using the semiclassical energies, but using our scheme
(cf. the right part of Fig. {\ref{pinn_schem_catania}).
The two schemes yield an almost equal result,
showing that the differences existing between the quantal and
the semiclassical calculations
are not due to the geometrical scheme used  to derive the pinning energy,
and that the contribution of the second nucleus plays in fact 
a marginal role, also following the formalism of ref. ~\cite{DonatiPizzo}. 
The situation may change at larger densities, where the vortex radius
becomes of the order of the internuclear distance, as we have discussed
in Section IV C.

\begin{figure}
\epsfig{file=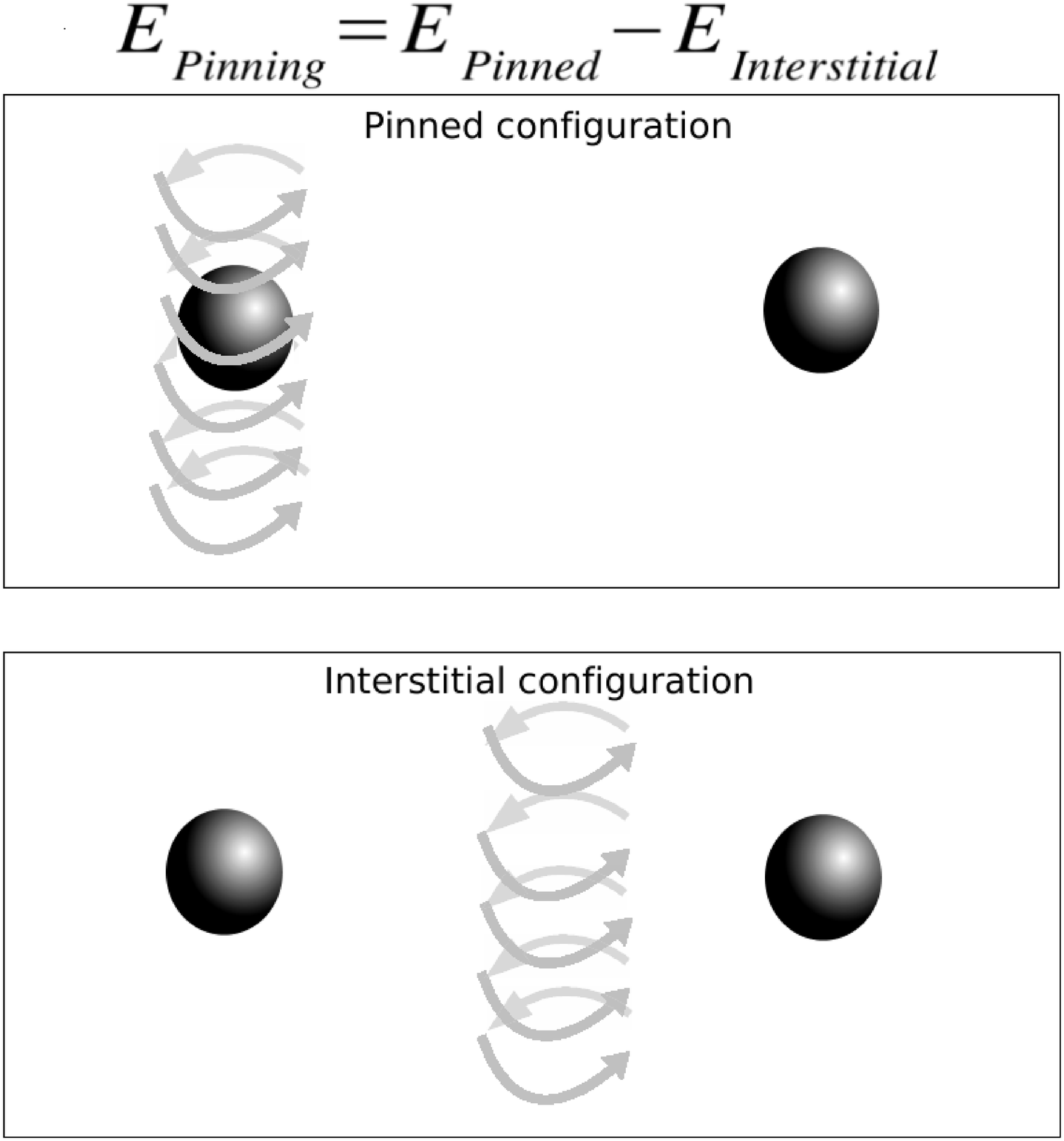, width=5cm, height=5cm}
\epsfig{file=bianco2, width=1cm}
\epsfig{file=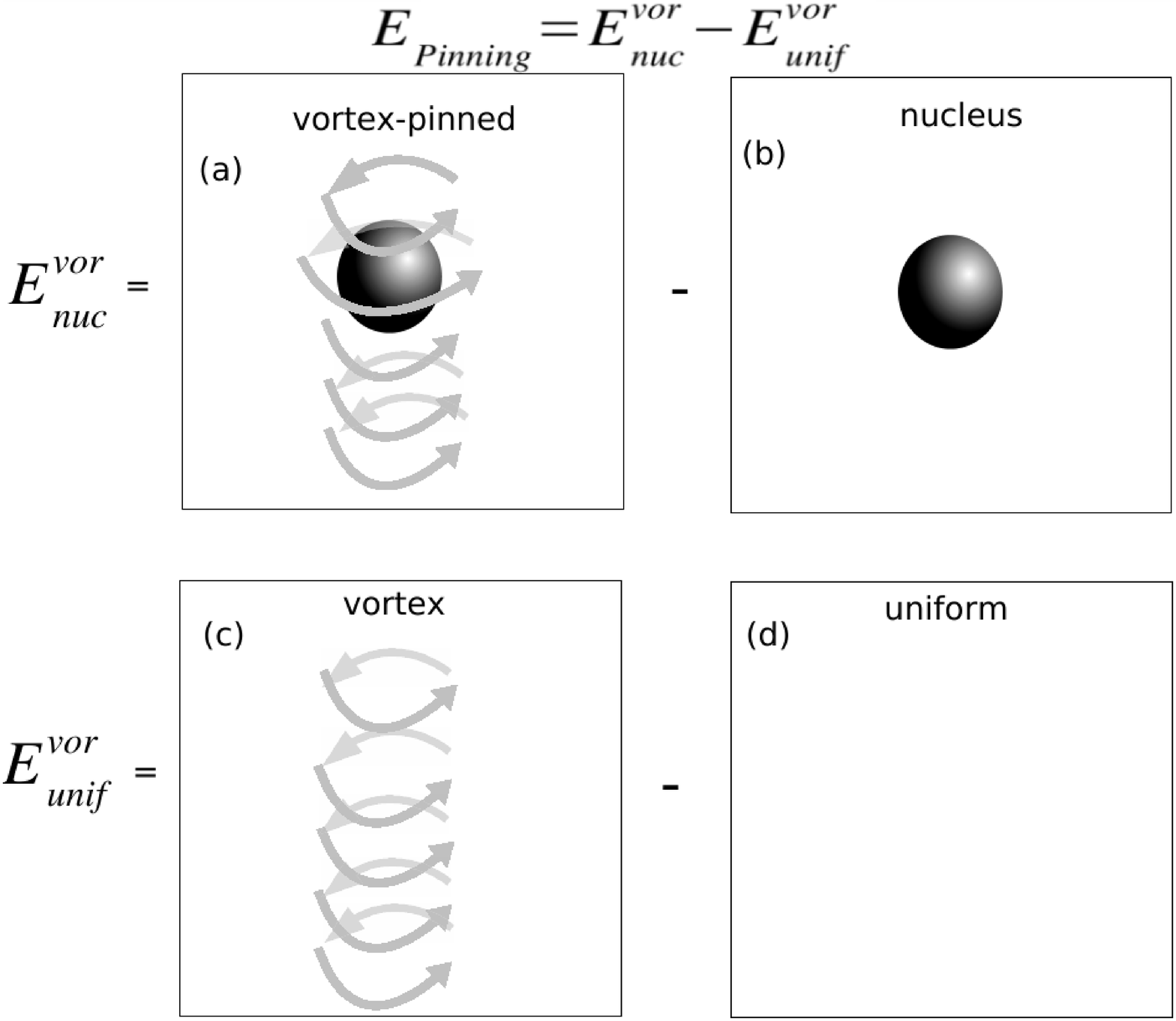, width=6cm}
\caption{(left) The scheme used by Donati and Pizzochero \cite{DonatiPizzo} where the pinning energy takes into account the presence of a neighboring nucleus; (right) the scheme we used, where the pinning energy is evaluated as the difference between the costs to
excite a vortex on a nucleus and in uniform matter.}
\label{pinn_schem_catania}
\end{figure}


\begin{figure}[h!]
\includegraphics{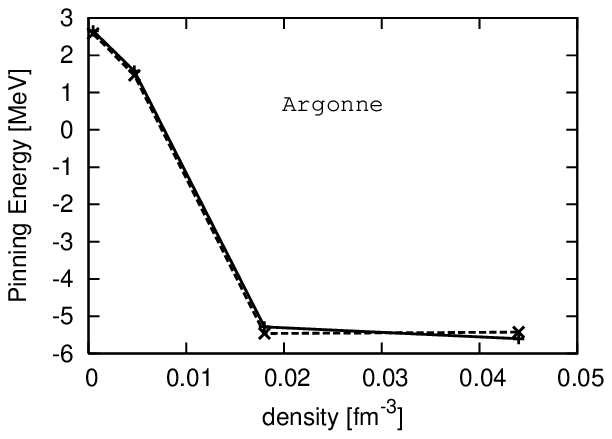}
\includegraphics{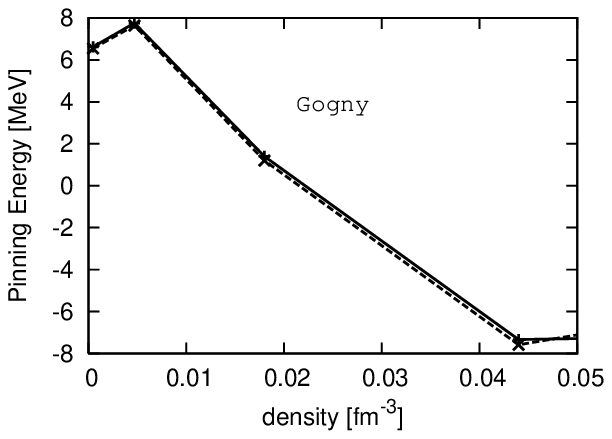}
\caption{Solid and dashed curves show respectively 
the pinning energies calculated using our scheme (Fig. \ref{pinn_schem}, right part),  
or the scheme of Ref.\cite{DonatiPizzo} (Fig. \ref{pinn_schem}, left part). 
In both cases, we have computed
the energies of the different configurations using the semiclassical model,
with the Argonne pairing interaction (left panel) 
or the Gogny pairing interaction (right panel).}   
\label{figSemi}
\end{figure}

\end{document}